\documentclass[journal,comsoc]{IEEEtran}

\usepackage[utf8]{inputenc}
\usepackage[english]{babel}
\usepackage[T1]{fontenc}% optional T1 font encoding

\usepackage{cite}
   \usepackage[pdftex]{graphicx}
   \DeclareGraphicsExtensions{.pdf,.jpeg,.png}
   \usepackage{everysel}
   \usepackage{keyval}
   \usepackage{ragged2e}
\usepackage{epstopdf}

\usepackage{amsmath}
\usepackage{mathtools}

\usepackage{textcomp}
\usepackage{siunitx}
\usepackage{gensymb}
\usepackage{balance}
\usepackage{bm}
\usepackage{xr}
\usepackage{array}
\usepackage{pbox}
\usepackage{subfigure}
\usepackage{booktabs}
\usepackage{xcolor}
\usepackage[font=footnotesize]{caption}
\usepackage[font=footnotesize]{subcaption}
\usepackage{soul}

\usepackage{colortbl}

\usepackage{algorithmic}

\usepackage{float}

\usepackage[acronyms,nonumberlist,nopostdot,nomain,nogroupskip]{glossaries}
\newacronym{rfi}{RFI}{Radio Frequency Interference}
\newacronym{lb}{LB}{Lower Band}
\newacronym{ub}{UB}{Upper Band}
\newacronym{awg}{AWG}{Arbitrary Waveform Generator}
\newacronym{dso}{DSO}{Digital Storage Oscilloscope}
\newacronym{3gpp}{3GPP}{3rd Generation Partnership Project}
\newacronym{adc}{ADC}{analog-to-digital converter}
\newacronym{1g}{1G}{first generation}
\newacronym{5g}{5G}{fifth generation}
\newacronym{6g}{6G}{sixth generation}
\newacronym{aimd}{AIMD}{Additive Increase Multiplicative Decrease}
\newacronym{am}{AM}{Acknowledged Mode}
\newacronym{amc}{AMC}{Adaptive Modulation and Coding}
\newacronym{aqm}{AQM}{Active Queue Management}
\newacronym{awgn}{AGWN}{Additive White Gaussian Noise}
\newacronym{balia}{BALIA}{Balanced Link Adaptation}
\newacronym{bdp}{BDP}{Bandwidth-Delay Product}
\newacronym{bf}{BF}{Beamforming}
\newacronym{cc}{CC}{Congestion Control}
\newacronym{cdf}{CDF}{Cumulative Distribution Function}
\newacronym{cn}{CN}{Core Network}
\newacronym{cqi}{CQI}{Channel Quality Information}
\newacronym{cp}{CP}{Control Plane}
\newacronym{csirs}{CSI-RS}{Channel State Information - Reference Signal}
\newacronym{dc}{DC}{Dual Connectivity}
\newacronym{dce}{DCE}{Direct Code Execution}
\newacronym{dci}{DCI}{Downlink Control Information}
\newacronym{dl}{DL}{Downlink}
\newacronym{dmr}{DMR}{Deadline Miss Ratio}
\newacronym{dmrs}{DMRS}{DeModulation Reference Signal}
\newacronym{e2e}{E2E}{End-to-End}
\newacronym{ecn}{ECN}{Explicit Congestion Notification}
\newacronym{edf}{EDF}{Earliest Deadline First}
\newacronym{enb}{eNB}{evolved Node Base}
\newacronym{epc}{EPC}{Evolved Packet Core}
\newacronym{es}{ES}{Edge Server}
\newacronym{fdma}{FDMA}{Frequency Division Multiple Access}
\newacronym{fdd}{FDD}{frequency division duplexing}
\newacronym{fft}{FFT}{Fast Fourier Transform}
\newacronym[firstplural=Radio Access Technologies (RATs)]{rat}{RAT}{Radio Access Technology}
\newacronym{fs}{FS}{Fast Switching}
\newacronym{ftp}{FTP}{File Transfer Protocol}
\newacronym{gnb}{gNB}{Next Generation Node Base}
\newacronym{harq}{HARQ}{Hybrid Automatic Repeat reQuest}
\newacronym{hetnet}{HetNet}{Heterogeneous Network}
\newacronym{hh}{HH}{Hard Handover}
\newacronym{hol}{HOL}{Head-of-Line}
\newacronym{ia}{IA}{Initial Access}
\newacronym{imt}{IMT}{International Mobile Telecommunication}
\newacronym{los}{LoS}{line-of-sight}
\newacronym{nlos}{NLoS}{non-line-of-sight}
\newacronym{lte}{LTE}{Long Term Evolution}
\newacronym{m2m}{M2M}{Machine to Machine}
\newacronym{mac}{MAC}{Medium Access Control}
\newacronym{mc}{MC}{Multi-Connectivity}
\newacronym{mcs}{MCS}{Modulation and Coding Scheme}
\newacronym{mec}{MEC}{Mobile Edge Cloud}
\newacronym{mi}{MI}{Mutual Information}
\newacronym{mimo}{MIMO}{multiple-input multiple-output}
\newacronym{mmwave}{mmWave}{millimeter wave}
\newacronym{mptcp}{MPTCP}{Multipath TCP}
\newacronym{mr}{MR}{Maximum Rate}
\newacronym{mss}{MSS}{Maximum Segment Size}
\newacronym{mtd}{MTD}{Machine-Type Device}
\newacronym{mtu}{MTU}{Maximum Transmission Unit}
\newacronym{nfv}{NFV}{Network Function Virtualization}
\newacronym{nr}{NR}{New Radio}
\newacronym{ofdm}{OFDM}{orthogonal frequency division modulation}
\newacronym{dftso}{DFT-S-OFDM}{discrete-Fourier-transform-spread OFDM}
\newacronym{otfs}{OTFS}{orthogonal time frequency space}
\newacronym{papr}{PAPR}{Peak-to-Average Power Ratio}
\newacronym{pdcch}{PDCCH}{Physical Downlink Control Channel}
\newacronym{pdcp}{PDCP}{Packet Data Convergence Protocol}
\newacronym{pdsch}{PDSCH}{Physical Downlink Shared Channel}
\newacronym{pdu}{PDU}{Packet Data Unit}
\newacronym{pf}{PF}{Proportional Fair}
\newacronym{pgw}{PGW}{Packet Gateway}
\newacronym{phy}{PHY}{physical}
\newacronym{pbch}{PBCH}{Physical Broadcast Channel}
\newacronym[plural=\gls{mme}s,firstplural=Mobility Management Entities (MMEs)]{mme}{MME}{Mobility Management Entity}
\newacronym{prb}{PRB}{Physical Resource Block}
\newacronym{pss}{PSS}{Primary Synchronization Signal}
\newacronym{pucch}{PUCCH}{Physical Uplink Control Channel}
\newacronym{pusch}{PUSCH}{Physical Uplink Shared Channel}
\newacronym{rach}{RACH}{Random Access Channel}
\newacronym{ran}{RAN}{Radio Access Network}
\newacronym{red}{RED}{Random Early Detection}
\newacronym{rf}{RF}{Radio Frequency}
\newacronym{rlc}{RLC}{Radio Link Control}
\newacronym{rlf}{RLF}{Radio Link Failure}
\newacronym{rrc}{RRC}{Radio Resource Control}
\newacronym{rrm}{RRM}{Radio Resource Management}
\newacronym{rr}{RR}{Round Robin}
\newacronym{rs}{RS}{Remote Server}
\newacronym{rsrp}{RSRP}{Reference Signal Received Power}
\newacronym{rss}{RSS}{Received Signal Strength}
\newacronym{rtt}{RTT}{Round Trip Time}
\newacronym{rw}{RW}{Receive Window}
\newacronym{rx}{RX}{receiver}
\newacronym{sa}{SA}{standalone}
\newacronym{sack}{SACK}{Selective Acknowledgment}
\newacronym{sap}{SAP}{Service Access Point}
\newacronym{sch}{SCH}{Secondary Cell Handover}
\newacronym{scoot}{SCOOT}{Split Cycle Offset Optimization Technique}
\newacronym{sdma}{SDMA}{Spatial Division Multiple Access}
\newacronym{sinr}{SINR}{Signal to Interference plus Noise Ratio}
\newacronym{sm}{SM}{Saturation Mode}
\newacronym{snr}{SNR}{signal-to-noise ratio}
\newacronym{son}{SON}{Self-Organizing Network}
\newacronym{ss}{SS}{Synchronization Signal}
\newacronym{srs}{SRS}{Sounding Reference Signal}
\newacronym{sss}{SSS}{Secondary Synchronization Signal}
\newacronym{tb}{TB}{Transport Block}
\newacronym{tcp}{TCP}{Transmission Control Protocol}
\newacronym{tdd}{TDD}{time division duplexing}
\newacronym{tdma}{TDMA}{Time Division Multiple Access}
\newacronym{tfl}{TfL}{Transport for London}
\newacronym{tm}{TM}{Transparent Mode}
\newacronym{trp}{TRP}{Transmitter Receiver Pair}
\newacronym{tti}{TTI}{Transmission Time Interval}
\newacronym{ttt}{TTT}{Time-to-Trigger}
\newacronym{tx}{TX}{transmitter}
\newacronym{ue}{UE}{User Equipment}
\newacronym{ul}{UL}{Uplink}
\newacronym{uml}{UML}{Unified Modeling Language}
\newacronym{um}{UM}{Unacknowledged Mode}
\newacronym{utc}{UTC}{Urban Traffic Control}
\newacronym{vm}{VM}{Virtual Machine}
\newacronym{rsrq}{RSRQ}{Reference Signal Received Quality}
\newacronym{rssi}{RSSI}{Received Signal Strength Indicator}
\newacronym{crs}{CRS}{Cell Reference Signal}
\newacronym{nsa}{NSA}{Non Stand Alone}
\newacronym{mrdc}{MR-DC}{Multi \gls{rat} \gls{dc}}
\newacronym{endc}{EN-DC}{E-UTRAN-\gls{nr} \gls{dc}}
\newacronym{5gc}{5GC}{5G Core}
\newacronym{si}{SI}{Study Item}
\newacronym{iab}{IAB}{Integrated Access and Backhaul}
\newacronym{wf}{WF}{Wired-first}
\newacronym{hqf}{HQF}{Highest-quality-first}
\newacronym{mlr}{MLR}{Maximum-local-rate}
\newacronym{wbf}{WBF}{Wired Bias Function}
\newacronym{mib}{MIB}{Master Information Block}
\newacronym{sib}{SIB}{Secondary Information Block}
\newacronym{kpi}{KPI}{Key Performance Indicator}
\newacronym{ppp}{PPP}{Poisson Point Process}
\newacronym{gtp}{GTP}{GPRS Tunneling Protocol}
\newacronym{amf}{AMF}{Access and Mobility Management Function}
\newacronym{dash}{DASH}{Dynamic Adaptive Streaming over HTTP}
\newacronym{http}{HTTP}{HyperText Transfer Protocol}
\newacronym{udp}{UDP}{User Datagram Protocol}
\newacronym{cu}{CU}{Central Unit}
\newacronym{du}{DU}{Distributed Unit}
\newacronym{mt}{MT}{Mobile Termination}
\newacronym{sdap}{SDAP}{Service Data Adaptation Protocol}
\newacronym{tdm}{TDM}{Time Division Multiplexing}
\newacronym{fdm}{FDM}{Frequency Division Multiplexing}
\newacronym{sdm}{SDM}{Space Division Multiplexing}
\newacronym{dag}{DAG}{Directed Acyclic Graph}
\newacronym{st}{ST}{Spanning Tree}
\newacronym{uav}{UAV}{unmanned aerial vehicle}
\newacronym{ber}{BER}{bit error rate}
\newacronym{fcc}{FCC}{Federal Communications Commission}
\newacronym{itu}{ITU}{International Telecommunications Union}
\newacronym{spdt}{SPDT}{Single-Pole Double Throw}
\newacronym{mls}{MLS}{Microwave Limb Scanner}
\newacronym{api}{API}{Application Programming Interface}
\newacronym{ntp}{NTP}{Network Time Protocol}
\newacronym{rest}{REST}{Representational state transfer}
\newacronym{lo}{LO}{local oscillator}
\newacronym{if}{IF}{intermediate frequency}
\newacronym{qam}{QAM}{Quadrature Amplitude Modulation}
\newacronym{ummimo}{UM-MIMO}{Ultra Massive \gls{mimo}}
\newacronym{cbrs}{CBRS}{Citizen Broadband Radio Service}
\newacronym{noma}{NOMA}{Non-Orthogonal Multiple Access}
\newacronym{rre}{RR}{Radio Regulations}
\newacronym{wrc}{WRC}{World Radio Conference}
\newacronym{cept}{CEPT}{European Conference of Postal and Telecommunications Administrations}
\newacronym{ehf}{EHF}{Extremely High Frequencies}
\newacronym{uhf}{UHF}{Ultra High Frequencies}
\newacronym{shf}{SHF}{Super High Frequencies}
\newacronym{ras}{RAS}{Radio Astronomy Service}
\newacronym{eess}{EESS}{Earth-Exploration Satellite Service}
\newacronym{ses}{SES}{Space-Exploration Service}
\newacronym{dsss}{DSSS}{Direct Sequence Spread Spectrum}
\newacronym{cdma}{CDMA}{Code Division Multiple Access}
\newacronym{sll}{SLL}{side-lobe-levels}
\newacronym{bs}{BS}{base station}
\newacronym{psd}{PSD}{Power Spectral Density}
\newacronym{fss}{FSS}{Frequency Selective Surface}
\newacronym{umts}{UMTS}{Universal Mobile Telecommunications System}
\newacronym{nrdz}{NRDZ}{National Radio Dynamic Zone}
\newacronym{thz}{THz}{terahertz}
\newacronym{sthz}{sub-THz}{}
\newacronym{ghz}{GHz}{gigahertz}
\newacronym{hemt}{HEMT}{high-electron mobility transistors}
\newacronym{2deg}{2DEG}{two-dimensional electron gas}
\newacronym{dsp}{DSP}{digital signal processing}
\newacronym{dac}{DAC}{digital-to-analog converter}
\newacronym{evm}{EVM}{error vector magnitude}
\newacronym{cmos}{CMOS}{complementary metal–oxide–semiconductor}
\newacronym{eth}{ETH}{Eidgenössische Technische Hochschule}
\newacronym{soi}{SOI}{silicon-on-insulator}
\newacronym{pa}{PA}{power amplifier}
\newacronym{pae}{PAE}{power added efficiency}
\newacronym{lna}{LNA}{low noise amplifier}
\newacronym{nf}{NF}{noise figure}
\newacronym{2d}{2D}{two-dimensional}
\newacronym{sige}{SiGe}{Silicon Germanium}
\newacronym{hbt}{HBT}{heterojunction bipolar transistors}
\newacronym{ingaas}{InGaAs }{indium gallium arsenide}
\newacronym{gaas}{GaAs}{gallium arsenide}
\newacronym{ic}{IC}{integrated circuit}
\newacronym{mmic}{MMIC}{\gls{mmwave} \gls{ic}}
\newacronym{gan}{GaN}{gallium nitride}
\newacronym{sic}{SiC}{silicon carbide}
\newacronym{inp}{InP}{indium phosphide}
\newacronym{rtd}{RTD}{resonant tunneling diodes}
\newacronym{impatt}{IMPATT}{impact ionization avalanche transit time}
\newacronym{gbps}{Gbps}{gigabits-per-second}
\newacronym{rfsoc}{RFSoC}{\gls{rf} \gls{soc}}
\newacronym{rfsocs}{RFSoCs}{\gls{rf} \gls{socs}}
\newacronym{sdr}{SDR}{software-defined radio}
\newacronym{fpga}{FPGA}{field-programmable gate array}
\newacronym{gsps}{GSps}{gigasamples-per-second}
\newacronym{qcl}{QCL}{quantum cascade laser}
\newacronym{dfg}{DFG}{difference-frequency generation}
\newacronym{pca}{PCA}{photoconductive antenna}
\newacronym{cpu}{CPU}{central processing unit}
\newacronym{ai}{AI}{artificial intelligence}
\newacronym{ml}{ML}{machine learning}
\newacronym{gpu}{GPU}{graphics processing unit}
\newacronym{soc}{SoC}{system-on-chip}
\newacronym{socs}{SoCs}{systems-on-chips}
\newacronym{wnoc}{WNoC}{wireless network-on-chip}
\newacronym{noc}{NoC}{network-on-chip}
\newacronym{ram}{RAM}{random access memory}
\newacronym{pcie}{PCIe}{Peripheral Component Interconnect Express}
\newacronym{usb}{USB}{Universal Serial Bus}
\newacronym{iot}{IoT}{Internet of Things}
\newacronym{iont}{IoNT}{Internet of Nano-Things}
\newacronym{wpan}{WPAN}{wireless personal area network}
\newacronym{xr}{XR}{eXtended Reality}
\newacronym{wlan}{WLAN}{ wireless local area network}
\newacronym{lidar}{LIDAR}{Light Detection and Ranging}
\newacronym{fmcw}{FMCW}{frequency-modulated continuous wave}
\newacronym{v2x}{V2X}{vehicle-to-everything}
\newacronym{leo}{LEO}{low-Earth orbit}
\newacronym{nasa}{NASA}{National Aeronautics and Space Administration}
\newacronym{jpl}{JPL}{Jet Propulsion Laboratory}
\newacronym{umi}{UMi}{urban microcell}
\newacronym{uma}{UMa}{urban macrocell}
\newacronym{rma}{RMa}{rural macrocell}
\newacronym{fspl}{FSPL}{free-space path loss}
\newacronym{irs}{IRS}{intelligent reflecting surfaces}
\newacronym{ris}{RIS}{reconfigurable intelligent surfaces}
\newacronym{aod}{AoD}{angle of departure}
\newacronym{aoa}{AoA}{angle of arrival}
\newacronym{pdp}{PDP}{power delay profile}
\newacronym{isi}{ISI}{inter-symbol interference}
\newacronym{qos}{QoS}{quality of service}
\newacronym{qoe}{QoE}{quality of experience}
\newacronym{fwa}{FWA}{fixed wireless access}
\newacronym{hbn}{h-BN}{hexagonal boron-nitride}
\newacronym{mos2}{MoS2}{molybdenum disulfide}
\newacronym{ict}{ICT}{information and communications technology}
\newacronym{ntn}{NTN}{non-terrestrial network}
\newacronym{nu}{NU}{Northeastern University}
\newacronym{oam}{OAM}{orbital angular momentum}
\newacronym{mmse}{MMSE}{minimum mean squared error}
\newacronym{ltcc}{LTCC}{low-temperature co-fired ceramic}
\newacronym{lcp}{LCP}{liquid crystal polymers}
\newacronym{cte}{CTE}{coefficient of thermal expansion}
\newacronym{tds}{THz-TDS}{\gls{thz} time-domain spectroscopy}
\newacronym{ap}{AP}{access point}
\newacronym{fma}{FMA}{frequency modulated array}

\usepackage{dblfloatfix}    % To enable figures at the bottom of page

\usepackage{multicol}
\usepackage{tabularx}
\usepackage{booktabs}

\begin{document}
%\title{The TeraNova Platform: An Integrated Testbed for Ultra-broadband Wireless Communications at \textit{True} Terahertz-band Frequencies}

\title{The Evolution of Applications, Hardware Design, and Channel Modeling for Terahertz (THz) Band Communications and Sensing: Ready for 6G?
%\\
%\hl{Alternative}\\
%Closing the Terahertz Technology and Channel Modeling Gap for 6G Systems\\
%\hl{OR any combination}
}

\author{
Josep M. Jornet,~\IEEEmembership{Fellow,~IEEE}, 
Vitaly Petrov,~\IEEEmembership{Member,~IEEE}, 
Hua Wang,~\IEEEmembership{Fellow,~IEEE},\\
Zoya Popovic,~\IEEEmembership{Fellow,~IEEE},
Dipankar Shakya,~\IEEEmembership{Student Member,~IEEE},
Jose V. Siles,~\IEEEmembership{Senior Member,~IEEE},
and Theodore S. Rappaport,~\IEEEmembership{Fellow,~IEEE}\vspace{-7mm}

\thanks{Josep M. Jornet is with the Institute for the Wireless Internet of Things, Northeastern University, USA.}%
\thanks{Vitaly Petrov is with the Division of Communication Systems, KTH Royal Institute of Technology, Sweden.}%
\thanks{Hua Wang is with the Department of Information Technology and Electrical Engineering, ETH Zürich, Switzerland.}%
\thanks{Zoya Popovic is with the Department of Electrical, Computer and Energy Engineering, University of Colorado Bouldar, USA.}%
\thanks{Theodore S. Rappaport and Dipankar Shkya are with NYU Wireless, New York University, USA.}%
\thanks{Jose V. Siles is with the NASA Jet Propulsion Laboratory, USA.}%

\thanks{The work has been supported in part by the National Science Foundation grants CNS-1955004, CNS-2011411, CNS-2117814, CNS-2216332, CNS-2037845, CNS-1909206, the NYU WIRELESS Industrial Affiliates program, and the
Commission of the European Union CHIPS Joint Undertaking SHIFT project Grant Agreement No. 101096256.}%
}

\bstctlcite{BSTcontrol}  % run this directive at start of document

\maketitle

% As a general rule, do not put math, special symbols, or citations
\begin{abstract}

For decades, the \gls{thz} frequency band had been primarily explored in the context of radar, imaging, and spectroscopy, where multi-\gls{ghz} and even \gls{thz}-wide channels and the properties of terahertz photons offered attractive target accuracy, resolution, and classification capabilities. Meanwhile, the exploitation of the terahertz band for wireless communication had originally been limited due to several reasons, including (i) no immediate need for such high data rates available via terahertz bands and (ii) challenges in designing sufficiently high power terahertz systems at reasonable cost and efficiency, leading to what was often referred to as ``the terahertz gap.'' Over the recent decade, advances on many fronts have drastically changed the terahertz landscape. First, the evolution from 5G-grade to 6G-grade wireless systems dictates the need to support novel bandwidth-hungry applications and services for both data transfer (i.e., \gls{xr}, the Metaverse, and vast modeling needs of \gls{ai} and \gls{ml}), as well as centimeter-precision sensing and classification (i.e., for standalone position location, \gls{v2x} or \gls{uav} tracking). Second, substantial progress in terahertz hardware has been achieved, offering promise that the terahertz technology gap will be closed. Hence, terahertz-band wireless communication seems inevitably an essential part of the future networking technology landscape in coming decades. To design efficient terahertz systems, the peculiarities of terahertz hardware and terahertz channels need to be understood and accounted for. This roadmap paper first reviews the evolution of the hardware design approaches for terahertz systems, including electronic, photonic, and plasmonic approaches, and the understanding of the terahertz channel itself, in diverse scenarios, ranging from common indoors and outdoors scenarios to intra-body and outer-space environments. The article then summarizes the lessons learned during this multi-decade process and the cutting-edge state-of-the-art findings, including novel methods to quantify power efficiency, which will become more important in making design choices. Finally, the manuscript presents the authors' perspective and insights on how the evolution of terahertz systems design will continue toward enabling efficient terahertz communications and sensing solutions as an integral part of next-generation wireless systems.

\end{abstract}

\glsresetall
% Note that keywords are not normally used for peer-reviewed papers.

%\glsresetall
\glsreset{ghz}
\glsreset{xr}
\glsreset{ai} 
\glsreset{ml}
\glsreset{v2x}
\glsreset{uav}

% For peer-reviewed papers, you can put extra information on the cover page as needed:
% \fi
%
% For peer-reviewed papers, this IEEEtran command inserts a page break and creates the second title. It will be ignored for other modes.
\IEEEpeerreviewmaketitle

\section{Introduction}
\label{sec:introduction}

Each generation of the global cellphone industry has produced technological innovations that brought forth unexpected new use cases that exploit greater channel bandwidths and data rates. From the early days of analog cellular using 25~kHz or 30~kHz channels in the \gls{1g} of cellphones, today’s 5G cellphone standards (e.g., the \gls{3gpp}) exploit \gls{ofdm}, \gls{mimo} antennas as well as massive \gls{mimo}, and use many concatenated channel bandwidths, in 20~MHz chunks, allowing over 100~MHz of usable bandwidth with hundreds of megabits-per-second per user. 

5G was the first global cellphone standard also to introduce spectrum above 6~GHz, and ushered in the era of \gls{mmwave} communications~\cite{rappaport2013millimeter}, where both \gls{fdd} and \gls{tdd} can implement multi-gigabits-per-second data rates in 200~MHz channel chunks and have enabled wireless carriers to implement both mobile services in urban cores with high pedestrian traffic loads and large venues (such as stadiums) as well as offering \gls{fwa} for homes and businesses in an unexpected and profitable way. Not all governments or carriers have adopted the \gls{mmwave} bands, leading to varying opinions about the efficacy of \gls{mmwave}. Yet, virtually all global cellphone makers and infrastructure vendors now ship \gls{mmwave} transceivers as part of their product offerings. In the unlicensed device arena,  mmWave has found Wi-Fi adoption through IEEE 802.11ad, and more recently, IEEE 802.11ay, which theoretically supports up to 100~Gbps data transfers. The move to greater bandwidths appears inevitable as more applications and services require massive data transfer rates. 

Immediately after the release of 5G~\cite{Vannithamby20205G}, both academia and industry started to envision what 6G should bring to the users and, accordingly, what technologies will be needed~\cite{rappaport2019wireless,akyildiz20206g,giordani2020toward,saad2019vision,tataria20216g}. The need to accommodate the exponentially growing number of wirelessly connected devices and their mounting data-rate requirements motivated the exploration of higher frequency bands, beyond \glspl{mmwave}, including the sub-terahertz (sub-THz, 100-300~GHz) and terahertz (THz, 300-10~THz) bands~\cite{akyildiz2014terahertz,rappaport2019wireless,chen2019survey,xing2019indoor,sarieddeen2020next, Xing2021iclModels, song2021terahertz, Ju2021Spatial, chaccour2022seven, akyildiz2022terahertz, rappaport2022cup, shafie2022terahertz,kurner2022book}. 

{This frequency range brings exciting opportunities to many applications across scales. The very small wavelength at sub-terahertz and terahertz frequencies (from 3~mm at 100~GHz down to 30~$\mu$m at 10~THz) enables the development of miniature antennas that can be potentially embedded everywhere, from nanosensors in the \gls{iont} to within computing processors for \glspl{wnoc}. This is not possible at lower frequencies (e.g., sub-6~GHz and even \gls{mmwave}, because of the much larger wavelengths). At the same time, the very small size of individual antennas allows for their integration in very large numbers over compact footprints. For a fixed antenna footprint (e.g., the size of current sub-6~GHz antennas on cellphones or in base stations), a sub-terahertz or terahertz antenna offers much higher directivity gains. This motivates, for example, the adoption of sub-terahertz frequencies for satellite communication networks, where the combination of high gain antennas in transmission and reception can facilitate closing the link compared to lower frequency bands. Across scales, this frequency range offers very large bandwidth (from tens to hundreds of GHz, if not more), which are only limited by hardware constraints or spectrum policy and regulations.}

Much has been said about what these frequencies might and might not bring to the table, both in terms of communications and sensing {(see Table~I in~\cite{akyildiz2022terahertz} for a summary of the existing works)}. Still, today, five years before the expected release of 6G, one question remains: {Will} terahertz technology be ready?

This paper aims to answer this question by providing an updated view of two critical aspects, namely, terahertz hardware technology and channel modeling. The former determines the type and properties of the signals that can be adopted (in terms of frequency, bandwidth, and power, among others), the hardware-induced phenomena that can alter them (including amplitude and phase noises or non-linear distortion) as well as the efficiency with which these can be processed at the transmitter and the receiver. The channel models describe the phenomena affecting the propagation of such signals from the transmitter to the receiver, which drastically depends on the application scenario (from indoors to outdoors, static or mobile, other emerging applications, and new opportunities). These two aspects set the cornerstone on which the communication techniques (e.g., time, frequency and phase synchronization, modulation and coding, MIMO strategies, etc.) and the networking protocols (e.g., user discovery, beam management, mobility support, etc.) need to be developed. 

The remainder of this paper is organized as follows. 
In Section~\ref{sec:use_cases}, we describe the envisioned use cases for terahertz communications in 6G, classified not only as a function of the targeted communication range or use case but also the need to support different levels of mobility. {This highlights the diversity in applications of terahertz technology while facilitating the identification of common aspects that guide the design of terahertz networks.}
In Section~\ref{sec:hardware}, we describe the evolution of terahertz hardware technology, highlighting the trends in analog front-ends, digital back-ends, and antenna systems. {Compared to existing works, updated and largely quantitative data is provided concerning state-of-the-art terahertz device technologies.}
In Section~\ref{sec:channel}, we present the key lessons learned from the extensive terahertz channel largely experimental modeling efforts in the last five years.
In Section~\ref{sec:challenges}, we discuss some of the critical aspects and opportunities relating to both hardware and channel modeling, {as well as how these impact and enable the design of physical and link layer technologies.}
We conclude the paper in Section~\ref{sec:conclusion}.

% \begin{figure*}
% \centering
% 	\includegraphics[width=\linewidth]{./Figures/example.pdf}
% 	\caption{Example of a two-column wide figure. Remove the star for a single column. Caption of the figure.}
% 	\label{fig:example}
% \end{figure*}

%----------------------------------------------------%
\section{Major Use Cases for Terahertz Communications and Sensing}
\label{sec:use_cases}

\begin{figure*}
  \centering
  \includegraphics[width=0.9\textwidth]{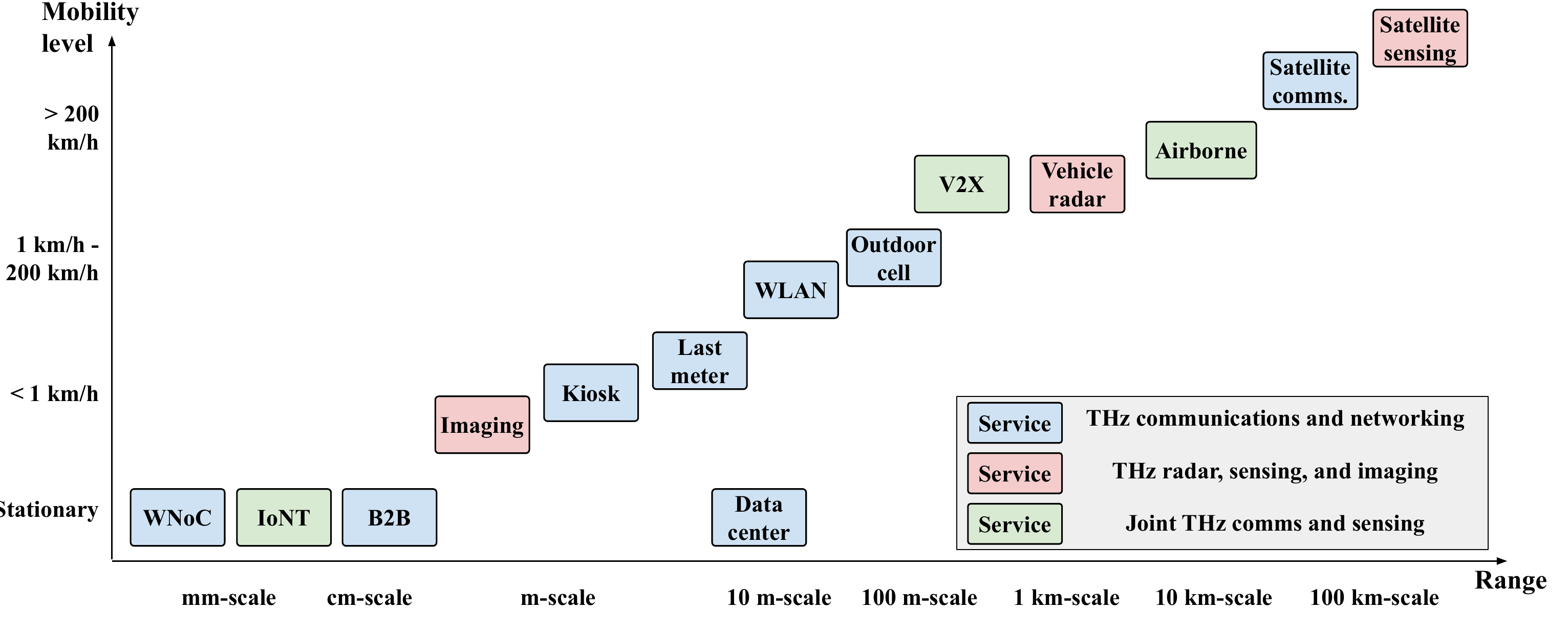}
  \caption{Classification of major use cases for terahertz communications and sensing.}
  \label{fig:use_cases}
\end{figure*}

Both the terahertz hardware design considerations and, especially, the terahertz channel modeling approaches heavily depend on the target environment and the target use case, as well as the carrier frequencies and form factors of the equipment allocated to such use. Therefore, in this section, we first present a harmonized vision of the major use cases for terahertz communications and sensing. Since supporting mobility is one of the key challenges for terahertz wireless networks, we arrange the use cases accordingly. {We go} from inherently stationary through low mobility all the way to highly mobile use cases, such as terrestrial wireless access, vehicular terahertz systems, and satellite-to-satellite terahertz communications. The proposed classification is also illustrated in Figure~\ref{fig:use_cases}.

The set of attractive use cases for terahertz communications and sensing is primarily determined by: (i) potentially (but not necessarily) millimeter-scale or sub-millimeter-scale size of terahertz antenna elements (see Section~\ref{subsec:antennas}); and (ii) potentially (but not necessarily) large system bandwidth in the order of up to several tens and even few hundreds of gigahertz. Another essential feature of the use cases here is that a combination of large-scale antenna systems with terahertz signal frequency leads to very directional transmissions that have both advantages (e.g., less interference) and challenges to overcome (beam alignment, node discovery, accurate position location, system clock synchronization, etc.). Whenever relevant, we point to the key distinct feature of terahertz communications for a discussed use case below.

%-------------------------------%
\subsection{Stationary Use Cases for Terahertz}

%--------%
\subsubsection{Wireless Networks on Chips}
The rapid increase in the characteristics, capabilities, and inline, complexity of the computer chips challenges the scalability of the state-of-the-art approach to building computer platforms. Interconnecting all the essential chip components via either a shared wired bus or a set of wired busses going through the central hub becomes extremely challenging to implement. Modern \glspl{cpu} already feature 16+ computing cores, multiple co-processors (e.g., tailored to \gls{ai} or other specific applications), and several layers of shared cache memory, while \glspl{gpu} or latest \gls{socs} feature even more individual elements. 

Hence, a novel use case arises, often referred to as \glspl{wnoc}. The core idea is to replace multiple individual wired connections among small-scale elements on the chip with high-rate optical or terahertz wireless links through a shared medium~\cite{abadal2014area,guirado2022wireless,sun2014mimo}. Implementing such a system immediately raises novel challenges (design, fabrication, and control of nano-scale terahertz radio modules). At the same time, successful adoption of such an approach allows addressing the scalability problem of future chips by pushing the possible number of interconnected elements from a few hundred to many hundreds or even thousands of cores~\cite{petrov2017terahertz}.

Simultaneously, \glspl{wnoc} theoretically feature greater energy efficiency and a more compact design than traditional wired \glspl{noc}. Both the inherently small antenna size and the large bandwidth play vital roles for this use case, enabling ultra-high-rate short-range wireless links among many miniature terahertz radios on a chip. Importantly, from the wireless networking point of view, \glspl{wnoc} present one of the simplest setups to {control -- a fixed} number of connected nodes with full control of their behavior and no node mobility.

%--------%
\subsubsection{Board-to-Board Communications}
\label{subsubsection:b2b}
This use case addresses the setup similar to the previous one but at a slightly larger scale. Specifically, board-to-board terahertz communications are envisioned as a possible alternative to space- and energy-consuming wired connections between the key elements in the computer, e.g., \gls{cpu} to \gls{gpu}, \gls{cpu} to \gls{ram}, and \gls{cpu} to the networking interface. Point-to-point terahertz links can replace or complement state-of-the-art solutions, including \gls{pcie}, high-rate \gls{usb}, and other wired options. More futuristic variations suggest integrating board-to-board terahertz communications with other essential elements, e.g., by utilizing cooling pipes between the \gls{cpu} and the \gls{gpu} as a waveguide for the terahertz signal~\cite{petrov2017enabling}. 

There are two key distinct features of board-to-board terahertz communications from the previous use case, \glspl{wnoc}. First, the communication normally happens at the centimeter, not millimeter (or even sub-millimeter) scale. Second, the communication system typically considers no more than a few (usually, two) connected elements, so there is no need for sophisticated control algorithms and protocols, as in the terahertz \gls{wnoc}. Similar to the previous item, both small-scale terahertz antennas and large terahertz bands are essential enablers for this use case. Notably, the size requirements are less strict than for the \gls{wnoc}, making not only terahertz, but sub-terahertz, or even \gls{mmwave} radios applicable.

%--------%
\subsubsection{Data Center Links}
Building terahertz-enabled high-rate wireless links in the data center is a further extrapolation of the previous use case (detailed in subsection~\ref{subsubsection:b2b}) to an order of magnitude larger distances. Specifically, the data center use case suggests interconnecting different server racks using point-to-point wireless terahertz links instead of Ethernet or optical fiber cables. This use case may sound counter-intuitive to some, as the data center is a perfect example of stationary deployment, where wired connections have an advantage. However, the envisioned replacement of rack-to-rack wired links with high-rate and low-latency wireless analog relaxes multiple restrictions essential for the data center layout (e.g., placement of racks in the room) and even cooling system design~\cite{Katz2009Tech,cheng2020thz,eckhardt2021realistic}.

Terahertz wireless links for data centers are also one of the key motivating use cases for the recently published IEEE standard for sub-terahertz connectivity, IEEE~802.15.3d--2017, and its upcoming revision just approved in 2023~\cite{802.15.3dstandard}. The wide bandwidth is the essential property of terahertz communications, while the scale already allows for larger and relatively simple horn or lens-type terahertz antennas~(see Section~\ref{subsec:antennas}).

%--------%
\subsubsection{Fronthaul/Backhaul}
The last but definitely not least stationary use cases for terahertz communications, are wireless fronthaul or wireless backhaul for 5G-Advanced and, especially, upcoming 6G cellular networks~\cite{rappaport2019wireless}. By design, multiplexing several (tens) of high-rate mobile access links (e.g., \gls{mmwave}) into a single wireless link requires an order of magnitude greater bandwidth than only the terahertz band or optical systems can offer. Meanwhile, sub-terahertz and terahertz links are relatively robust to adverse weather conditions~\cite{amarasinghe2020scattering,sen2022terahertz,li2023performance} which gives them a notable advantage from the reliability point of view.

Same as for the data centers above, the large bandwidth of the terahertz signal is the essential feature for the required high data rates, while the use of large-scale antenna systems on both sides addresses the spreading loss issue. Tentatively, wireless fronthaul/backhaul data links are likely one of the first use cases for terahertz communications to be adopted in 6G-grade networks.

%-------------------------------%
\subsection{Low Speed Mobility}
\setcounter{subsubsection}{4}

%--------%
\subsubsection{Nano Sensor Networks}
One of the first use cases originally considered for wireless communications in the terahertz band is related to data exchange between micro- and nano-scale machines~\cite{akyildiz2010electromagnetic,jornet2023nanonetworking}. These envisioned interconnected small-scale robots are to perform various tasks ranging from environmental sensing up to in-body medical invasions. The proposal partially stems from an earlier concept of the \gls{iot} pushing it further to the so-called \emph{\gls{iont}}~\cite{akyildiz2010internet}.

The miniature (millimeter-scale or sub-millimeter-scale) size of terahertz antennas and attracted terahertz transceivers is the key inherent property of terahertz communications making it not only suitable for this use case, but even further, making it one of only a few possible communication methods at this scale (together with e.g., optical wireless systems, molecular communications, and bacterial nanonetworks)~\cite{yang2020comprehensive}.

%--------%
\subsubsection{Imaging}
Historically, terahertz imaging was one of the first-ever use cases for terahertz wireless systems. Specifically, terahertz radiation is used to identify materials composing the scanned object, as different materials show different electromagnetic signatures (e.g., frequency-dependent transmissivity or reflexivity) when illuminated by the terahertz signal. Introduced and implemented several decades ago (e.g., terahertz communication systems are still in the prototype stage today), terahertz imaging facilitates efficient and reliable scanning in many areas. These include, among others, luggage and postal mail screening, 2D and 3D biological and chemical examination~\cite{mittleman2018twenty,valuvsis2021roadmap}.

Importantly, terahertz radiation is non-ionizing -- not enough energy carrier per quantum to remove an electron from the atom or molecule. Therefore, it is notably safer in practical scenarios (not fully safe though~\cite{reddy2023photothermal}), especially when it comes to medical scanning. Two major distinct properties of the terahertz band play vital roles for this use case: (i)~large bands available (hence, facilitating the scanner resolution) and (ii)~the fact that many common materials have notably different properties in reflecting, diffracting, and passing over the terahertz signal. Last but not least, the terahertz signal better penetrates certain obstacles (e.g., human skin) than optical signals, making it a promising option when it comes to multi-later 3D scanning.

%--------%
\subsubsection{Kiosk Download / Data Shower}
Another low-mobility use case for terahertz communications is typically referred to as one of the three similar terms: (i)~kiosk download; (ii)~data shower; and (iii)~information shower~\cite{Rappaport2015book,song2018prototype}. While the exact definitions of these vary from source to source, the general idea is almost the same. A kiosk (or a data/information shower) is an ultra-small (a few meters in coverage or less) and, importantly, an ultra-high-rate wireless cell that allows quasi-instant downloads/uploads of huge chunks of data (data rates of tens/hundreds of gigabits to a few terabits per second).

The core idea of this proposal is to complement the existing coverage-centric wireless networks (microwave or even recent \gls{mmwave} ones) with \emph{strategically-placed} high-rate terahertz data showers, thus allowing the users to exchange large amounts of cached data when in their range. Such locations may be the entrance of the metro station or a large shopping mall, an airport corridor, a busy intersection, or any other place with a high volume of humans passing per minute. While these few short-range showers are not able to completely replace existing networks, they can assist them notably by offloading a significant portion of heavy traffic~\cite{petrov2016applicability}. Here, the large bandwidth of the terahertz signal is the main properly facilitating high-rate data exchange but simultaneously limiting the coverage of the terahertz data shower.

%--------%
\subsubsection{Last Meter Interconnect}
The last-meter terahertz connection is a variation of the terahertz kiosk for office or on-body environments. This use case suggests applying terahertz connectivity as the last hop between a distant remote server and the user terminal. Specifically, a single one-meter range terahertz cell (backed to the office wired network) can replace several Ethernet wired connections in a typical office desk or {cubicle, e.g., desktop, laptop, high-resolution display, among others}~\cite{petrov2018last}.

A more futuristic version of this use case describes a set of wearable (or eventually implantable) devices interconnected with terahertz wireless links, thus featuring extremely high rates and ultra-low latencies~\cite{akyildiz2010internet}. These formed \glspl{wpan} will include but are not limited to smart glasses (e.g., \gls{xr}~\cite{gapeyenko2023standardization}), on-body or in-body sensors, and even (eventually) Internet-to-brain and brain-to-Internet interfaces~\cite{jornet2019optogenomic}. Similar to the previous use case, the high rates and low latencies are enabled primarily by wide terahertz bands, while the same wideband nature of transmissions limits the coverage of these envisioned terahertz \glspl{wpan}.

%-------------------------------%
\subsection{Medium Speed Mobility}
\setcounter{subsubsection}{8}

%--------%
\subsubsection{Femto Cells}
Terahertz-enabled ultra-high-rate femtocells and/or terahertz \glspl{wlan} present a decisive use case for \emph{mobile} terahertz communications. Specifically, an evolution of state-of-the-art IEEE 802.11ad/ay \glspl{wlan} operating at 60~GHz is envisioned at the next Wi-Fi development cycle within 6G or 6G-Advanced timeline. In parallel, the design of cellular-controlled indoor sub-terahertz/terahertz cells is of interest to further boost the performance of cellular wireless networks indoors~\cite{rappaport2019wireless,akyildiz2022terahertz}.

This setup presents one of the most challenging use cases for terahertz communications to implement, as both the complex wireless channel with a lot of potential obstacles (furniture pieces, human bodies, walls, etc.) and unpredictable mobility of the user terminals (e.g., smartphones or \gls{xr} glasses) must be overcome. Importantly, not only canonical macro-scale mobility (large-scale movements of the user terminal) should be accounted for but also micro-scale (or so-called small-scale) mobility comprising unpredictable shifts and rotations of the device itself. It has been shown that such rotations often lead to unexpected misalignment of the narrow terahertz beams, thus compromising network reliability and performance. Same as above, wideband terahertz signals enable high-rate data exchange, while large-scale (thus directional) terahertz antenna systems are needed to maintain the desired coverage of several (tens of) meters.

%--------%
\subsubsection{Micro Cells}
This use case presents a futuristic extension of the previous one, suggesting true terahertz radio to be used for cellular access links (50~m to 200~m coverage). While it has been experimentally shown that wideband terahertz signals can be reliably received from a large distance of up to several kilometers~\cite{sen2023multi}, maintaining the 100~m+ coverage for mobile terahertz access links is a challenge, as such distances will inherently demand even narrower terahertz beams. Consequently, all the supporting control algorithms and protocols must be capable of operating over narrow (less than a few degrees) terahertz beams. While there are several promising solutions presented in this area recently~\cite{boulogeorgos2021directional,morales2021adapt,xia2019expedited,xia2020routing}, the set of unsolved research problems to address remains large. These include among others, reliable, efficient, and low-overhead beam tracking algorithms, fast node discovery protocols, novel interference management techniques, and intelligent time-frequency-space resource allocation solutions. Still, a theoretical possibility to leverage an order of magnitude larger bands than those available at \gls{mmwave} frequencies makes this use case tempting to continue research and engineering activities.

%--------%
\subsubsection{Automotive Radars}
Utilizing terahertz signals for automotive (e.g., vehicle-mounted) radars is the second sensing-centric use case in our list. In contrast to terahertz imaging discussed above, automotive radars primarily target revealing the relative distance to the target and the relative velocity of the target, not the target's composition~\cite{norouzian2019next}. Modern vehicle manufacturers already integrate up to a dozen sensors into their latest models, from ultrasonic parking sensors through \gls{mmwave}/sub-terahertz cruise radars to optical cameras to monitor the surroundings. Selected prototypes of autonomous driving vehicles get equipped with \gls{lidar} solutions, which are effectively ``light-based laser radars''.

While existing \gls{mmwave} and sub-terahertz cruise radars already enable decisive applications, such as adaptive cruise control and semi-autonomous driving, further development of these systems using wider true terahertz signals will notably contribute to their angular and distance resolution~\cite{jiang2022lidar}. Pulse-based radars and \gls{fmcw} radars are two widely spread approaches to estimating the distance and velocity of the target~\cite{vasilyev2012terahertz}. The first approach primarily relies on the round trip time of the reflected/scattered radar signal to estimate the distance, while the Doppler shift in the signal frequency will reveal the relative velocity. \gls{fmcw}-type radars achieve a similar goal by comparing the received signal with its original shape in time and frequency domains. Here, both the narrow terahertz beam and the wide bandwidth of the terahertz radar signal naturally contribute to the performance of the terahertz-enabled automotive radars.

%--------%
\subsubsection{Vehicular Communications}
In parallel to improving automotive radars, the terahertz community is currently exploring the possibility of partially reusing selected elements (e.g., terahertz antenna arrays) for terahertz \gls{v2x} communications. While one may argue that there is typically not enough traffic generated by a single vehicle to deploy a high-rate terahertz link, this is not always true. First, there are use cases, where a single terahertz vehicle-to-infrastructure connection may relay several active mobile links between user devices in the car (smartphones, tablets, XR glasses, etc.) and in-vehicle entertainment systems~\cite{Kanhere2011performance}. Hence, instead of serving several mobile links, the network only has to serve one, while the vehicle-mounted access point serves the rest. This usage scenario is especially practical for public transport, where e.g., the entire bus or the entire tram/train coach can be multiplexed into a single terahertz access link~\cite{guan2019measurement}.

Another decisive application motivating terahertz \gls{v2x} is autonomous collective driving. Latest studies show that cooperation among the vehicles is one of the factors both: (i)~simplifying control of the vehicle swarms; and (ii)~improving the system performance. For instance, a group of connected vehicles can drive at high speeds on a highway with very low distance between each other, thus increasing the road capacity. For this scenario, wideband terahertz signals enable high rates and low latencies, while using steerable directional beams facilitates little to no interference with other terahertz-capable cars, pedestrian users, and infrastructure nearby. Notably, connected smart vehicles are also one of the promising scenarios for joint terahertz communications and sensing, where the same hardware components are used for both radar sensing and data exchange~\cite{petrov2019unified}.% (see Section~\ref{subsec:jcsh}).

%-------------------------------%
\subsection{High Speed Mobility}
\setcounter{subsubsection}{12}

%--------%
\subsubsection{Connectivity for Airborne Nodes}
The list of use cases with high mobility of nodes starts with exploiting terahertz communications for airborne nodes. These include both human-operated airplanes and \glspl{uav}~\cite{Xia2020multi}. Within the first group, one of the primary targets is to improve connectivity to passenger airplanes. A commercial passenger airplane is a very expensive device whose cost may easily exceed one hundred million US dollars. A one-way ticket for a transatlantic flight is over several hundred US dollars for economy and up to several tens of thousands for business and first class.

Despite these costs, a flying airplane is currently one of the world's worst connected places, with no more than a couple of megabits per second available per passenger over existing Ku (12-18~GHz), K (18-27~GHz) and Ka (26.5-40~GHz) bands (if all the passengers get connected). Improving this situation with mobile terahertz communication links between an airplane and a \gls{leo} satellite or between an airplane and the ground network (either directly or through another airplane) is a decisive practical usage scenario. Another vector of interest comes from primarily military-type use {cases -- connecting} two or more airplanes, drones, or other flying objects (e.g., missiles) into a single low-latency terahertz network. While wideband terahertz signals are good for bandwidth-hungry data exchange, highly directional terahertz transmissions also facilitate covert and secure airborne communications.

%--------%
\subsubsection{Satellite Remote Sensing}
Remote sensing is historically one of the first groups of use cases associated with terahertz radio systems that debuted decades ago. Until now, the US \gls{nasa} has been the creator of some of the highest power terahertz front-ends~\cite{siles2018new} Today, there are dozens of satellite-based terahertz sensors deployed already and more currently in development due to several major reasons~\cite{siegel2007thz}. First of all, as discussed above in relation to terahertz imaging, the band is very useful for multi-layer imaging of different objects. A combination of these unique properties with the wide band of the terahertz signal and low interference with Sun radiation make terahertz waves good candidates for remote sensing of planets. Further, as water is one of the key absorbers at terahertz frequencies, there are many satellite-based terahertz sensors deployed around the Earth used for environmental sensing, including weather forecasts.

\begin{figure*}[!b]
    \centering
    \includegraphics[width=0.85\textwidth]{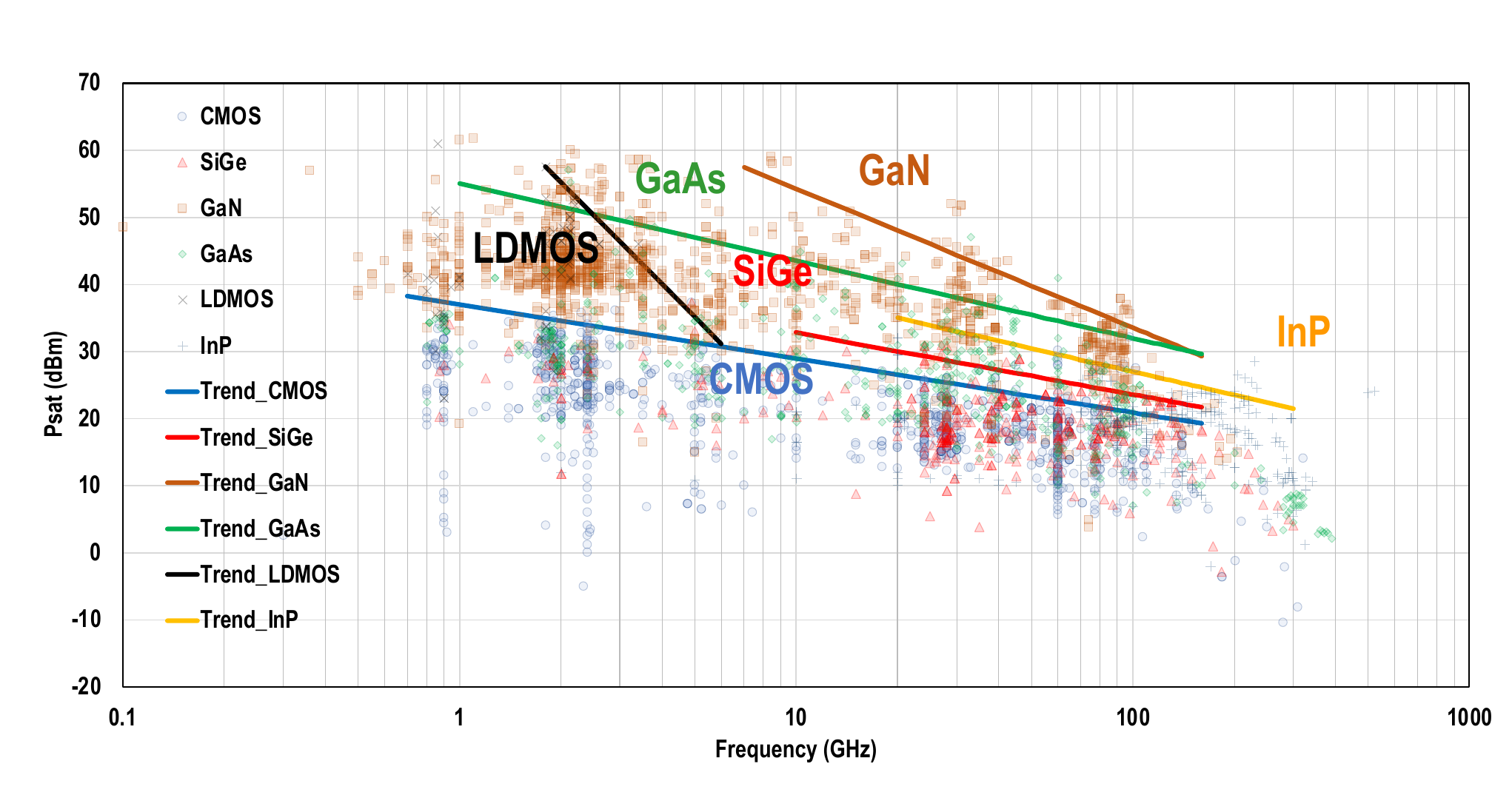}
    \caption{Power amplifier {survey. Saturated} output power vs frequency for different electronic front-end technologies~\cite{wang2023power}.}
    \label{fig:amplifiers}
\end{figure*}

%--------%
\subsubsection{Low Earth Orbit Satellite Communications}
Stemming from the previous two, terahertz band communications present an attractive alternative to \gls{mmwave} and optical systems when it comes to satellite communications (especially, building large-scale mega constellations at the \gls{leo}). On one side, the terahertz band features larger bands than those currently used in the K band or around for satellite-to-airplane and satellite-to-ground connectivity~\cite{alqaraghuli2021performance,Yang2024universal}. On the other, relatively wider terahertz beams demand less precision from the beam-pointing mechanism than laser-based inter-satellite cross-links~\cite{alqaraghuli2023road}. Importantly, terahertz signals are also less affected by the Sun radiation and atmospheric turbulence than optical signals~\cite{gao2023scintillation}.

Last but not least, while \gls{mmwave} bands may be too narrow for cross-links and optical links may be not powerful enough to facilitate Earth-to-satellite connectivity, the terahertz band may serve as a sweet spot in the middle~\cite{alqaraghuli2023road,aliaga2024modeling}. Prospective \gls{leo} satellites may be equipped with a single terahertz radio module for cross-links, access links, and environmental sensing (e.g., using joint communications and sensing) instead of three separate radio modules (optical, \gls{mmwave}, and terahertz), all operating at different bands and consisting of different radios. Hence, the stringent space and weight restrictions for spacecrafts become less crucial with the use of terahertz wireless systems.

\setcounter{subsubsection}{0}

\section{Evolution of Terahertz Hardware Technology}
\label{sec:hardware}
The terahertz band was called the terahertz gap for decades because of the lack of device technologies to support communication and sensing applications at these frequencies. However, the situation is much different today. As we summarize in the next sections, there are multiple solutions to realize the critical hardware building blocks of a terahertz wireless system. These include the analog front-ends, the antenna systems, and the \gls{dsp} back-ends. Importantly, while many times developed independently by the communications and the sensing communities, these hardware blocks are fundamentally the same, and their joint design can only benefit communications, sensing, and joint communication and sensing applications.

\subsection{Analog Front-ends}
\label{subsec:analog}
The analog front-end generates, modulates, filters, and amplifies signals at terahertz frequencies with multi-gigahertz bandwidths. The key performance metrics of a front-end include the frequency bands of operation, modulation bandwidth, transmission power, receiver sensitivity, amplitude and phase noise, and power efficiency.

There are three approaches to building terahertz front-ends. 
\subsubsection{Electronic Approach}
This approach pushes the limits of the devices and designs used in microwave and millimeter-wave frequency systems toward the terahertz-band frequencies. The general challenges for terahertz high-performance solid-state front-ends are mostly caused by the limited performance of electronic devices at the terahertz frequency spectrum, including limited device power gain, output power, energy efficiency, noise figure, and the circuit footprint. Moreover, there are considerable challenges in accurately measuring the characteristics of the fabricated electronic terahertz front-ends with micron and sub-micron device areas and contact geometries. Considering the device parasitics, signal transitions, and power limitations in terahertz circuits, detailed calibration and measurement techniques become paramount to ensure the desired performance of terahertz devices \cite{popovic2011thz}.
In such direction, new capabilities such as the multi-user terahertz measurement facility at NYU provides no-cost open access to test and measurement hardware and capabilities for collaborating researchers across the US~\cite{shakya2023exploring}.

% CMOS
\paragraph{Silicon Technology} Currently, most commercially available \gls{cmos} technologies offer device unit-gain at frequencies $f_{max}$ around 300-350~GHz, which is subject to further degradation due to layout parasitics. Considering a perfectly neutralized differential device pair, the resulting device unilateral power gain ($U$) is inversely proportional to the square of the operating frequency, and its unit-gain frequency $f_{max}$ stays almost the same as that of the native device~\cite{pozar2011microwave}. Therefore, most reported \gls{cmos} amplifiers operate at 150~GHz or below, with a theoretically limited device gain of 6~dB per stage before accounting for any passive matching network losses. Although there are reported techniques to boost the device gain beyond $U$~\cite{bameri2017high,park2019230,tang2021140,de2023sige}, tradeoffs with device stability and bandwidth typically need to be made.

\begin{figure}
    \centering
    \includegraphics[width=\columnwidth]{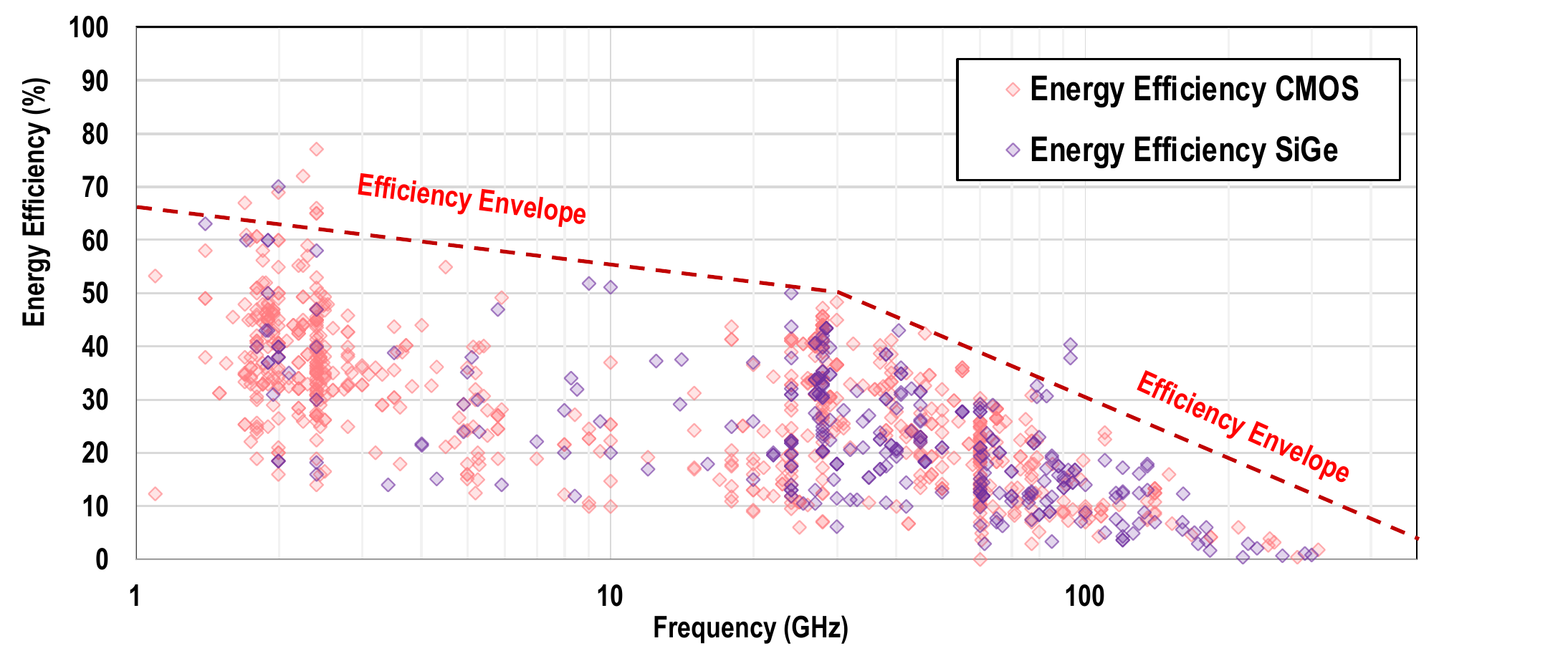}
    \caption{Energy efficiency of \gls{cmos} and \gls{sige} \glspl{ic}.}
    \label{fig:energy}
    \vspace{-5mm}
\end{figure}

\begin{figure}[!b]
    \centering
    \includegraphics[width=\columnwidth]{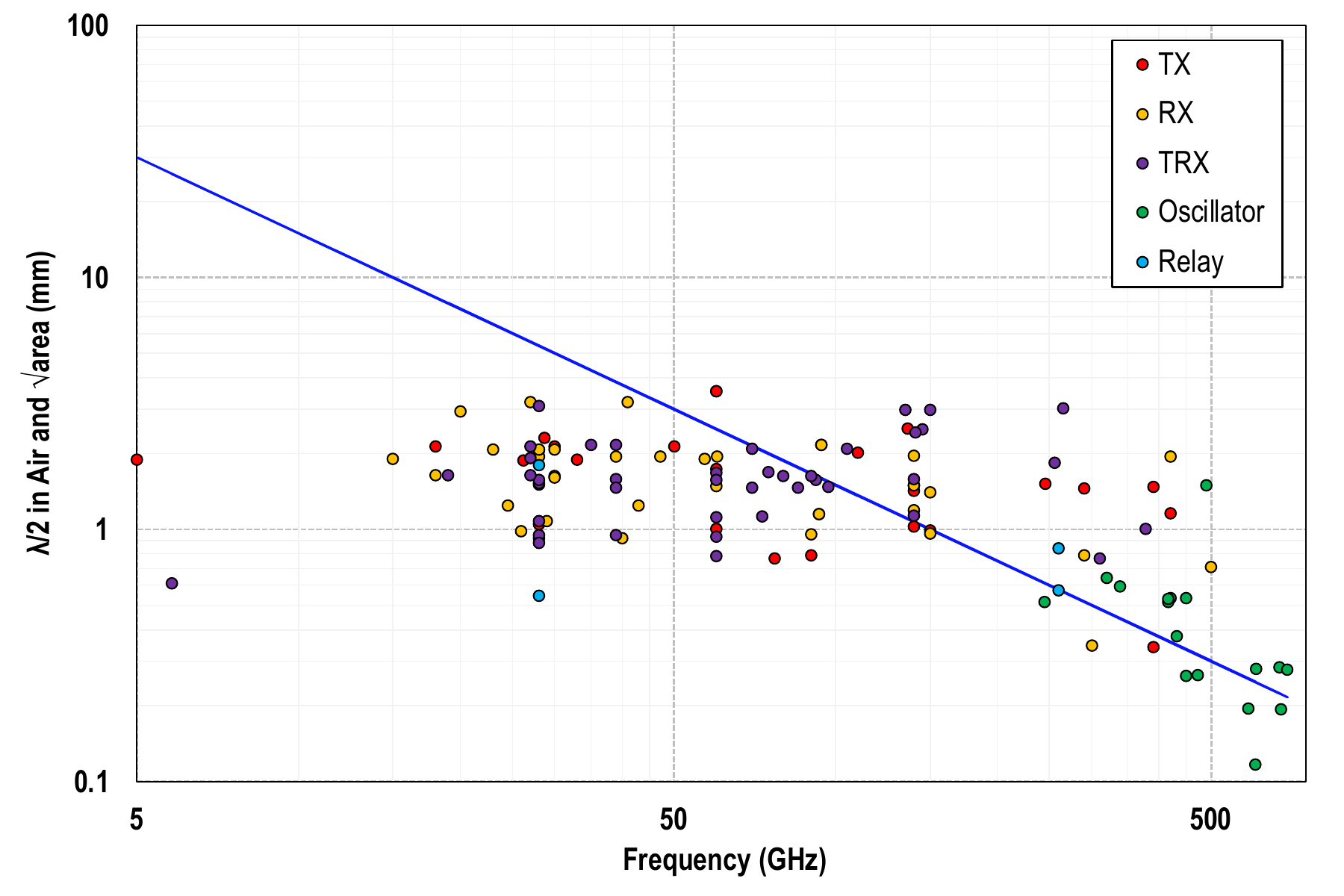}
    \caption{Chip element area for different circuits/system types.}
    \label{fig:footprint}
\end{figure}

While the \gls{cmos} device power gain is one root cause for its limited terahertz performance, the device Johnson limit \cite{johnson1966physical} also predicts the diminishing output power capability of \gls{cmos} devices at terahertz frequencies, which matches well with reported data based on \gls{eth} \glspl{pa} Survey~\cite{wang2023power} (Figure~\ref{fig:amplifiers}). Further, the limited device gain and output power capability require more amplifier stages in cascade and in power combining, both degrading the energy efficiency (Figure~\ref{fig:energy}). In addition, the low device power gain directly compromises the achievable noise figure, while using more amplifier stages to achieve the target gain results in large front-end transceiver circuit footprints that exceed the standard $\lambda/2$ antenna array grid size beyond 150~GHz (Figure~\ref{fig:footprint})~\cite{wang2023array}.

Recently reported D-band \gls{cmos}/\gls{cmos} \gls{soi} \glspl{pa} mostly adopt class-A biasing to maximize the device gain at the expense of energy efficiency~\cite{kim2023broadband,tang2021140,li2022high}. They achieve saturated output power ($P_{sat}$) of +18~dBm, 1~dB-gain-compression output power ($OP_{1dB}$) of +14~dBm, and peak \gls{pae} of 10-14\%. However, the modulation energy efficiency is typically below 2\%. Recent D-band \gls{cmos}/\gls{cmos} \gls{soi} \glspl{lna} report \gls{nf} of 4.7~dB to 6~dB and often employ device gain boosting techniques~\cite{hetterle2022design,lee2023d,lee2023d2,yun2021d}. 

In addition, for frequency generation for D-band wireless systems, the popular approach is to generate the \gls{lo} signal first at lower \gls{mmwave} frequency and then multiply the \gls{lo} frequency up to the D-band by frequency multipliers~\cite{callender2022fully,agrawal2023128,park2022d,mock2022high,park2022d2}, which exhibits a balanced trade-off of phase noise, frequency tuning range, and power consumption. At the system level, multiple D-band \gls{tx} and \gls{rx} designs have been reported with typical 10-14\% carrier bandwidth to support high-speed modulations~\cite{chou2022low,hamani2020108,li2021eight,wang20212,hamani202084,park2022d2,hamani2022d,li2022eight,gonzalez2023d,callender2022fully}. While the single-channel \gls{tx}/\gls{rx} performance is fundamentally governed by the \gls{pa}/\gls{lna} performance, realizing \gls{2d} scalable arrays for practical mobile wireless applications (e.g., \gls{tx}/\gls{rx} co-channel or dual-polarization) will entail other challenges on circuit footprint and thermal density. They will require other techniques, such as heterogeneous integration.         

To operate close to or beyond device $f_{max}$, one should resort to device nonlinearity and harmonic generation, leading to even poorer output power capability and energy efficiency. For example, \gls{cmos} circuits are used to build systems at frequencies approaching 300~GHz, but with transmit power in the order of only 1~mW, capable only for short-range communication and sensing~\cite{sengupta2018terahertz}. 
Another option to increase the performance of silicon-based devices, such as the \gls{cmos} devices described this far, is to explore the integration of silicon with other materials in new structures, such as with {advanced \gls{sige} \glspl{hbt} from Globalfoundries, STMicroelectronicsics, and IHP.} Such devices can achieve $f_{max}$ over 500~GHz and are less sensitive to layout parasitics. However, similar design tradeoffs still exist, showing the limitation of silicon-based devices and the need for III-V compound technologies and heterogeneous integration. {On the other hand, silicon-based technologies offer unparalleled integration density, signal processing, and controls/reconfigurations, which are essential for terahertz front-ends and systems.}  %Now the writing has a natural transition to III-V contents by Zoya.%  
%\hl{@Hua: here, it would be good to add specific references to the most recent CMOS-based works, including your own}% 

\paragraph{III-V Semiconductor Technologies}
% To increase the power output to tens of milliwatts, amplifiers based on III-V semiconductor materials such as InP can be utilized~\cite{deal2017660}. %\hl{@Zoya: here, it would be good to talk about the most recent III-V amplifiers at these frequencies, including your own work. Probably this is a good place to bring up the challenges in heterogeneous integration.} 
%
%%%%%%%%%%  ZOYA   %%%%%%%%%%%
%%%%%%%%%%%%%%%%%%%%%%%%%
%
%{Solid-state power amplifiers (PAs) are limited in total output power, bandwidth, and efficiency, and are expensive due to the need for power combining and low-loss packaging of the analog front end. 
%
Although silicon-based circuits have shown high-frequency operation with medium level power from power-combined devices, for watt-level power, III-V semiconductors are needed. To increase the power output to tens of milliwatts, amplifiers based on III-V semiconductor materials can be utilized~\cite{deal2017660}.

\glspl{hemt} using indium gallium arsenide (InGaAs) developed in the 1980s~\cite{cohen2012} allowed a path for \gls{mmwave} solid-state amplifiers and other components at W-band (75-110\,GHz). The power of an individual solid-state device at high frequencies is limited, and traditional binary corporate combining is dominated by loss before substantial power levels can be reached. Spatial power combining techniques introduced using \gls{gaas} technology include, e.g., a 272-element lens array using the same number of \gls{gaas} \glspl{mmic} with a total output power of 36\,W at V band (40-75\,GHz)~\cite{Sowers1999}. With research over the past few decades bringing wide band-gap semiconductors to the stage, \gls{gan} has been improving high-frequency performance with operating frequencies in the hundreds of gigahertz~\cite{Shinohara2013,Bozanic2019}. Although vacuum tubes have demonstrated several hundreds of kW at W-band, e.g., \cite{Felch2008-nuclear}, they are narrowband and require large power supplies and magnets. The wide bandgap of \gls{gan} and high associated operating voltages make it the semiconductor of choice for high-power solid-state transmitters. The \gls{sic} substrate additionally offers good thermal properties compared to \gls{gaas} and silicon. %Figure~\ref{fig:comparison} shows a useful qualitative comparison of different semiconductor technologies as a function of $f_T$ and $f_{max}$ up to 1.5\,THz. 
A number of recent millimeter-wave \gls{gan} processes with gate lengths in the 20--90\,nm range have shown high performance across V and W bands \cite{QorvoEuMC2022,HRLEuMC2022,Fraunhofer_GaN_W_band_switch_2018,Rocchi2022}.

% \begin{figure}[t]
%   \centering
%   \includegraphics[width=1\columnwidth]{Figures/SScomp.png}
%   \caption{Comparison of semiconductor processes for scaled frequency performance, as a function of transition frequency over maximal frequency. (Reproduced from \cite{Bozanic2019}.) \hl{Do we need both figures} {Zoya: feel free to delete. I do not know what you mean by ``both''.} Josep: I refer to the now Figure 2}
%   \label{fig:comparison}
% \end{figure}

Various \gls{gan} processes on \gls{sic} substrates currently achieve cutoff frequencies above 200\,GHz. For example, an $f_T$ up to 275\,GHz is shown in a 40-nm \gls{gan} on \gls{sic} \gls{hemt} process \cite{HRLEuMC2022}. Power densities as high as 3\,W/mm are shown at W-band in \gls{gan} on \gls{sic}~\cite{HRLEuMC2022}. Device efficiency is also increasing at W band, with 45\% \gls{pae} reported for a 3\,W/mm high power device at 94\,GHz, while the same device can reach 56\%~\gls{pae} with an output power density of 780\,mW/mm~\cite{HRLEuMC2022}. The highest published \gls{gan} W-band transmitter, intended for an active denial weapon, produces around 6.8\,kW~\cite{kW} by spatially combining over 8,000 \gls{gan}-on-\gls{sic} \gls{mmic} \glspl{pa}, each with over 1\,W of power and \gls{pae} $>20\%$ around 93\,GHz. This approach is modular and therefore scalable.

Advanced \gls{gan} processes also achieve minimum noise figures below 2~dB at W~band~\cite{QorvoEuMC2022, HRLEuMC2022}, and about 7.6~dB above 100~GHz \cite{Fraunh-LNA}. The ability to design with multiple gate lengths allows for increased complexity, such as low noise/high gain stages followed by high linearity stages in a single \gls{lna}~\cite{FraunhoferEuMC2022}. A comprehensive review of \gls{gan} \glspl{mmic} up to the 110-GHz range is given in~\cite{sonnenberg2022v}. Despite the impressive results in \gls{gan}, the lowest noise figures and highest frequencies of operation are obtained in \gls{inp}, e.g., \cite{leong2017850, cooke2021670, deal2017660, chattopadhyay2017340}.

\gls{gan} \glspl{mmic} with various functionality through W band have been demonstrated. This includes switches with high IP3 (over 30~dBm) and isolation (over 40~dBm)~\cite{sonnenberg2023v}, continuous 90$^\circ$ and 10$^\circ$ phase shifters from 50--110~GHz~\cite{romano202350}, active frequency doubling and tripling with conversion gain \cite{sonnenberg2023w}, active circulators~\cite{romano2023v} and a 50--110~GHz amplifier-isolator with 60\,dB isolation \cite{romano202346}.

%%%%%%%%%%%%%%%%%%%%%%%%%%%%%%%%%%%%%

\begin{figure}
    \centering
    \includegraphics[width=\columnwidth]{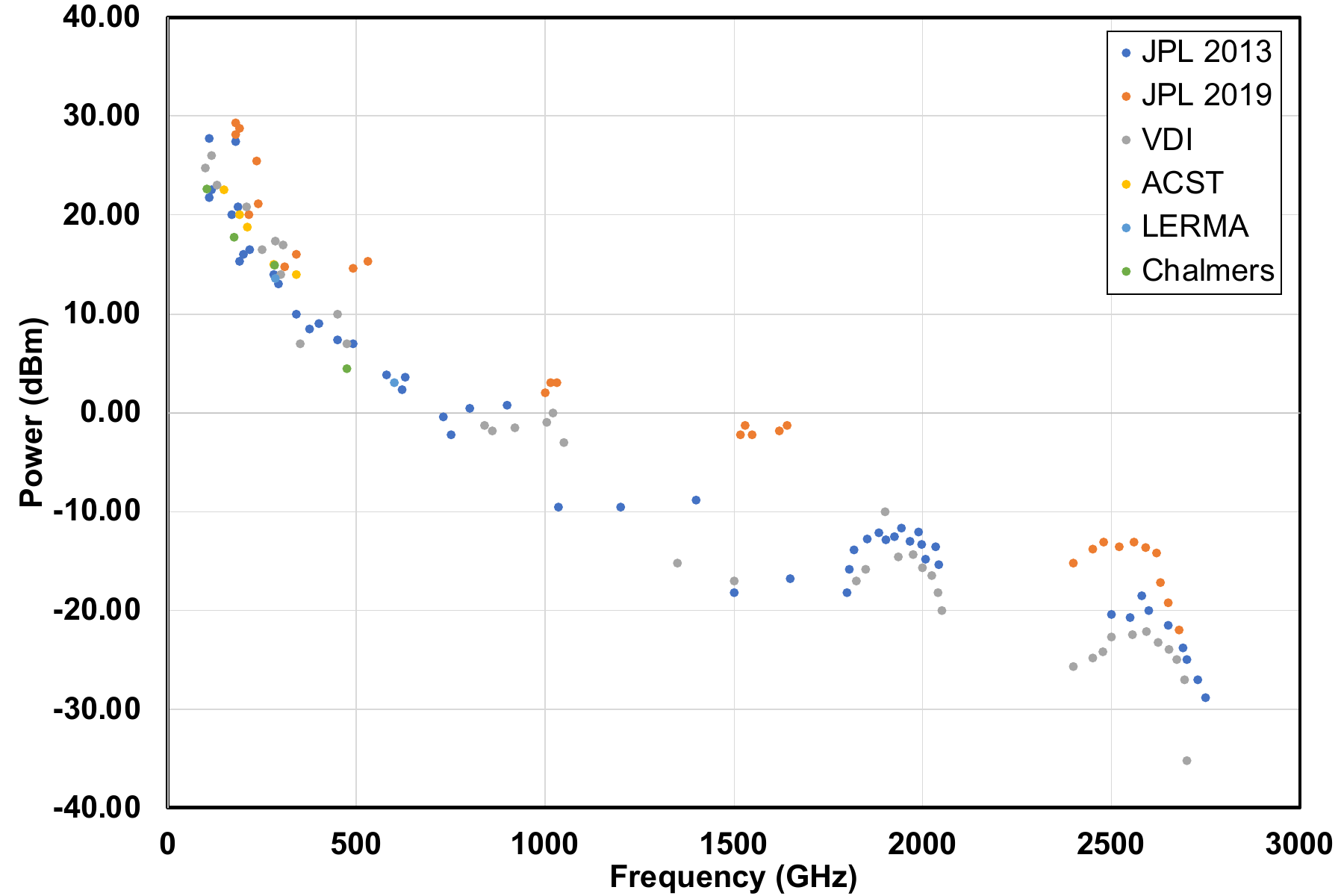}
    \caption{Schottky-diode-based frequency multipliers {survey. Output} power vs frequency for achieved by different groups, including the NASA JPL, Virginia Diodes Inc (VDI), ACST. LERMA at the Observatoire de Paris/C2N, and Chalmers University.}
    \label{fig:multipliers}
\end{figure}

III-V semiconductor materials can also be utilized to build different types of diodes with applications in terahertz signal generation and detection, including Schottky diodes, \glspl{rtd}, and \glspl{impatt}. For example, \gls{gaas}-based Schottky diodes are commonly used as frequency multipliers and mixers in frequency up- and down-converting chains~\cite{mehdi2017thz}. For example, in~\cite{siles2018new}, frequency-multiplied terahertz sources providing up to ten times more output power at room temperature than the previous state of the art are prototyped at 180~GHz, 240~GHz, 340~GHz, 530~GHz, 1~THz, and 1.6~THz. Using such technology, a frequency tripler with 200~mW of output power at 225~GHz is built and utilized to establish a 2-km-long link carrying multi-\gls{gbps} in~\cite{sen2023multi}. A survey of the most recent works demonstrated based on this technology is shown in Figure~\ref{fig:multipliers}.

% \hl{@All: where do we talk about packaging and all the thermal issues?}

% In general terms, electronic terahertz devices can provide high transmission power, but at the cost of non-linearities and phase noise.

\subsubsection{Photonic Approach}
%\hl{To be expanded} In this approach, the limits of devices and architectures used in optical communication systems are pushed down in frequency towards the terahertz band. There are mostly three fundamental techniques to do so. First, quantum cascade lasers (QCLs) are designed to emit terahertz photons, by leveraging intra-band transitions in III-V semiconductor materials~\cite{williams2006high}. When cooled to cryogenic temperatures, these can emit powers approaching 10~mW. Second, photoconductive antennas~\cite{berry2014generation}, which consist of a metallic terahertz antenna with a photoconductor material at the gap, can be utilized to downconvert an optical pulsed signal to terahertz frequencies. Such devices can be commonly found in time-domain spectroscopy platforms, with output powers approaching 1~mW. Last but not least, the use of frequency difference generation is becoming increasingly popular~\cite{sung2021design}. In this case, a photomixer is used to multiply a modulated optical laser with a second non-modulated laser at a second wavelength shifted from the first laser by the target terahertz carrier signal. While the generated terahertz power can still not meet that of electronic frequency-multiplied sources, the higher linearity, the much lower phase noise, and the possibility to leverage optical modulators make this a promising future path.

In this approach, the limits of devices and architectures used in optical communication systems are pushed down in frequency towards the terahertz band. The devices in this category are very diverse, ranging from electrically pumped to laser pumped, pulsed-wave to continuous-wave, and many even include electronic and/or plasmonic components. While offering faster, lower phase noise, and high Q-factor, this approach has a few disadvantages such as lower power, difficulty in implementation, cryogenic temperature operation, bulky, low efficiency, minimal tunability, and often only emits short pulses. 

An example of an electrically-pumped optoelectronic device is the \gls{qcl}~\cite{kazarinov71,williams2006high,sirtori10}, a type of semiconductor laser with emissions through intersubband transitions in a repeated stack of alternating semiconductors, forming multiple quantum well heterostructures. Laser-pumped devices~\cite{fulop20} include \glspl{dfg}~\cite{stohr06}, optical rectification, optical parametric oscillators, and \glspl{pca}~\cite{burford17}. The \gls{dfg} achieves optical heterodyne signal generation with a beam combiner of two optical lasers of different frequencies, $\omega_1$ and $\omega_2$, followed by a photomixer to output a single laser beam of frequency $\omega_2 - \omega_1$, in this case in the terahertz range. 

\glspl{pca}~\cite{berry2014generation} use laser pulses instead of continuous-wave lasers, which are incident on a highly resistive direct semiconductor thin film with two electric contact pads. The incident laser has higher photon energy than the semiconductor energy gap and is absorbed in the film, creating short-lasting electron-hole pairs until recombination.

\begin{figure*}
    \centering
    \includegraphics[width=\textwidth]{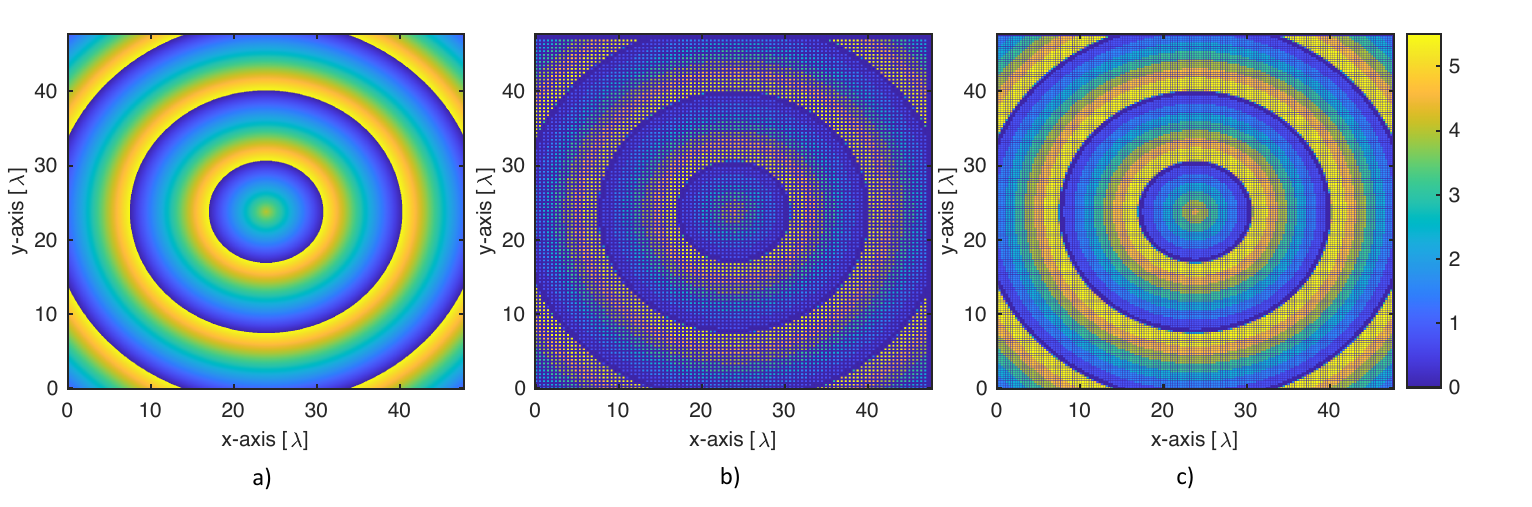}
    \caption{Generation of a Bessel beam through a) an ideal lens, b) a conventional antenna array with $\lambda/2$ element size and spacing, and c) a plasmonic antenna array with $\lambda_{spp}$$=$$\lambda/20$ element size and spacing. Note that at 300~GHz, $\lambda=$1~mm, and, thus, the theoretical footprint of these structures is only 5~cm.}
    \label{fig:antenna_resolution}
\end{figure*}

\subsubsection{Plasmonic Approach}
This approach aims to leverage the properties of plasma waves and surface plasmon polariton (SPP) waves to build devices that intrinsically operate at terahertz frequencies. Among others, plasma waves --or oscillations of electrical charges in a material-- at terahertz frequencies can be excited in an asymmetric plasmonic cavity formed in the \gls{2deg} channel of a \gls{hemt}, due to the so-called Dyakonov-Shur instability~\cite{dyakonov1993shallow}. Such a transistor can be built with III-V semiconductor materials~\cite{mikhailov1998plasma,knap2004terahertz,ryzhii2005transit,petrov2017amplified,nafari2017modeling} and/or with two-dimensional materials such as graphene~\cite{crabb2021hydrodynamic,aizin2023plasma}. 

Besides signal generation, the same structure can provide high-speed modulation of the generated plasma wave in frequency and amplitude~\cite{crabb2022amplitude}. Amplitude modulation is achieved by varying the DC bias current passing through the transistor channel. Alternatively, frequency modulation can be achieved by varying the gate voltage. In other words, this device can act as a direct amplitude or frequency up-converter from baseband or an \gls{if} to terahertz frequencies. Moreover, direct phase modulation can be achieved by means of a tunable plasmonic waveguide built again with graphene~\cite{singh2016graphene}. In this setup, by varying the bias voltage of the plasmonic waveguide, the speed of the SPP wave can be changed, and accordingly, the phase at the output of the waveguide can be modulated. In addition to other on-chip graphene-based modulators~\cite{island2020chip}, off-chip modulating structures based on graphene and other two-dimensional materials have been proposed~\cite{gopalan20202d}. These structures interact with the radiated signals and should not be confused with on-chip structures integrated with the transmitter that interact with the signal before being radiated. In this sense, they are closer to the antenna systems described in the next section.

In broad terms, the size of plasmonic devices is proportional to the plasmonic wavelength, which is generally much smaller than the free-space wavelength of a signal at the same frequency. The parameter that measures the ratio between the free-space and plasmonic wavelengths is the plasmonic confinement factor, and can easily be between 10 and 100 in graphene~\cite{grigorenko2012graphene}. As a result, the generated power of graphene-based plasmonic devices is generally very low (approaching 1~$\mu$W). However, their very small footprint supports both their embedding in nanomachines to enable the nanoscale applications of the terahertz band and their integration in larger numbers to build highly functional on-chip arrays~\cite{singh2020design}. Compared to the electronic and photonic approaches, the plasmonic approach has been much less explored, resulting in a high-risk, high-reward opportunity~\cite{shur2020terahertz}.

\subsection{Antenna Systems}
\label{subsec:antennas}

As in any wireless communication system, antennas are needed to convert on-chip signals into free-space propagating electromagnetic waves at the transmitter and perform the reciprocal function at the receiver. Moreover, antenna systems can also be found along the channel, acting as surfaces that can manipulate electromagnetic radiation in different ways (e.g., as fixed or programmable reflectors, focusing transmission lenses, and polarization filters, to name a few), {as we further elaborate in Sec.~\ref{sec:challenges}.} The key performance metrics of an antenna system include radiation efficiency, directivity gain, and beamwidth.

Fundamentally, classical antenna theory remains valid at terahertz frequencies, but there are some caveats. First, the very small wavelength of terahertz radiation leads to tiny terahertz antennas. For example, a resonant dipole antenna at 1~THz is approximately 150~$\mu$m and has the conventional doughnut-shaped radiation diagram. Correspondingly, in reception, the tiny size of the dipole results in a very small effective area or aperture, leading to very high spreading losses (more in Sec.~\ref{sec:channel}). Making the antenna larger to increase its effective area automatically leads to a more directional radiation pattern. This is why directional antennas are commonly used at terahertz frequencies. For example, horn antennas or even small dish antennas with gains ranging from 20~dBi to 55~dBi are commercially available today.

Another way to achieve high gain is by building antenna arrays. Antenna arrays offer the advantage of being able to program the radiation diagram by controlling at least the phase, if not also the amplitude, at every antenna. The very small size of individual terahertz antennas allows their dense integration in very small footprints. Besides the radiating elements, an array needs to integrate at least the control elements (e.g., phase shifters/time delays and amplitude controllers). Moreover, suppose the antenna needs to support multiple input multiple output (MIMO) communications. In that case, it will require numerous front-ends, up to one per antenna, but usually one per sub-group of antennas~\cite{lin2016terahertz}. Currently, designs with up to 16 ~streams, each feeding either one or a small group of antennas, have been demonstrated when following an electronic approach~\cite{abu2021end}.

In addition to antennas, lenses can be utilized to control the radiated terahertz signals. Lenses can be used to focus the signal at a distance or to generate different types of wavefronts, such as non-diffracting Bessel-beams~\cite{durnin1987exact, durnin1988comparison} and self-accelerating Airy beams~\cite{siviloglou2007observation}. Moreover, besides dielectric lenses, these functionalities can also be implemented using metasurfaces. Metasurfaces are arrays of meta-atoms or custom-designed electromagnetic elements whose size is much smaller than the wavelength~\cite{tao2009reconfigurable}. As with antenna arrays, programmable metasurfaces outperform fixed dielectric lenses as they can be tuned to implement different functionalities by inputting specific magnitude and phase distributions.

\begin{figure*}
    \centering
    \includegraphics[width=\textwidth]{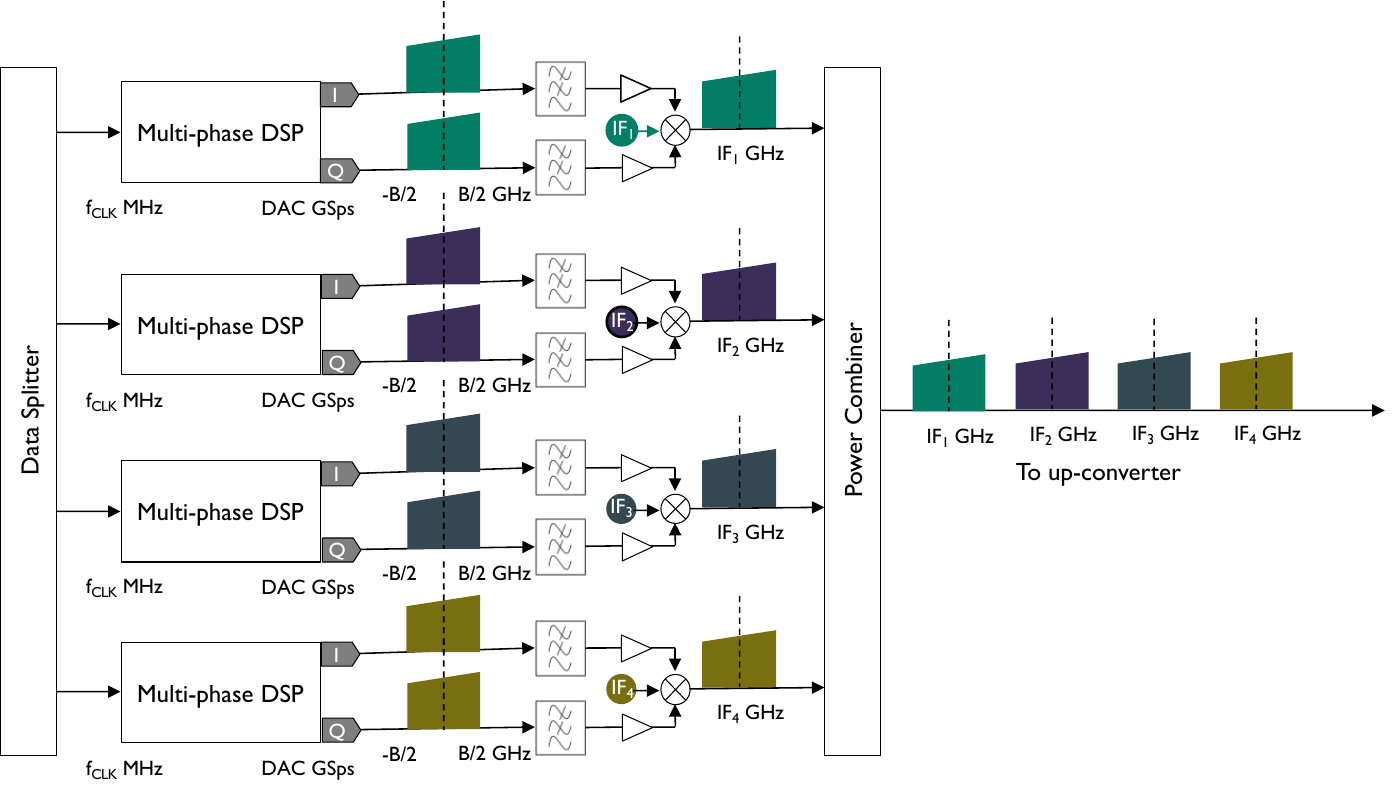}
    \caption{A multi-channel \gls{dsp} engine. A large bandwidth is digitally processed by splitting it into four broadband channels, and each channel can be processed in real-time through a multi-phase parallel design.}
    \label{fig:multi-channel}
\end{figure*}

\subsubsection{A Note on Plasmonic Structures} When adopting the plasmonic approach to build analog front-ends, plasmonic nano-antennas are needed~\cite{jornet2010graphene,tamagnone2012reconfigurable,jornet2013graphene,gric2018tunable}. The main advantage of plasmonic nano-antennas is that they are significantly smaller than the wavelength. While this leads to a smaller effective area, it also opens the door to their very dense integration~\cite{akyildiz2016realizing,yardimci2017high,hosseininejad2019digital}. Moreover, due to the small size of the plasmonic front-end itself, arrays of integrated plasmonic front-ends and plasmonic nano-antennas can be made~\cite{singh2020design}. The fact their size is smaller than the wavelength and that each array element provides its own power and independent phase and amplitude control results in the possibility of performing not only beamforming as with metallic antenna arrays but also wavefront engineering as with programmable metasurfaces. All these structures can be designed to operate in transmission, reception, or reflection~\cite{singh2022wavefront}.

\subsubsection{The Role of Reflecting Surfaces} As mentioned at the beginning of this section, in addition to the transmitter or the receiver, electromagnetic structures, including antennas, arrays, lenses, and metasurfaces, can be placed along the channel to enhance the propagation of terahertz signals. Particularly, \gls{irs} and \gls{ris} have become extremely popular across all frequencies~\cite{Wu2011intelligent,Liu2021reconfigurable,Huang2020holographic}. At lower frequencies, \gls{irs} can increase the spectral efficiency of existing networks~\cite{Yu2019miso,Yang2020coverage,Di2020hybrid,Almohamad2020}. As we move towards sub-terahertz and terahertz frequencies, \gls{irs} become a critical technology to increase coverage and, among others, overcome the impact of blockage by engineering \gls{nlos} paths around obstacles~\cite{nie2020beamforming,Hao2022ultra,chen2021intelligent}. Focusing on the hardware aspects, \gls{irs} at terahertz frequencies are usually based on programmable reflect-arrays or programmable metasurfaces~\cite{Liaskos2018wireless}. At lower frequencies, PIN diodes or varactors are utilized to change the reflection phase of each reflectarray or metasurface element. However, the design and fabrication of such devices are challenging at terahertz frequencies. {In~\cite{lan2023real}, a \gls{gan}-based reconfigurable metasurface has been designed and experimentally demonstrated. The metasurface implements an array-of-sub-arrays architecture with sub-wavelength spacing and 1-bit control per sub-array, and supports wavefront engineering at 0.34~THz.} Alternatively, the use of graphene as the tunable element in terahertz \gls{irs} has been theoretically proposed~\cite{singh2020hybrid,taghvaee2022multiwideband}. Lastly, non-reconfigurable reflecting surfaces with minimal complexity have been recently demonstrated~\cite{ju2023icc,badran2023design}.

\subsubsection{A Discussion on the Near Field of Large Radiating Structures} At this point, it is relevant to remember that the far field of a radiating structure depends on the antenna size as $2D^2/\lambda$, where $D$ is the largest antenna dimension~\cite{balanis2016antenna}. Accordingly, for example, the far field of a 20-cm antenna array at 120~GHz starts at 32~m. The far field of the same antenna size at 1.05~THz (i.e., the center frequency of the first absorption-defined transmission window above 1~THz) does not start until 280~m. If, instead, a much larger antenna structure, such as a 2~m dish or surface, is used, the far field of the same communication and sensing system at 120~GHz and 1.05~THz is 3.2~km and 28~km, respectively. As a result, in many applications, terahertz communications and sensing will happen within the near-field of the antenna. Unfortunately, traditional beam management strategies, which imply a plane wave assumption with a uniform phase and where the spreading effect results in a Gaussian intensity~\cite{balanis2016antenna}, including those proposed for terahertz systems, are inaccurate~\cite{singh2023wavefront}.

\subsection{Digital Back-ends}
\label{subsec:digital}
The main motivation to move to terahertz frequencies is its enormous bandwidth. Accordingly, a \gls{dsp} engine able to exploit such bandwidth is needed. The key performance metrics include the sampling frequency and the sampling resolution.

Common to optical (wired and wireless) systems, a major bottleneck in the \gls{dsp} engine is posed by the \glspl{dac} and the \glspl{adc}. As of today, data converters with sampling frequencies of up to 256~Gigasamples-per-second (GSaps) can be found in commercial laboratory-grade equipment (e.g., Keysight M8199B). As per the Nyquist sampling theorem, such data converters could operate with signals with analog bandwidths of up to $B=f_s/2$, i.e., ideally close to 128~GHz. Practically, this is less than that (e.g., 80 GHz for the aforementioned device). While this is remarkable, the size, cost, and thermal requirements of such data converters limit their application to very specific setups, far from what a handheld device could afford. In addition, such very high sampling frequencies come at the cost of lower resolutions (e.g., 8 bits).  A lower resolution leads to a higher \gls{evm} right from the start, i.e., at the transmitter, and thus, impacts the performance of high-order modulations.

Alternatively, highly parallelized \gls{dsp} engines are being developed. In particular, separate sub-channels can be independently processed by much slower data converters and (orthogonally) multiplexed in frequency (see Figure~\ref{fig:multi-channel})~\cite{ariyarathna2023toward}. Following this approach, we have recently demonstrated what, as of today, is the fastest \gls{sdr} platform for wireless communications, able to process in real-time 8~GHz of bandwidth by multiplexing four 2-GHz-wide channels in frequency~\cite{abdellatif2022real}. More specifically, this platform leverages the state of the art in \gls{rfsoc} and multi-phase processing strategies to generate four parallel IQ streams with baseband bandwidth of 1.25~GHz, each; and an \gls{if} multiplexing/demultiplexing custom board to generate/split a single ultra-broadband signal with 8~GHz of bandwidth. Multi-phase processing strategies are needed to match the difference in speed between the \gls{fpga} clock (e.g., up to 512~MHz in the most optimistic case) and the speed of the data-converters on the \gls{rfsoc} (e.g., a few \gls{gsps}).

{

\begin{table*}[ht]
\centering
\caption{ Summary of the State of the Art in THz Front-ends.}
\label{tab:front-ends}
\begin{tabular}{|p{1.4in}||p{1.4in}|p{1.4in}|p{1.4in}|}
\hline
\textbf{ Property} & \multicolumn{3}{c|}{\textbf{Technology Pathways}} \\
\cline{2-4}
 & \textbf{Electronic} & \textbf{Photonic} & \textbf{Plasmonic} \\ \hline\hline
\textbf{Frequency Range} &  Easily < 1~THz, potentially up go 10~THz &  Easily > a few THz, potentially < 1~THz &  1~THz and up \\ \hline
\textbf{Bandwidth} &  Up to tens of GHz & { > 10~GHz} & {> 10~GHz} \\ \hline
\textbf{Transmit Power} &  100s of mWs < 300 GHz, few mWs at 1 THz & { <10~mW} & <1~m$W$ \\ \hline
\textbf{Amplitude and Phase Noise} & High & Low & Unknown \\ \hline
\textbf{Technology Maturity} & High & Medium & Low \\ \hline
% \hline
% \emph{Main references} & \cite{wang2023power,siles2018new} & \cite{} & \cite{} \\ \hline
\end{tabular}
\end{table*}
}

\subsection{Optimizing Power Waste for Green Communications} Green \gls{ict} has become an important topic for sustainable development as energy demand for \gls{ict} soars, and the impacts of anthropogenic climate change become unavoidable \cite{matinmikko2022wiopt}. ICT is estimated to consume over one-fifth of the global electricity supply by 2030, equivalent to 8000 TWh/year \cite{hoefflinger2020spr}. While limiting the growth of \gls{ict} is impractical, improving efficiency and reducing power waste of communications devices certainly alleviates the energy burden. As 6G networks are estimated to serve millions of connected devices with smaller cell sizes, potentially leveraging the terahertz and sub-terahertz spectrum, a metric to quantify the power waste of individual devices, cascaded communication systems, and networks is urgent. The \textit{power waste factor, W,} denotes the amount of wasted power for devices, systems, or networks \cite{Kanhere2022iwc,Ying2023gc}. $W$ can be used by circuit and system designers to make informed decisions about design choices for devices based on the amount of power wasted. $W$ provides an intuitive mathematical framework for power waste that closely resembles Harald Friis' noise figure \cite{friis1944ire}. 

{
In particular, the power waste factor, $W$, for any device or linear system (including any active or passive device or channel) is defined as
\begin{equation}
    W = \frac{P_{consumed,path}}{P_{signal}}
\end{equation}
where $P_{consumed,path}=P_{signal}+P_{non-signal}$~\cite{Ying2024mw,rappaport2024waste}. The waste factor efficiency, $\eta W$, is the reciprocal of $W$.
The additive wasted power of any device is given by
\begin{equation}
    P_{waste} = P_{non-signal} = (W-1)P_{signal}
\end{equation}
where it is clear that any power that is consumed but not used in the output signal (e.g., $P_{non-signal}$) is wasted power.
}

Considering a cascade of $N$ devices having a device gain, $G$, $W$ for the device cascade can be obtained as~\cite{Ying2023gc}:
\begin{equation}\label{WF1}
	W={{W}_{N}}+\frac{\left(W_{N-1}-1\right)}{G_N}
	+\frac{\left(W_{N-2}-1\right)}{G_N G_{N-1}}
	+\ldots+\frac{\left({W_1}-1\right)}{\prod_{i=2}^N G_i}.
\end{equation}
With \textit{W}, circuit designs can be optimized for minimal energy waste; individually optimized devices can then be cascaded to build energy-efficient communication systems that form an energy-optimized network together. Further, W can also be implemented in data centers to minimize their power waste \cite{Ying2023gc}. The wasted power determined using $W$ can be elegantly tied to the data rate delivered by the cascaded communication system to evaluate the Consumption Efficiency Factor (CEF) \cite{murdock2013jsac,Kanhere2022iwc}. Mathematically, CEF is expressed as 
\begin{equation}\label{CEF1}
	\text{CEF}~\left[\frac{bps}{Watt}\right]=\frac{\text{Data rate }[bps]}{W\times\text{Output Signal Power }[Watt]}
\end{equation}
and signifies the data throughput per watt of power consumed. Therefore, with $W$ and $CEF$, minimizing energy waste becomes an intrinsic part of the design process, creating the pathway for realizing the ``green-G" future.

{
\subsection{Summary}
In a nutshell, these are the key takeaways relating to the state of the art in THz device technologies:
\begin{enumerate}
    \item Terahertz analog front-ends are quickly becoming available. Today, electronic frequency-multiplied transceivers at sub-terahertz frequencies are commercially available for equipment testing while the research community shifts towards developing silicon/III-V heterogeneous systems, photonics, and, eventually, plasmonic-based configurations. A summary of the technologies is provided in Table~\ref{tab:front-ends}.
    \item The small wavelengths of terahertz signals enable the development of (sub) millimetric omnidirectional antennas as well as very high gain directional antennas with compact footprints (e.g., 20~dBi in 1~cm$^2$). The latter can be in the form of fixed directional antennas (e.g., horn antennas, commercially available) or antenna arrays (still in development).
    \item In addition to antennas, compact lenses, and metasurfaces can be utilized to engineer the terahertz radiation at the transmitter, at the receiver, and along the channel in the form of \glspl{ris}. The main challenge today is to make such surfaces programmable at terahertz frequencies due to the challenges in developing tunable elements in this band.
    \item While the main motivation to move to terahertz frequencies is the availability of larger bandwidths, today, one of the main technology challenges relates to the ability to digitally process them, mainly due to the cost, size, power requirements, and energy consumption of high-speed data-converters. This motivates both the adoption of largely parallelized signal processing techniques (e.g., multi-channel systems), as well as the development of hybrid analog \& digital processing techniques. 
    \item Across all the building blocks of a terahertz radio, energy efficiency becomes critical. The waste factor is a new metric that allows end-to-end optimization of energy consumption. 
\end{enumerate}
}

% \subsection{Hardware for Communications and Sensing}
% \label{subsec:jcsh}
% As illustrated in Section~\ref{sec:use_cases}, the terahertz band provides equally exciting opportunities for both communications and sensing. 

%----------------------------------------------------%
\section{Key Lessons from\\Terahertz Channel Modeling}
\label{sec:channel}

\subsection{Overview}
Channel measurements and channel modeling for terahertz communications in various environments have been one of the primary research targets in the community until very recently. To date, hundreds of measurement-based and simulation-based terahertz channel models have been presented in parallel to tens of analytical terahertz channel model studies~\cite{han2022terahertz,Ghosh2023THz}.
{Because of the high close-in free space path loss at greater frequencies (e.g., smaller wavelengths), virtually all propagation measurements and communications in the terahertz bands (perhaps with the exception of \gls{wnoc} and \gls{iont}) require directional antennas, from which omnidirectional channel models are created~\cite{sun2015synthesizing,rappaport2012cellular}.
Omnidirectional channel models are used by industry within standards bodies such as the \gls{3gpp} global cellphone standards body so that any antenna pattern may be implemented in simulation~\cite{samimi2015statistical,xing2021millimeter}.
}

Notably, as we argue in this article, the evolved understanding is that there will not likely be any single widely adopted terahertz channel model. Instead, there is a need to develop a range of terahertz channel models tailored to a specific use case and a particular propagation environment. {The reason here is that while the physics of the terahertz wave propagation does not change much among the use cases, different distinct features of terahertz propagation have different weights when it comes to their contribution to the structure of the received signal.} Therefore, it is more beneficial to work on a set of terahertz channel models for different use cases rather than aim to deliver a single unified model (with enormous complexity) that suits them all.

Below, we briefly outline and review the latest findings in the field going from simpler to more complex setups that are potentially within the 6G timeline: outdoor terahertz links, indoor terahertz access, and vehicular terahertz systems. We also mention the specific use cases from the list above for which the given modeling approach is the most applicable. We start with reviewing outdoor setups, where the line-of-sight (LoS) terahertz path is the dominant factor determining the performance in subsection~\ref{subsec:outdoor_channel}. We then proceed with subsection~\ref{subsec:indoor_channel} detailing the peculiarities of terahertz channel indoors. We later discuss vehicular terahertz setups in subsection~\ref{subsec:vehicular_channel}. {
The key takeaways are given in subsection~\ref{sec:channel_summary}. Table~\ref{tab:channel} illustrates and complements our discussion.
}

%---------------%
\subsection{Outdoor Terahertz Channel}
\label{subsec:outdoor_channel}

\begin{figure}
 \centering
 \includegraphics[width=\columnwidth]{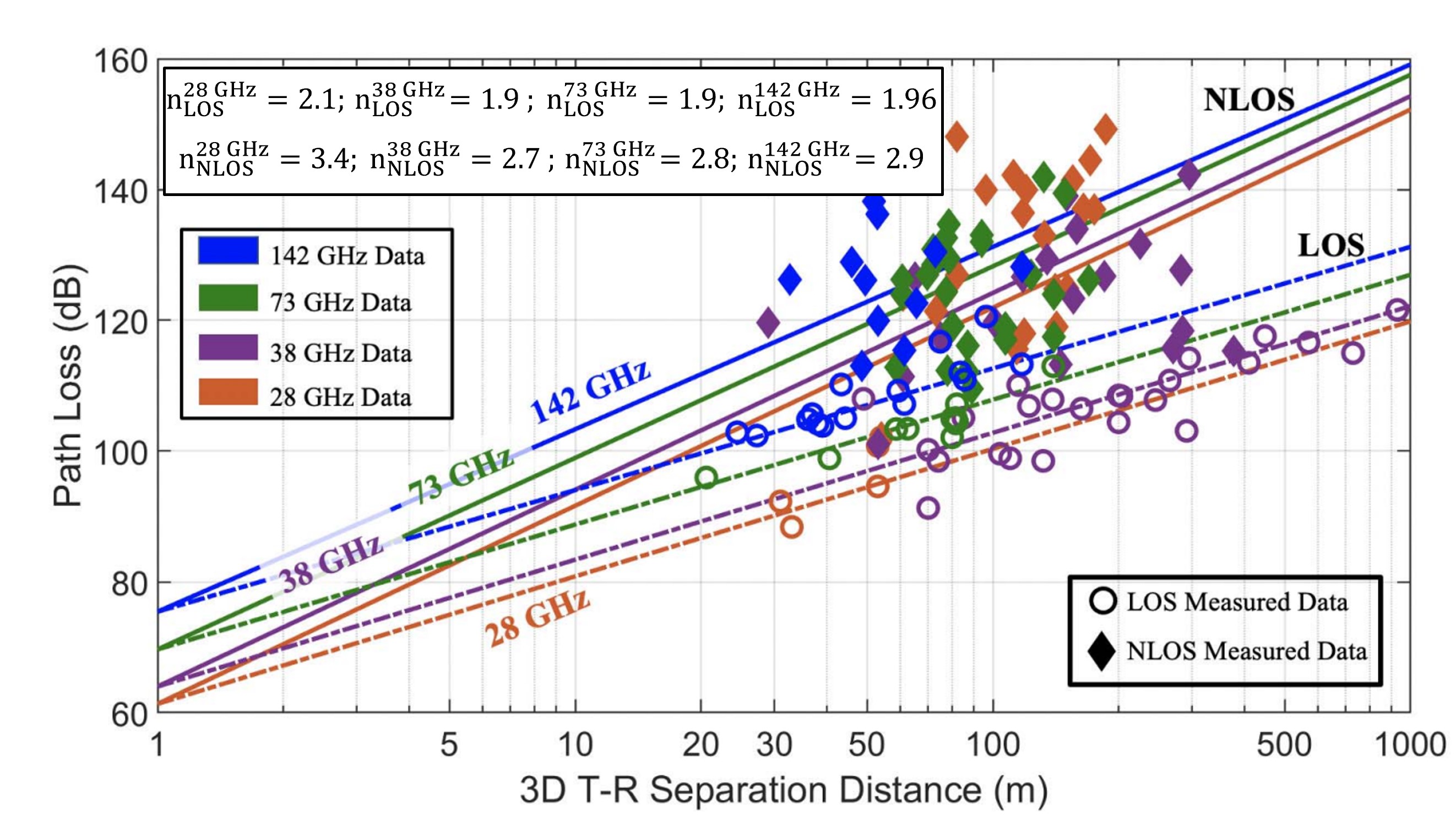}
 \caption{\textcolor{black}{Outdoor UMi omnidirectional CI path loss models with a reference distance of 1~m and without antenna gain~\cite{Xing2021iclModels}. 28, 73, and 142 GHz measurements were conducted in New York and 38 GHz in Austin, TX.}}
 \label{fig:outdoor_channel}
\end{figure}

The non-occluded LoS terahertz channel typical for various outdoor environments is one of the most in-depth studies to date. During the last two decades, numerous measurement-based and modeling-based studies have explored various aspects of outdoor terahertz channels. We aim to summarize the key findings from those below. In brief, the impact of five different effects have been studied to date: (i)~spreading loss; (ii)~molecular absorption loss; (iii)~molecular absorption noise; (iv)~scattering; and (v)~diffraction~\cite{han2022terahertz}\footnote{In addition to these five, outdoor terahertz channel may also feature non-negligible impact of human-body blockage and the multi-path components coming from, e.g., non-line-of-sight paths, reflected or scattered from buildings or other obstacles, as further discussed in the next subsection.}.

%-----%
\subsubsection{Spreading Loss}
Propagation path loss over distance is based upon the spherical spreading of the radiated wavefront (e.g. $1/d^2$) as modeled by H. Friis in the 1940s~\cite{rappaport2002ph}. The distance-squared path loss is well known to represent free space loss due to spreading, with the ``$2$'' often referred to as the path loss exponent~\cite{Rappaport2015book}. Today, the general understanding is that the spreading loss is the dominant factor when it comes to LoS outdoor sub-terahertz and terahertz channels contributing the most to the overall attenuation in the majority of conditions. This conclusion has been confirmed in various theoretical~\cite{kurner2007propagation,jornet2011channel}, simulation-based~\cite{tarboush2021teramimo,yi2022ray,zhang2023deterministic}, and measurement-based studies for a wide range of scenarios including \gls{umi}~\cite{Xing2021iclModels}, \gls{uma}~\cite{Samimi2016tmtt}, and \gls{rma}~\cite{Mac2017jsac}, among others\cite{Rappaport2017tapOverview, Han2018icm, Han2022icst, Shafi2017jsac, Abbasi2023twc, Sheikh2023ijm, Guan2019tvt}. \textcolor{black}{Based on the close-in path loss model with a reference distance of 1 m, measurements conducted for UMi scenario in New York and Austin, TX, Figure 8 plots the path loss with distance at mmWave and sub-terahertz frequencies in Line-of-Sight (LoS) and Non-Line-of-Sight (NLoS) \cite{Xing2021iclModels}. The CI model characterizes the path loss behavior with a single parameter, the path loss exponent (n), beyond the free-space reference distance and was found to be the most robust using a vast range of measurement databases across the mmWave and the sub-THz spectrums~\cite{sun2018propagation}. To overcome the increasing free-space path loss, wireless systems have adopted highly directional antennas and antenna arrays with narrow beamwidths at both ends of the link\cite{Mac2017jsacChSdr,rappaport2022cup}.} As the physics of terahertz spreading is not principally different from lower frequencies (including \gls{mmwave}), most of the measurement-based works agree on the fact that the average distance-dependent path loss exponent for sub-terahertz and terahertz channels is around 2 (with slight deviations depending on the scenario)~\cite{Poddar2023comst}, as illustrated in Figure~\ref{fig:outdoor_channel}. Hence, the canonical \gls{fspl} equation originated by Friis holds for \gls{los} terahertz propagation as well.

However, one essential element to highlight here is that the \gls{fspl} equation is based on the Friis transmission equation, which is only valid in the far field~\cite{balanis2016antenna}. Therefore, the \gls{fspl}-based model is only valid for long-range stationary terahertz communications (e.g., for fronthaul and backhaul) employing small-scale horn or lens antenna. On the contrary, \emph{mobile} terahertz communications for 6G-grade and 7G-grade wireless access links will likely utilize notably larger phased terahertz arrays or, even leverage terahertz \gls{irs}~\cite{chen2021intelligent,singh2022wavefront,dovelos2021intelligent}. The distinct feature of the latter is a non-negligible near-field zone of the terahertz antenna system that can be comparable to the range of mobile terahertz communications, as discussed in Section~\ref{subsec:antennas}. %For instance, the near-field zone of 12.5~cm-wide 140~GHz antennas employed in the TeraNova terahertz platform is slightly over 12~m, while scaling the system to true terahertz frequencies beyond 1~THz will push the near-field zone over 50~m. 
The near-field region for commonly used terahertz antennas can be of tens to hundreds of meters or even more as we move to true terahertz frequencies. Such values are comparable to the coverage range of prospective terahertz outdoor cells and also exceed the target coverage range of indoor terahertz \gls{wlan} access points discussed in the next subsection. The fact that the receiver is in the near field zone of the transmitter lowers the distance-dependent exponent far down from 2 and also leads to multiple research challenges summarized in the next section.

%-----%
\subsubsection{Molecular Absorption}
The second key feature often mentioned in terahertz outdoor channel studies is the additional signal attenuation caused by molecular absorption. While molecular absorption loss is not exclusive to terahertz frequencies (e.g., some \gls{mmwave} sub-bands, such as around 60~GHz, are notably affected as well), it can be orders of magnitude stronger than at lower frequencies~\cite{jornet2011channel}. Molecular absorption loss does not exist in a vacuum but in other environments typical for terahertz communications. Specifically, when the terahertz wave propagates through a gas, some portion of the wave energy gets converted into the kinetic energy of some of the environment molecules that have their resonant frequencies next to the ones of the terahertz wave itself~\cite{goody1995atmospheric}. The presence of molecular absorption loss (i)~makes terahertz channel (even in the simplest \gls{los} case) frequency-selective; and (ii)~adds an additional distance-dependent exponent to the aggregated path loss equation that notably complicates further analysis~\cite{petrov2016interference,Kokkoniemi2016frequency,petrov2017interference}.

\begin{figure}
 \centering
 \includegraphics[width=0.95\columnwidth]{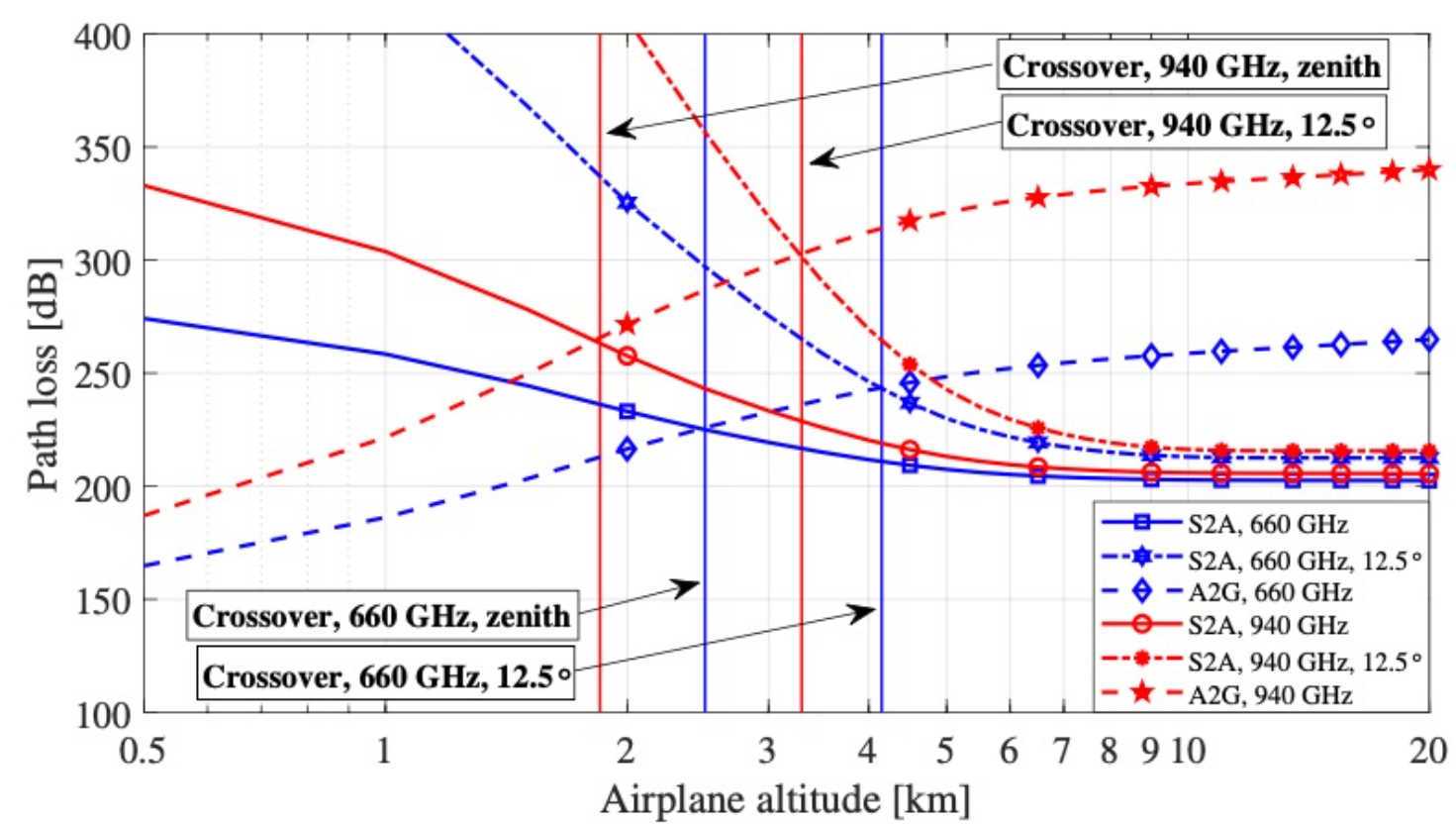}
 \caption{\textcolor{black}{Airplane terahertz communications measurements and results~\cite{kokkoniemi2021channel}.}}
 \label{fig:airplane_channel}
\end{figure}

However, the impact of molecular absorption is of secondary importance and can be neglected in many typical use cases and environments under two important conditions. The first condition is that the entire spectrum of the transmitted signal lies within the so-called \emph{terahertz transparency window}~\cite{boronin2014capacity,petrov2015efficiency}. For any sub-terahertz transmission, this one is fulfilled automatically, as the first transparency window is up to 540~GHz, while sub-terahertz transmissions are limited to 300~GHz~\cite{marcus2022millimeter}. The second important condition comes from the distance-dependency of the molecular absorption loss. Here, it has been revealed both theoretically and experimentally that (under the first condition as well) the molecular absorption loss has an impact of less than 2~dB for any distance shorter than a few hundred meters. Therefore, the impact of molecular absorption can be ignored in the first-order analysis of sub-terahertz channels for outdoor cellular access, \gls{v2x}, vehicular radars, and even short-range fronthaul and backhaul terahertz links. The only exception here is substantially wet weather (e.g., heavy rain) when an additional 2-3~dB should be subtracted from the received power value estimated by the terahertz channel model~\cite{sen2022terahertz}.

Airborne terahertz communications (including airplane-to-satellite links) are a special case for terahertz channel modeling. Particularly, as terahertz molecular absorption is not only distance but also a pressure- and temperature-dependent variable, the value changes at different altitudes. As both higher temperature and greater pressure contribute to the terahertz molecular absorption loss, its value decreases rapidly with altitude (lower pressure and colder air). Specifically, terahertz molecular absorption loss becomes almost negligible at altitudes greater than 10~km making this effect not only distance- but also altitude-dependent. For selected terahertz sub-bands this comes to an interesting trade-off as recently revealed in~\cite{kokkoniemi2021channel} and illustrated {in Figure~\ref{fig:airplane_channel}. Specifically, starting from a given altitude, a notably longer (e.g., 200~km) link between an airplane and a \gls{leo} satellite features \emph{lower} loss than a comparable link between the same airplane and the ground station (no more than 10~km away).}

{
Notably, molecular absorption in terahertz communications is not always an exclusively negative effect to avoid. In contrast, this effect can also be leveraged to e.g., boost the secrecy and security of the data exchange over the terahertz channel. Specifically, as discussed in~\cite{Han2023molecular} among other works, the transmission over the terahertz band can be limited in range by sensing the signal in the sub-band close to one or several ``absorption lines'' -- frequencies features by a notably greater molecular absorption loss. Hence, if the separation distance between the transmitter and the receiver is known, the transmitter becomes capable of ``cutting'' the message propagation beyond the receiver, so any other nodes staying further away cannot reliably decode the message.
}

%-----%
\subsubsection{Molecular Absorption Noise}
The third distinct feature studied for terahertz communications is the molecular absorption noise. {This effect is caused by the fact that a fraction of the kinetic energy absorbed by the molecules from the terahertz signal is reemitted back to the environment as a standalone signal at the same/similar frequency as per the so-called emissivity of the channel~\cite{goody1995atmospheric,jornet2011channel}.} Canonical 5G-grade channel models typically decouple noise-related effects from channel-related effects, as most noise sources are present at the receiver and are thus not part of the channel model itself. This is different for terahertz communications, as this specific type of noise -- molecular absorption noise --- comes not from the receiver but from the environment itself. It has been further revealed that this noise source is also correlated with the transmitted signal (e.g., greater transmit power leads to greater noise level)~\cite{jornet2014femtosecond,kokkoniemi2016discussion}. Hence, accounting for the presence of molecular absorption noise properly is extremely {challenging. E.g., }even the use of canonical Shannon capacity and \gls{snr} formulas implies that the noise is independent of the signal, which is no longer true at terahertz frequencies. The resulting analytical channel and interference models in the presence of molecular noise get notably more complicated compared to the ones in the very same setup but ignore this effect~\cite{petrov2017interference}.

Fortunately, the total power of molecular absorption noise theoretically cannot exceed the power captured from the molecular absorption loss discussed above. Hence, the latest studies conclude that (while the molecular absorption noise is likely present in most of the setups) its contribution to the received signal power and shape is of secondary importance compared to (i)~spreading loss and (ii)~molecular absorption loss itself. Therefore, a common trend recently is to start ignoring this effect in complex first-order studies for the sake of analytical tractability~\cite{Petrov2020Capacity,Poddar2023comst}. This choice is recommended for channel modeling targeting most practical terahertz use cases from Section~\ref{sec:use_cases} and most environments. While the accuracy is affected only slightly, the overall complexity of the analysis decreases dramatically, allowing for more sophisticated models to be built.

%-----%
\subsubsection{Scattering and Diffraction}
The remaining two items related to terahertz channel modeling in open environments are scattering and diffraction. Both can be theoretically characterized by relatively complex expressions, as in, for example, in~\cite{kokkoniemi2014frequency}, among other works. The good news here is that scattering loss in terahertz communications (in contrast to optical wireless systems) in typical homogeneous environments (e.g., air) is extremely low and can be ignored in first-order analysis in most conditions. Two exceptional cases here are: (i)~adverse weather conditions (specifically, snow), where scattering from snow particles may lead to additional 5-8~dB losses and extra multi-path created even in a pure LoS channel~\cite{amarasinghe2020scattering}; and (ii)~Scintillation/atmospheric turbulence effects. Therefore, as discussed further in~\cite{sen2022terahertz}, different terahertz channel models for the fronthaul/backhaul terahertz wireless link are needed for summer and winter conditions.

Regarding diffraction, the effect is especially visible when the terahertz signal meets a sharp object (e.g., a side of the building or furniture element). One of the essential findings here is that the terahertz link can be established even in partial blockage in the ``obstacle's shadow'' due to the diffraction effect~\cite{kokkoniemi2016diffraction}. However, this finding is primarily related to \gls{nlos} terahertz communications discussed in the next subsection. On the contrary, diffraction from the environment molecules themselves (e.g., different molecules present in the air) is almost always negligible when it comes to terahertz channel modeling.

%---------------%
\subsection{Indoor Terahertz Channel}
\label{subsec:indoor_channel}
Another active area of research related to terahertz channel modeling primarily targets indoor terahertz channels. Indoor terahertz small cells or terahertz-empowered \glspl{wlan} are among the target use cases and deployment scenarios for terahertz communications, as discussed in Section~\ref{sec:use_cases}. Complementing the findings on \gls{los} terahertz propagation discussed above, indoor terahertz channel measurement and models primarily focus on the impact of the terahertz wave interaction with typical obstacles, such as building walls, furniture elements, and human bodies\footnote{As noted in the previous subsection, selected findings on the human-body blockage and reflection/scattering from building walls may also be relevant to outdoor terahertz channel models, e.g., for outdoor mobile access links.}.

%---%
\subsubsection{Room environment: Home or Office}
Home or office is one of the most present environments when it comes to research works on indoor terahertz channels (both measurement-based and modeling-centric). Several key building blocks are needed to enable accurate terahertz channel modeling for indoor environments. The first one is an accurate and flexible terahertz LoS propagation model capturing all the essential features from the previous subsection. The second building block in an accurate characterization of the major objects present in the environment (particularly their penetration, reflection, and scattering properties at the frequencies of interest must be revealed). 

Here, an extensive set of studies has been delivered at the early stages of 5G standardization for \gls{mmwave} communications, specifically focusing on 26~GHz--30~GHz and 52~GHz--71~GHz frequency bands. The key findings from these studies are well summarized in earlier tutorials on \gls{mmwave} indoor propagation~\cite{electronics10141653,Rappaport2017tapOverview,Wang2018Survey}. After a decade of measurement and modeling of indoor sub-terahertz and terahertz communications, we may conclude that there is a great similarity between the \gls{mmwave} and the terahertz channel when it comes to indoor environments~\cite{xing2019indoor,chen2021channel,chen2023channel}. For both frequency bands, the presence or absence of the \gls{los} component is the primary factor determining the link performance. Further, both bands feature strong reflections from flat surfaces, such as office desks and primarily walls, glass windows, and even ceilings~\cite{petrov2018last}.

The key difference of terahertz communications here is not qualitative but quantitative. Specifically, many typical indoor surfaces are good reflectors for low-\gls{mmwave} signals (e.g., 28~GHz). Therefore, many indoor \gls{mmwave} channel models assume them perfectly flat and act as reflectors with a certain loss of usually no more than a few dB. The difference here comes from the fact that for sub-terahertz and especially true terahertz frequencies above 1~THz (hence, over 30 times shorter wavelength than at 28~GHz), the same typical home or office surfaces (e.g., a painted wall or a wooden door) are not flat anymore with non-negligible roughness~\cite{jansen2011diffuse}. \textcolor{black}{As an example, 142 GHz scattering measurements in~\cite{Ju2019iccScatter} identified rough surfaces with the Rayleigh Criterion \cite{rappaport2002ph}, and found relatively high scattered power in non-specular directions for incident angles below 30 degrees for drywall.} As the variations in the surface level become comparable to the sub-millimeter wavelength of terahertz communications, the reflected signal is weaker, while additional scattered signal copies are created. These scattered signals do not follow the law of reflection, so they propagate in many directions. Further, these scattered components get reflected and scattered further from other objects in the environment, creating additional signal copies. Hence, while many \gls{mmwave} models give sufficiently accurate predictions indoors by purely modeling LoS and reflected paths (e.g., via ray-tracing methods), an accurate channel model for terahertz communications benefits from also including at least the first-order scattered paths (transmitter-obstacle-receiver) into the analysis~\cite{petrov2018last}.

%---%
\subsubsection{Corridor and Data Center}
The key peculiarity of indoor terahertz channel modeling for corridors is that a long corridor may act as a ``waveguide'' for terahertz signals. Notably, as the length of the corridor is typically at least several times greater than its width and height, the \gls{aod} and \gls{aoa} for most of the signal paths are relatively close to each other. This is a distinct feature of the corridor environment from a regular office or homeroom, where multipath comes from almost any angle due to pseudo-random reflection and scattering from the surfaces surrounding the transmitter and the receiver. Therefore, many of these multipath copies of the transmitted signal will be successfully received even when using narrow-beam terahertz antennas. Hence, in contrast to an indoor setup that features \emph{occasional} multipath that can be ignored in first-order studies, the terahertz channel model for corridors must capture the multipath components coming at least from the first- and second-order reflections, as they may be not notably weaker than the LoS component.

\if 0
\textcolor{black}{Industrial factories were proposed as a distinct scenario in the \gls{3gpp} Release 15~\cite{38901} for wireless propagation from other indoor environments such as offices and shopping malls due to the presence of various metallic machines, shelves, and robots. A comprehensive survey on the \gls{3gpp} standardized channel model for industrial Internet of Things (IIoT) scenarios is provided in~\cite{Jiang2021iotj}. Propagation measurements conducted at 142 GHz in four factories found similar LoS path loss exponent of $\sim$ 1.8 in indoor rooms and offices with corridors, but increased NLoS path loss exponent of $\sim$ 3.1 compared to $\sim$ 2.8 for an indoor room or office~\cite{Ju2023twc}.
}
\fi

In general, modern channel models developed for terahertz communications in data centers (one of the distinct use cases for terahertz, as discussed in Section~\ref{sec:use_cases}) have large similarities to the ones used for corridors. {This comes from the similarity of the environments and propagation conditions, as most data centers exploit corridor-based layouts with several rows of racks staying parallel to each other with a certain (usually fixed) separation distance.} The key difference here comes primarily from the materials used for the server rack blocks (usually flat aluminum, steel, or plastic covers). Hence, a series of server racks acts similar to a flat wall with perfect reflection capabilities. As a result, the formed corridor may also act as a waveguide leading to the \gls{pdp} featuring a single strongest LoS component that comes first. The \gls{pdp} continues by a few visible slightly weaker first- and second-order reflections followed by numerous higher-order reflections and scattered components that are typically too much delayed to contribute to the received power of the given symbol and primarily contribute to the \gls{isi}~\cite{Cheng2020Characterization}.

%---%
\subsubsection{Human-body Blockage}
One of the essential features of \gls{mmwave} and, especially, terahertz communications is substantial penetration loss when propagating through the human body. According to medical studies, up to 60\% of the adult human body consists of water, which is a strong absorber of terahertz radiation. Consequently, the human-body penetration loss at \gls{mmwave}, sub-terahertz, and terahertz frequencies ranges on the conditions (angle of incident, part of the body affected, beamwidth, etc.) but is in the order of 20~dB up to 35~dB~\cite{Slezak2018Empirical}. Such great losses not only challenge the overall average system performance (e.g., the capacity of the terahertz link) but also lead to frequent and unexpected outage events caused by dynamic human-body blockage.

An important aspect here is that these dynamic blockage events happen rapidly (e.g., the power degrades fast when the human crosses the communication path), so the communication system often does not have sufficient time to detect the event and react accordingly when the blockage happens. In normal conditions, these dynamic blockage events may last from hundreds of milliseconds up to several seconds, thus violating almost the \gls{qos} requirements for almost any traffic category besides the background file transfer~\cite{gapeyenko2017temporal}.

\if 0
\begin{figure}
 \centering
 \includegraphics[width=0.95\columnwidth]{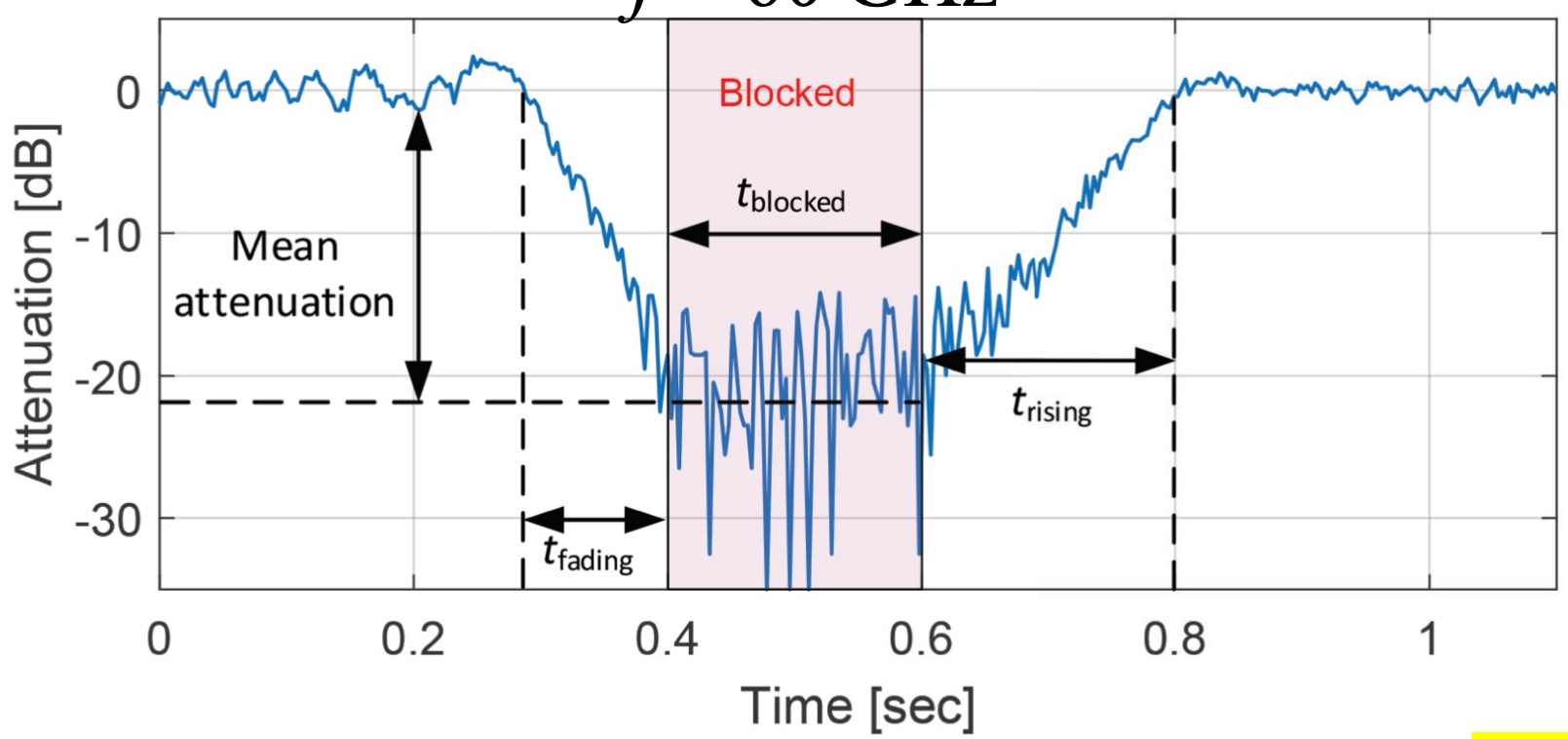}
 \caption{\textcolor{red}{Dynamic human-body blockage in terahertz communications (reproduced from~\cite{Slezak2018Empirical}).}}
 \label{fig:blockage_channel}
\end{figure}
\fi

Therefore, the latest studies agree that the human-body blockage must be accounted for in terahertz channel modeling for use cases and environments that involve human users (or even the pedestrians passing by who are not users). A consensus approach here is to model the human body as a cylinder with a given height and width representing a typical person (e.g., 1.7~cm in height and around 0.4~cm in width). Whenever a certain part of the signal passes through the human body, the corresponding signal is attenuated by a given number of decibels (usually, no less than 20~dB).

A more advanced option is to model diversity of humans in the environment by defining the cylinder height and width as stochastic variables (e.g., following a Normal distribution with a given mean and variance~\cite{gapeyenko2016analysis}). This approach acknowledges the fact that humans are different from each other but requires greater computational resources (for simulation-based models) as well as leads to more sophisticated analytical expressions (for theoretical models). Finally, the most in-depth approach suggests replacing a first-order cylinder model with more accurate models for different human body parts~\cite{Mokhtari2019Human}. This approach improves the accuracy of the resulting models but is exploited relatively rarely due to its notably greater complexity.

Another essential aspect of human-body blockage for indoor/outdoor terahertz access links is \emph{self-blockage} caused by the body of the human user itself. Here, the most widely adopted approach is to model a fixed separation distance between a terahertz mobile device and the person holding it. Hence, a certain 2D or 3D angle sector gets blocked by the user body, as revealed in~\cite{bai2014analysis} and other works on the topic.

\subsection{Vehicular Terahertz Systems}
\label{subsec:vehicular_channel}
Following the discussion in Section~\ref{sec:use_cases}, the vehicular environment is one of the promising use cases for terahertz communications, terahertz radar, and joint terahertz communications and sensing solutions. terahertz hardware components have made significant progress over the recent decade. Still, modern sub-terahertz (and especially terahertz) hardware is not sufficiently compact, cost-efficient, and energy-efficient to aim for its successful adoption in next-generation handheld personal devices (such as smartphones, tablets, laptops, or \gls{xr} glasses). On the contrary, vehicle-mounted systems feature notably less stringent restrictions on their weight (compared to the weight of the vehicle itself), cost, and power budget~\cite{samy2021power}.

\if 0
\begin{figure*}[!t]
  \centering
  \subfigure[Vehicular Measurement Setup]
  {
    \includegraphics[width=0.45\textwidth]{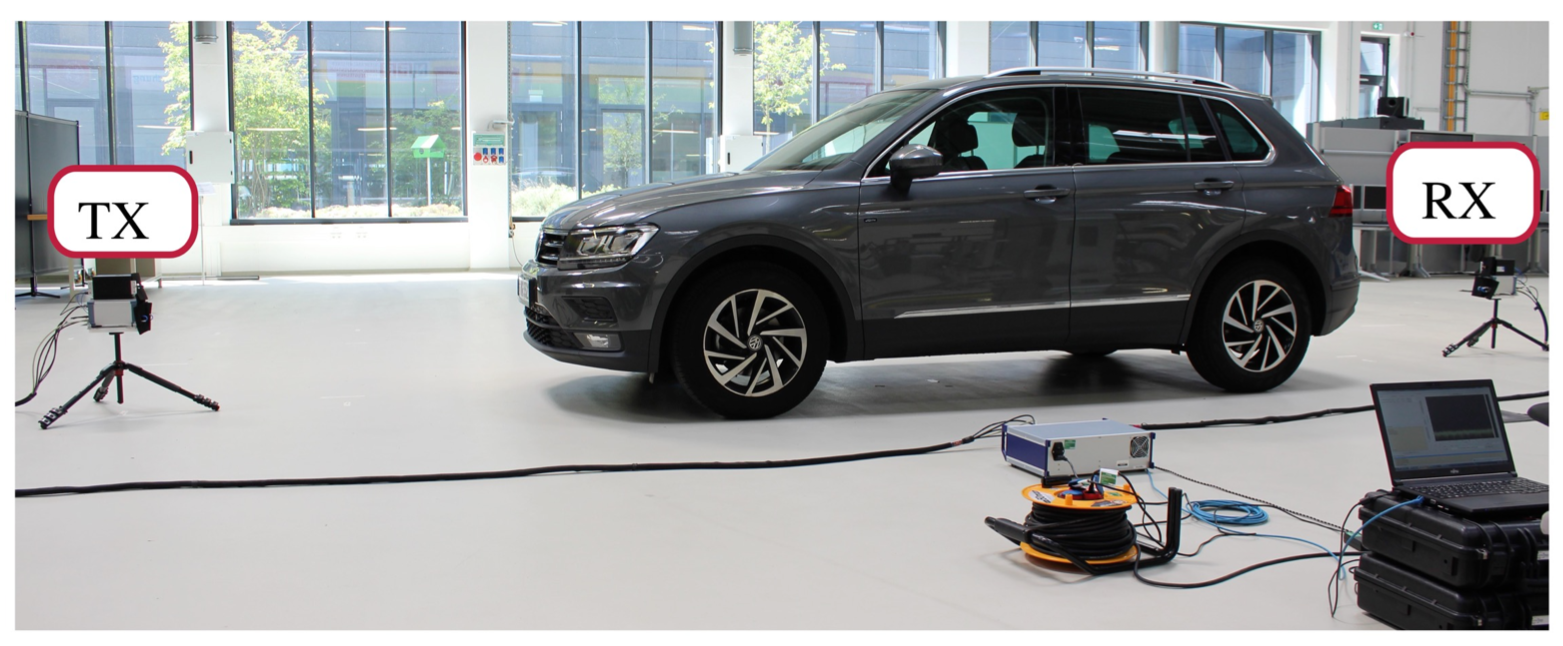}
    \label{fig:vehicular_channel_setup}
     
  }
  %\hfill
  \subfigure[Measurement Results]
  {
    \includegraphics[width=0.45\textwidth]{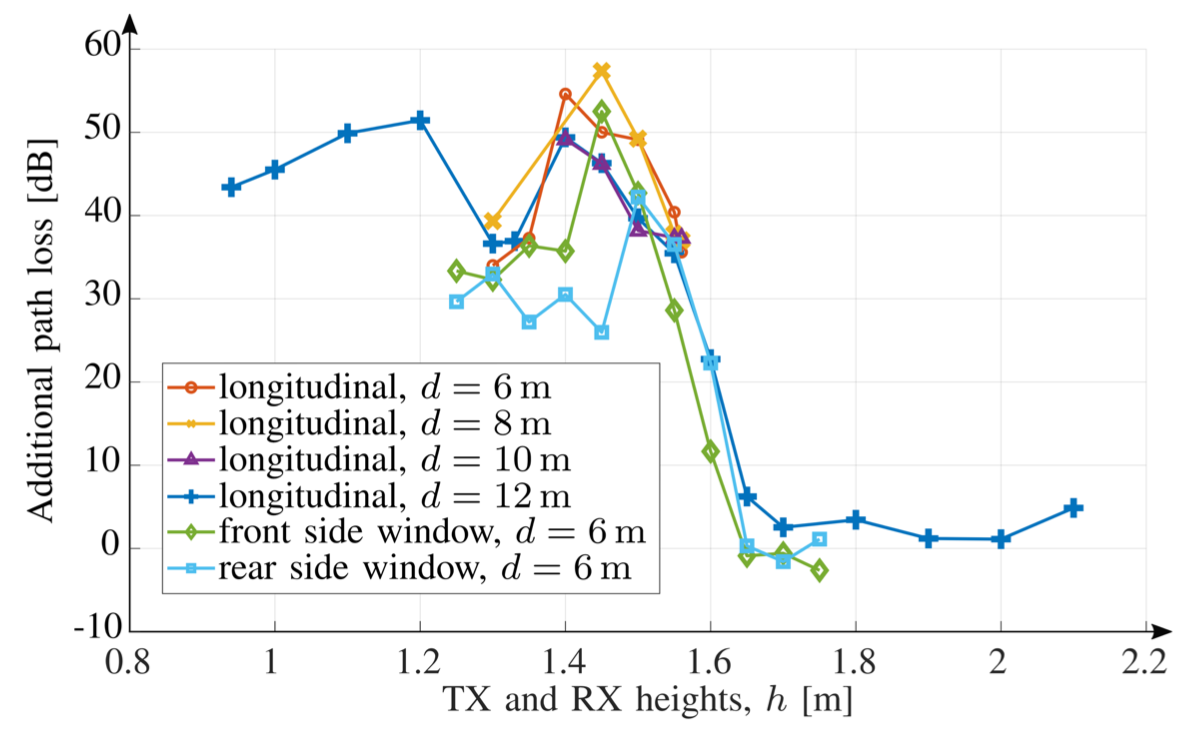}
    \label{fig:vehicular_channel_results}
  }
  \captionsetup{justification=centering}
  \caption{\textcolor{black}{Vehicular terahertz communications measurements and results~\cite{eckhardt2021channel}.}}
  \label{fig:vehicular_channel}
\end{figure*}
\fi

However, introducing vehicles to outdoor terahertz channels leads to several distinct features that require additional efforts in terahertz channel modeling. First and foremost, the vehicle body itself is a complex obstacle that is neither transparent for terahertz radiation nor absorbs/reflects it back in every configuration. Specifically, the latest studies show that a promising approach to model the vehicle body in terahertz channel modeling is by following the layered \emph{sandwich}-type path. The measurements reported in~\cite{eckhardt2021channel} show that the vehicle body is almost non-transparent at the engine level (up to 0.8~m, on average), while only decreasing the power of the terahertz signal passing at the windows level (0.8~m--1.5~m, depending on the vehicle) by a few decibels. Hence, a two-layer model (vehicle representation with a semi-transparent parallelepiped on top of a non-transparent parallelepiped) is the minimal set needed for terahertz channel modeling, while additional layers (roof, wheel, etc.) can be added as needed to improve the accuracy further.

The second important distinct feature of vehicular terahertz setups that should not be neglected in channel modeling is the signal propagation under the vehicle. Depending on the car model, the vehicle clearance (empty space under the vehicle body) may vary from typically 10~cm to 22~cm. As noticed in many works (from~\cite{Schneider2000Impact} down to~\cite{Petrov2020Measurements}, among others) and illustrated in~\ref{fig:vehicular_channel}, this ``tunnel'' under the vehicle often acts as a waveguide when it comes to vehicular communications (especially, direct vehicle-to-vehicle links), featuring relatively low signal attenuation~\cite{eckhardt2021channel}. Therefore, for more accurate modeling, the two-layer obstacle representation from above should be converted into a three-level model with another 10~cm--20~cm transparent layers added to the bottom.

Last, the human-body blockage mentioned in the previous subsection is often modeled with a group of identical (or at least statistically identical) human-body models, as humans are not several times different in size. These models may include non-transparent cylinders of a given height (average human height with zero or non-zero variance) and radius (average human width with zero or non-zero variance) or more sophisticated models of several cylinders/spheres representing different parts of the human body. On the contrary, there are notably different categories of vehicles present on a typical road: from small city cars to trucks, busses, and trams that can be longer than five cars combined. As noted in several studies (including, among others,~\cite{Tunc2021Mitigating} and~\cite{petrov2019analysis}) the presence of such large vehicles may impact a lot the performance of \gls{mmwave} and terahertz wireless links and thus must be accounted for in terahertz channel modeling. One of the feasible approaches here recommended is to develop a set of statistical models for the most common vehicle types (e.g., car, bus, and truck) and then deploy them on-site with a certain proportion following the percentage of those on a typical road.

\begin{figure}
 \centering
 \includegraphics[width=0.9\columnwidth]{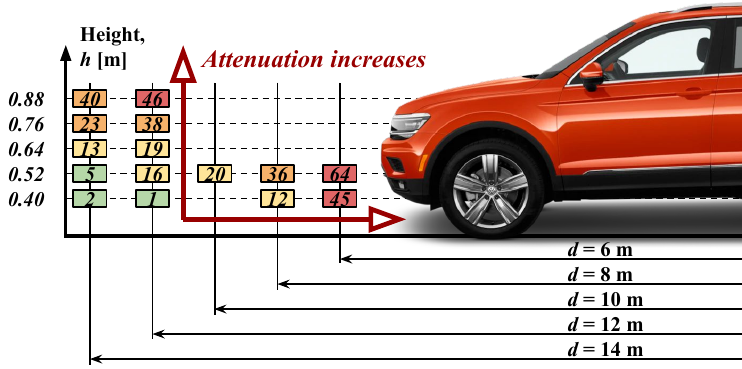}
 \caption{Non-negligible under the vehicle propagation in terahertz vehicular communications at 300~GHz.}
 \label{fig:vehicular_channel}
  \vspace{-5mm}
\end{figure}

{
\begin{table*}[ht]
\centering
\caption{Importance of key propagation effects in terahertz channel modeling for different use cases.}
\label{tab:channel}
\begin{tabular}{|l||c|c|c|c|}
\hline
\textbf{Key effects} & \multicolumn{4}{c|}{\textbf{Deployment scenarios}} \\
\cline{2-5}
 & \textbf{Indoor} & \textbf{Outdoor} & \textbf{Vehicular} & \textbf{Vehicular}\\ \hline\hline
\textbf{Spreading} & \textbf{High} & \textbf{High} & \textbf{High} & \textbf{High}\\ \hline
\textbf{Absorption} & Low & \emph{Medium} & \emph{Medium} & \textbf{High}\\ \hline
\textbf{Reflection} & \textbf{High} & \emph{Medium} & \textbf{High} & Low\\ \hline
\textbf{Scattering} & \emph{Medium} & Low & \emph{Medium} & Low\\ \hline
\textbf{Diffraction} & Low & Low & Low & Low\\ \hline
\textbf{Blockage} & \textbf{High} & \emph{Medium} & \textbf{High} & Low\\ \hline
\hline
\emph{Example models} & \cite{Ju2021jsac,Doeker2022Channel,han2014multi,chen2021channel,bodet2023characterizing,jansen2011diffuse,Ju2023twc} & \cite{amarasinghe2020scattering,li2023performance,Abbasi2023twc} & \cite{guan2019measurement,eckhardt2021channel} & \cite{kokkoniemi2021channel,gao2023scintillation}\\ \hline
\end{tabular}
\end{table*}
}

{

\subsection{Summary}
\label{sec:channel_summary}
In the authors' view, the following are the key takeaways from the latest progress on terahertz channel modeling:
\begin{enumerate}
\item The terahertz channel model must be used case-specific and deployment-specific to be sufficiently accurate.
\item The characteristics of the terahertz channel are heavily dependent on the presence/absence of the clear LoS between the transmitter and the receiver; however, concluding that terahertz links can \emph{only} work within LoS conditions is wrong and simplistic, there are numerous deployment configurations, where NLoS terahertz links provide sufficient link budget for communication purposes (e.g., through a clear firs-order or second-order reflection). 
\item Unlike prior approaches (e.g., \gls{3gpp}-driven for 5G-grade systems~\cite{38901}) it is not always possible to decouple the antenna-related effects from the channel-related effects due to the large size of the terahertz near field zone.
\item When operating in transparency windows~\cite{boronin2014capacity} or via relatively narrow terahertz sub-bands far from the absorption lines, for indoor setups, molecular absorption effects are of secondary importance compared to blockage, reflection, and scattering from obstacles present in the environment. The situation is the opposite in airborne setups, where absorption plays a non-negligible role, while the impact of blockage/reflection is often negligible. 
\item Under-vehicle propagation should not be ignored when modeling terahertz channels in the presence of cars and other vehicles in the environment.
\end{enumerate}

Concluding two in-depth discussions on terahertz hardware and terahertz channel in the present and the previous section, we now proceed with outlining some of the pressing challenges and open research problems in the field below.
}

%----------------------------------------------------%
\section{Challenges and Open Research Directions}
\label{sec:challenges}

\subsection{Next Steps in Terahertz Hardware}

%\subsubsection{From Terahertz Devices to Densely Integrated Terahertz Circuits}
%\hl{I'm not sure I see the difference between what we say here in this first point and what we say in Sec. III.A. Maybe we can talk about going to ``true" terahertz frequencies?}
%\textcolor{red}{1) Devices..}

\subsubsection{Towards True Terahertz Ultrabroadband Devices} 
While the so-called ``terahertz Gap" describes the lack of electronics or photonics means to generate and receive terahertz signals, one of its fundamental causes is the lack of terahertz electronic, photonic or plasmonic devices. Therefore, there is a perennial need for electronic or photonic devices that operate efficiently and effectively in the terahertz frequency bands. % Examples of state-of-the-art devices include IHP SiGe HBT SG13G2 (ft/fmax = 350/450GHz)~\cite{HWV3} and Teledyne 130nm InP HBT with fmax > 1THz~\cite{HWV4}
%
%Moreover, most existing terahertz front-ends rely on electronic frequency up-converting devices or photonic frequency-downconverting systems, whereas the least mature plasmonic path explores the direct generation of terahertz waves by leveraging the physics of plasma waves. 
%
In all three approaches, there is a need to increase the power, frequency of operation, and/or bandwidth. Besides the adoption of non-CMOS or beyond-CMOS technologies, including vertical \gls{sige} \gls{hbt} devices and \gls{inp} or \gls{gan} devices~\cite{kurner2022book}, new materials enter the game. In addition to graphene, other \gls{2d} nanomaterials, such as \gls{hbn} or \gls{mos2}, and few-atom-thick heterostructures open the door to new physics and properties that can be leveraged for terahertz wireless applications~\cite{latini2017interlayer,burdanova2020ultrafast}. While these technologies will not be ready for 6G, they are likely to be found at the basis of future generations.

\subsubsection{Circuits and System Architectures} As we discussed in Section~\ref{subsec:antennas}, the very small size of individual sub-terahertz and terahertz antennas allows their integration in large numbers in very small footprints. However, fitting the frontend electronics and antennas into the small $\lambda/2$ by $\lambda/2$ array pitch has become increasingly challenging. One possible solution is to explore ultra-compact circuit topologies~\cite{HWV5,HWV6,HWV7}. Alternatively, one may explore new packaging techniques and new \gls{2d} array architectures. For example, a popular architecture is the array of sub-arrays~\cite{lin2016terahertz,ning2023beamforming}, in which multiple digital channels control multiple analog chains each. Each analog chain, in turn, might be a fixed array or might incorporate amplitude, phase, or delay controllers. 

Alternatively, the adoption of graphene-based plasmonic technologies can lead to the design of fully digital arrays with direct modulation and beamforming weight control per antenna, potentially opening the door to ultra-massive MIMO systems~\cite{akyildiz2016realizing,singh2020design}. {As discussed in Sec.~\ref{subsec:analog}, sub-micrometric on-chip plasmonic terahertz sources and terahertz modulators together with micrometric plasmonic antennas can lead to extremely compact front-ends. Moreover, these front-ends leverage the concept of direct RF or antenna modulation, where, as opposed to traditional digital systems in which the in-phase (I) and quadrature (Q) components are generated in baseband digitally before being converted to the analog domain, a few or even a single digital line is utilized to control the amplitude and/or phase of the signal directly at RF~\cite{wang2019analysis,chien2022110,d2022135}. The possibility of replacing large, power-demanding data converters with individual digital lines might lead to new, more compact, and energy-efficient array architectures. While plasmonic technology is relatively at a very early stage, the concepts of direct RF or antenna modulation are also actively explored at \gls{mmwave} and sub-terahertz frequencies with CMOS technology.}

\subsubsection{Packaging Technologies}
Advanced packaging technologies are foundational to 6G electronics since they enable heterogeneous integration of different components, e.g., antennas, THz frontend circuits, analog baseband circuits, and digital backends, using different process technologies for system-level optimization. The key considerations on packaging technologies include RF performance (loss tangent), pin pitches, fabrication tolerance, thermal handling, thermal expansion, and mass producibility. Widely used in existing RF products, \gls{ltcc} technology offers low-loss tangent and hermetic properties. However, \gls{ltcc} is limited to thicker substrates, large feature sizes, and smaller panels, which are not compatible with 6G THz electronics that require fine signal pitches and large-scaled array integrations~\cite{HWV8}. \gls{lcp} offers low loss tangent and low moisture absorption but exhibits a large \gls{cte} and difficulty in creating precise cavities and fine resolution traces~\cite{HWV8}. Recently, there has been an increasing interest in employing glass substrate for 6G sub-terahertz/terahertz packaging due to its ultra-low loss tangent, precise metalization, fine signal pitch/density, and low cost. Example D-band radio-on-glass and phased-array-on-glass have been demonstrated with excellent\,performance~\cite{HWV9,HWV10}.

\subsection{Next-Generation Terahertz Channel Models}
Summarizing the discussion in the previous section on the progress in channel modeling for terahertz, there are three key challenges ahead toward the design of accurate, flexible, and useful next-generation channel models for terahertz communications:

%-----%
\subsubsection{Near-field Effects} The major distinct feature of terahertz communications in comparison to existing 5G-grade \gls{mmwave} solutions is the non-negligible near-field zone of terahertz antennas~\cite{Cui2023Near} that lasts for several (tens) of meters. As illustrated in multiple studies (~\cite{Bjornson2020Power}, \cite{reddy2023ultrabroadband}, and~\cite{Petrov2023mobile}, among many others), the near-field effect cannot be neglected for terahertz frequencies, as the received signal gets a notably different structure than with the canonical far-field assumption, Specifically, the antenna gain function in the near field becomes not only angle but also distance-dependent. As discussed further in~\cite{singh2022wavefront}, canonical far-field beamforming can demonstrate up to 7--10~dB difference in the terahertz near field versus the expected values from existing propagation models. Therefore, accurate and flexible extensions are to be introduced into the next-generation terahertz channel models to account for this important effect.

Another essential feature of near-field terahertz communications impacting next-generation channel models is the fact that the antenna-related effects cannot be that easily decoupled from the environment-related propagation effects. The key reason here is that the length of the near-field propagation zone (and also the impact on the received signal) heavily depends on the selected antenna configuration~\cite{petrov2023near}. Hence, a commonly-used (e.g., in \gls{3gpp} TR 38.901~\cite{38901}) two-stage approach, where the channel model is first derived assuming omnidirectional propagation and then tailored independently to different antenna configurations, is not directly applicable anymore and has to be modified accordingly~\cite{sen2024impact}. Last but not least, the community also recently started actively exploring alternative \emph{wavefronts} to complement or replace beamforming for near-field terahertz communications. These include, among others, beamfocusing, self-healing Bessel beams, and curved-shape Airy beams, all demanding different extensions in next-generation terahertz channel models~\cite{petrov2024wavefront,singh2023wavefront}.

%-----%
\subsubsection{Mobile Terahertz Users} Another inherent limitation of existing 5G-grade \gls{mmwave} and state-of-the-art terahertz channel models is the fact that the overwhelming majority of them are designed exclusively in stationary conditions, where both communicating nodes do not move during the entire duration of the data exchange. While this is a valid assumption in most cases at lower frequencies, for terahertz communications with an order of magnitude shorter wavelength, this is not 100\% applicable, as even minor changes in the node's location may lead to drastic variations in the received signal. This is especially crucial in the terahertz near field (as discussed above) and in indoor environments featuring rich multipath. Specifically, a slight change in the node location may change the sim phase/counter phase arrival of multipath components or even lead to a given strong multipath component being added/removed completely from the channel frequency response.

Last but not least, terahertz communications are envisioned to exploit extremely directional beams. Hence, in combination with large-scale mobility (movements of the user node itself), terahertz communications are notably affected by micro-scale (or small-scale) mobility of the mobile device itself (e.g., random displacements and rotations). As illustrated in a few recent studies (including but not limited to~\cite{SinghR2019Parameter} and~\cite{Petrov2020Capacity}), these small-scale movements (and especially small-scale rotations) may have an even greater impact on the performance of the terahertz link. Hence, while developing next-generation channel models for stationary terahertz communications is a valid approach, designing novel types of terahertz channel models already accounting for possible small-scale and large-scale mobility of one or both communicating nodes is a much more valuable option and a tempting research direction.

%-----%
\subsubsection{Generalized Statistical Channel Models} Developing generalized statistical channel models for terahertz communications is the third key research direction from today's terahertz channel models to next-generation approaches suitable for 6G standardization and beyond. recall that one of our key lessons discussed above is that there is likely no possibility to develop a ``common'' terahertz channel model suitable to all the environments, use cases, and node configurations (e.g., antennas). Therefore, the state-of-the-art vision is that the channel model must be tailored to a specific environment and a specific use case. Indeed, when comparing channel modeling (and especially channel measurement) contributions for sub-terahertz and terahertz frequencies, they evolved from general channel models for a wide range of scenarios back in the 2000s and 2010s (e.g.,~\cite{jornet2011channel}) all the way very in-depth studies on the peculiar effects in a particular corridor or an office environment with this exact layout of furniture~\cite{Doeker2022Channel,Yuanbo2022Channel}.

On one hand, this leads to notably more accurate results. On the other, the applicability of these environment-specific terahertz channel models is ultimately limited to exactly the same environment and the same hardware used for the data exchange. Following the discussion in the two items above, even slight variations introduced into the setup may lead to a notably different picture at the receiver. Hence, there is a clear gap here to address when developing next-generation channel models for terahertz communications -- designing a sufficiently general (thus widely applicable) while still sufficiently accurate terahertz channel model. Some essential research questions here to answer are: (i)~what is the required level of detail in the environment that is sufficient for terahertz channel modeling (e.g., resolution of the obstacles and their materials) and (ii)~what kind of small-scale and large-scale mobility can be tolerated without rapid deviations in the channel frequency response.

As of today, one of the promising directions here seems to be following the \gls{3gpp}-style approach (e.g., as in TR 38.901 and earlier documents \textcolor{black}{for channel modeling up to 100 GHz}), where the model is developed not for a given environment, but for a given \emph{class of environments} (e.g., any outdoor city street, or any indoor office building with typical characteristics). Maintaining the balance between accuracy and flexibility is one of the major challenges on the way from existing general terahertz channel models (not applicable to any given non-trivial scenarios) and existing scenario-specific terahertz channel models (not applicable to any other specific scenario) toward next-generation terahertz channel models for various typical classes of use cases and deployment scenarios (e.g., indoor office, outdoor cell, vehicular, WNoC, etc.).

However, the \gls{3gpp} modeling approach identifies a large number of ``clusters'' with each cluster having several multipaths traveling close in both space and time. Empirical observations show that multipaths in the same time cluster can arrive at a receiver from different spatial directions and multipaths in a spatial cluster can arrive at distinct times\cite{Rappaport2107vtcComps}. Moreover, multipaths in outdoor and indoor wireless channels are observed to become sparser when transitioning from mmWave to sub-terahertz and beyond \cite{Yu2020iccw}. With the increased sparsity, deterministic channel modeling using ray tracing tools, such as NYURay \cite{Kanhere2023icc}, can prove valuable for gaining insight into the propagation behavior at terahertz frequencies where accurate maps of the specific environment are available.

The time-cluster spatial-lobe approach models multipath propagation behavior through independent time and spatial clusters and forms the basis for the NYUSIM simulator for channel modeling up to 150~GHz in indoor, outdoor, and factory scenarios\cite{Ju2021jsac, Rappaport2017tapOverview}. Similarly, tools like TeraSim~\cite{hossain2018terasim}, conceived for the simulation of more general applications of terahertz systems (beyond cellular networks) are also currently being adopted to perform full-stack performance analyses of next-generation cellular networks~\cite{gargari2021full,polese2020toward,gargari20236g,gargari2023sebasi}. Tables~I and~II in \cite{Poddar2023comst} showcase other popular channel models and simulators for 5G and beyond, many of which will keep evolving to incorporate terahertz communications in the future.

\subsection{Building the Physical and Link Layer of Terahertz Networks}
The design of the physical and the link layers of terahertz communication systems need to capture both today's and the envisioned capabilities of terahertz radios as well as the peculiarities of the terahertz channel. In this section, we discuss how different state-of-the-art communication and networking technologies can achieve this goal.

{

\subsubsection{Bandwidth vs Beamwidth: Exploiting the Trade-off}
As discussed in Sec.~\ref{sec:use_cases}, the applications of terahertz communications are very diverse, ranging from \gls{wnoc} to \glspl{ntn}. As a result, it is difficult to define typical values for the transmit power, antenna configuration, bandwidth, and, ultimately, achievable bit rate.

The \textit{transmit power} of a terahertz radio drastically changes across frequencies. For example, as discussed in Sec.~\ref{subsec:analog}, the \gls{nasa} \gls{jpl} has demonstrated world-record high-power frequency multipliers with nearly 200~mW in the sub-terahertz range~\cite{siles2018new}. The same technology at 1~THz exhibits only a few milliWatts. 
To compensate for the relatively low power and increase the signal strength at the receiver, \textit{high-gain directional antennas} are commonly utilized. Again, as discussed in Sec.~\ref{subsec:antennas}, the small wavelength of terahertz signals allows for high directivity antennas in a very small footprint. %Compact antennas with gains ranging from 10~dBi up to 38~dBi are commercially available.
Besides this, is it relevant to note that the ability to close a link at the receiver depends not only on the received signal strength but on the total noise at the receiver and, ultimately, on the \gls{snr}. The noise power itself depends on the technology being used as well as on the total bandwidth. Therefore, there are many cases in which reducing the bandwidth is needed to close the link. 

The \textit{bandwidth} of a terahertz communication system depends on multiple factors, including the hardware capabilities, the channel peculiarities, and the legal limitations. 
As discussed in Sec.~III, current sub-terahertz and terahertz transceiver and antenna architectures can easily support 10~GHz of bandwidth and more. For example, in the TeraNova platform at \gls{nu}, front-ends with 20~GHz of bandwidth at a tunable center frequency between 110 and 170~GHz, 30~GHz between 210 and 240~GHz, and up to 50~GHz between 1 and 1.05~THz~\cite{sen2021versatile}. This limit is primarily set by the mixer at the transmitter and the receiver. 
Regarding the channel, and as presented in Sec.~IV, the available bandwidth is significantly larger. In the sub-terahertz range, there are only two absorption peaks (at 119~GHz and 183~GHz, respectively), theoretically enabling very large transmission windows. Above 300~GHz, there are many more absorption lines, but the separation between them is still of tens and even hundreds of GHz~\cite{boronin2014capacity}.
However, it is first relevant to note that the sub-terahertz spectrum up to 275~GHz is already allocated to different services, including fixed, mobile, and satellite communications, \gls{eess}, and space research~\cite{fcc_table_online}. Today, between 100 and 200~GHz, only 12.5~GHz of contiguous bandwidth is allocated to communication services. If sharing with space services is allowed, this value can increase to 32.5~GHz. A similar situation is found between 200 and 275~GHz. An extensive discussion on the coexistence and spectrum sharing issues at frequencies above 100~GHz is given in~\cite{polese2023coexistence}. 

To be able to provide quantitative data, next, we focus on indoor \gls{wlan} and outdoor cellular applications. First, in Fig.~\ref{fig:antenna_footprint}, we illustrate the achievable directional gain for a fixed antenna footprint of 10~cm$^2$ (comparable to that of a current smartphone) as a function of frequency, highlighting the ability to achieve very high gains with a very compact structure. Second, in Fig.~\ref{fig:bandwidth_beamwidth}, we illustrate the trade-off between antenna beamwidth or directivity and system bandwidth. More specifically, we consider a 30~meter \gls{los} link, with a transmitter delivering 100~mW of output power and 20~dBi of antenna gain, and a receiver with a noise figure of 20~dB and an antenna gain ranging from 0 to 40~dBi. 

Increasing the bandwidth while maintaining the \gls{snr} and, thus, increasing the bit rate requires a major increase in directivity gain, which requires larger radiating structures (though still compact, as per Fig.~\ref{fig:antenna_footprint}), leading to further near-field effects (with the challenges and opportunities that they bring, as previously discussed), and increased beam management complexity (as we elaborate later in this section). These trade-offs need to be captured when designing the physical and link layers of terahertz networks. For example, not all transmissions might need extremely high bit-rates and, thus, reducing the bandwidth can automatically relax many other requirements in the system.

\begin{figure}
  \centering
  \includegraphics[width=0.5\textwidth]{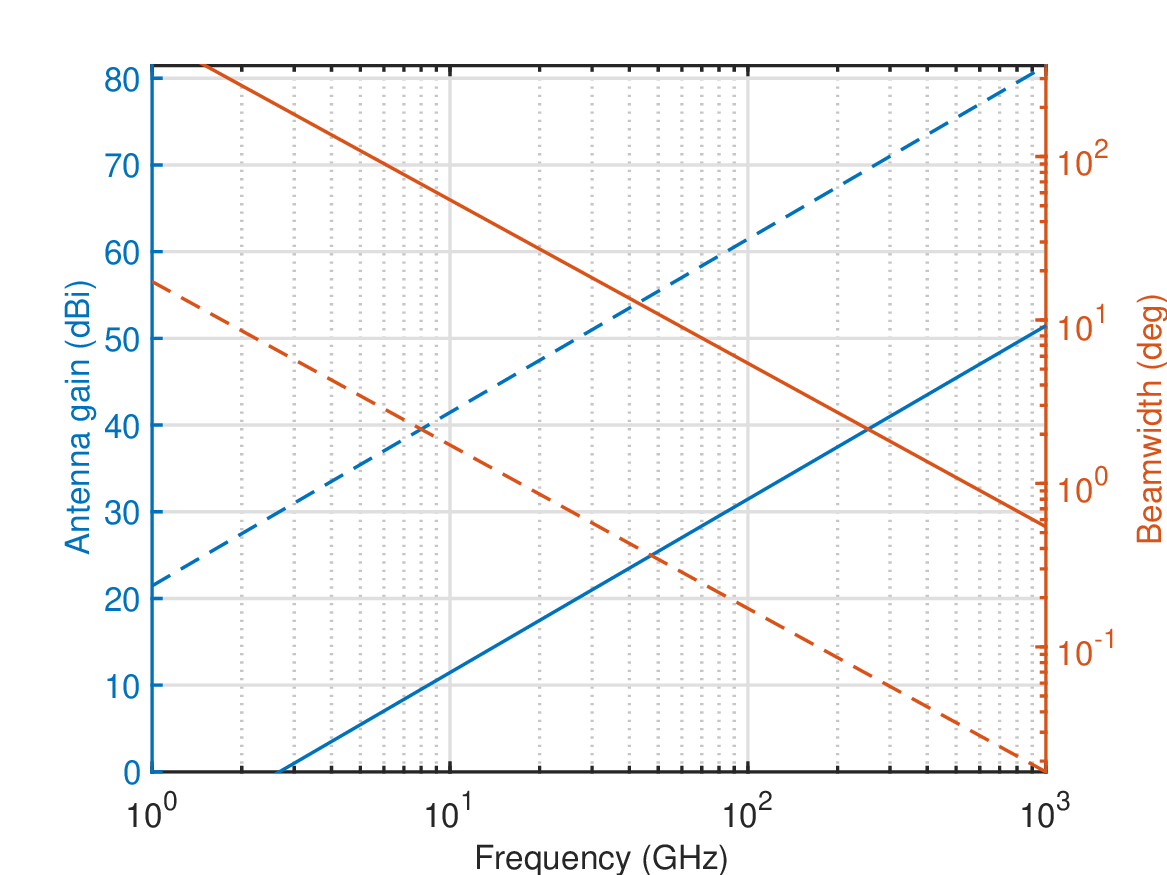}
  \caption{Antenna gain (blue) and beamwidth (red) as functions of frequency for a fixed antenna footprint of 10~cm$^2$ (solid lines) and 1~m$^2$ (dashed lines).}
  \label{fig:antenna_footprint}
\end{figure}

\begin{figure}
  \centering
  \includegraphics[width=0.5\textwidth]{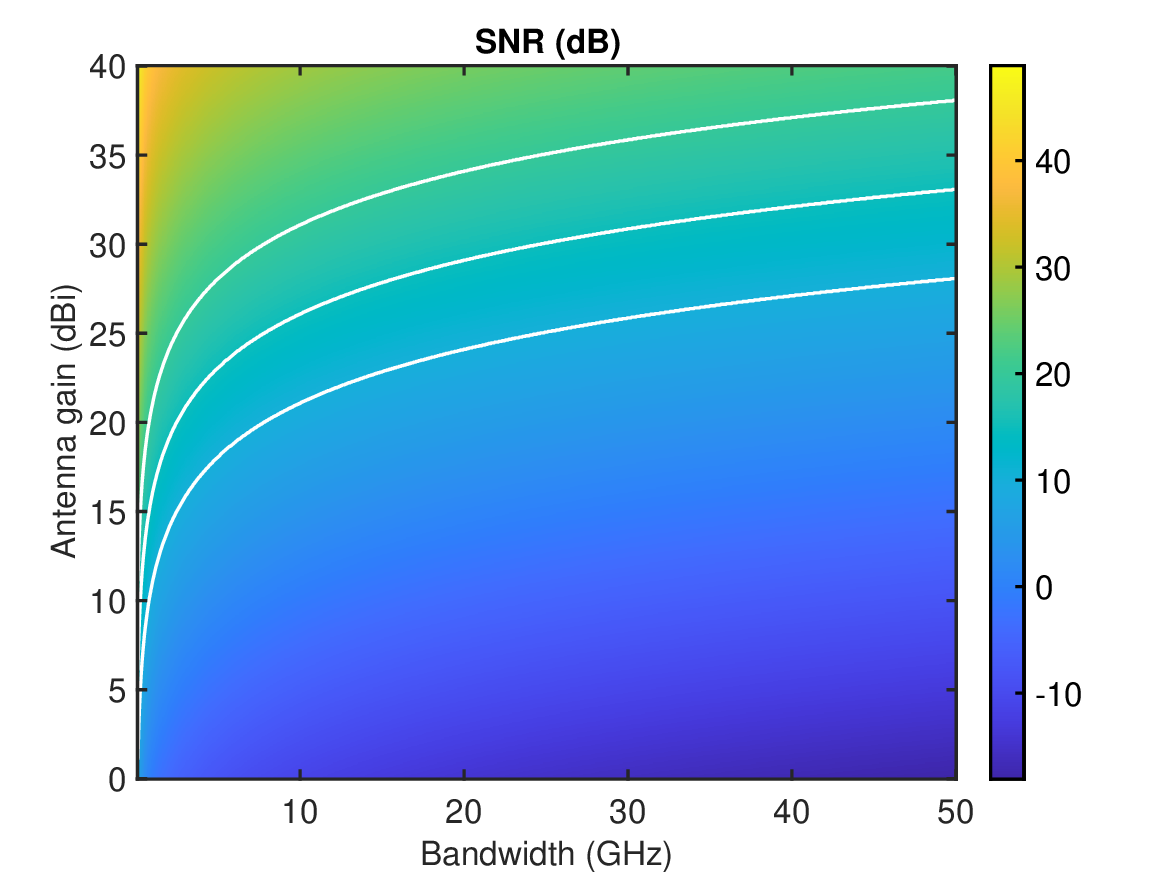}
  \caption{Required SNR as a function of antenna gain and bandwidth, for a 30-m LoS link at 140~GHz, with a transmitter with 100~mW of output power and 20~dBi of antenna gain, and a receiver with a noise figure of 20~dB, and an antenna gain ranging from 0 to 40~dBi. The white solid lines indicate the minimum antenna gain to achieve an SNR of at least 10, 15, and 20~dBi, respectively.}
  \label{fig:bandwidth_beamwidth}
\end{figure}

Once the feasibility of the link is established, the achievable bit rate with a maximum tolerable \gls{ber} depends on the specific modulation technique being used, as we discuss next.

\subsubsection{Ultra-broadband Waveform, Modulation and Coding}
The modulation techniques for terahertz signals have drastically evolved in the last decade. Due to their simplicity, the first terahertz systems considered only non-coherent modulations, such as on-off keying modulation~\cite{moeller20112} and, today, this is still one of the two physical layers supported by the only standard for terahertz systems, the IEEE 802.15.3d~\cite{802.15.3dstandard}. This early approach, together with the very large bandwidth supported by the terahertz channel, has often resulted in the misleading belief that there is no need for spectrally efficient modulations at terahertz frequencies. However, as discussed in Sec.~\ref{sec:hardware}, current terahertz transceivers exhibit low transmit power and potentially very high amplitude and phase noise. While antenna gains play a key role in the system performance, the need for spectrally efficient modulations that can maximize the bandwidth utilization and, ultimately, reach terabits per second becomes evident.

In this direction, different strategies exist. On the one hand, one can leverage the state-of-the-art modulations in 5G and 5G advanced systems and scale them up in bandwidth. For example, while the use of \gls{ofdm} might be discouraged due to its high \gls{papr}, \gls{dftso} can be adopted. Similarly, \gls{otfs} modulation could be utilized to compensate for the frequency offsets resulting not only from mobility in some applications but also from the phase noise of terahertz oscillators. There have been several papers thoroughly comparing these modulations~\cite{tarboush2022single,desombre2021performance,jian2021baseband}. Most recently, in~\cite{parisi2023modulations}, we have studied the joint impact of \gls{papr} and phase noise on single and multi-carrier modulations as well as on ultra-broadband spread-spectrum techniques~\cite{bosso2021ultrabroadband}. Based on newly developed experimentally-driven phase noise models at three different sub-terahertz and terahertz bands and introducing the concept of \gls{papr} penalty, we have concluded that \gls{dftso} offers a fair trade-off for data-rate and \gls{ber}. 

Besides traditional modulations, new waveforms are enabled by the behavior of the terahertz channel. For example, for very short-range applications, very short pulses, just a few hundred femtoseconds long, can be utilized~\cite{jornet2011channel}. The power spectral density of such pulses, commonly used in sensing applications including \gls{tds}, spans a few THz. Their very short duration enables very high symbol rates. For longer communication distances, the molecular absorption broadening effect and the distance-dependent bandwidth discussed in Sec.~\ref{sec:channel} motivate and enable new modulation techniques that facilitate the multiplexing of users in space. For example, in~\cite{han2014distance}, a modulation scheme that leverages multiple absorption-defined transmission windows and dynamically allocates a different number of sub-carriers per window based on the users distance is presented. In~\cite{bodet2022hierarchical}, a hierarchical-bandwidth modulation scheme that multiplexes in a single stream data with different modulation orders and symbol durations is proposed as a way to transform molecular absorption into an ally to spatially multiplex users within the same transmission beam. 

Finally, in terms of error control coding, much less has been done. For the time being, the only discussions on error control are focused on the nanoscale applications of the terahertz band and advocate for the use of low-weight codes, i.e., codewords with more binary zeros than ones, as a way to simultaneously minimize molecular absorption noise and multi-user interference~\cite{jornet2011low,kocaoglu2013minimum,akkari2016joint}. New error control strategies tailored to the imperfections of terahertz hardware, including device-induced frequency selectivity and large amplitude and phase noises, and the behavior of the channel, such as the frequency selectivity of molecular absorption, are needed, all while keeping an eye towards low computational complexity, so to meet the latency requirements of 6G. 

\subsubsection{Channel Estimation and Beam Management}
As just discussed, the high gain and narrow beamwidths are crucial in enabling terahertz communications. To implement mobile sub-terahertz and terahertz wireless communications, it is necessary to implement beamforming to steer the radiated signal or to control the direction of reception while communicating with multiple transceivers~\cite{rappaport2015pearson}. Hybrid beamforming architectures that use digital beamforming at the baseband with analog beamforming at the RF are already used in today’s \gls{mmwave} 5G cellphones and are anticipated to be widely utilized for THz transceivers. Digital beamforming facilitates spatial multiplexing gains, while analog beamforming--through phase shifters at the RF--offers directionality gains from the radiating antenna~\cite{Sun2018twc}. As per the discussion in Sec.~V.A., if fully digital antenna arrays become available, both multiplexing and beamforming gains can be combined and digitally implemented.

Channel estimation is pivotal to provide the proper channel state in order for the analog and digital beamforming vectors to be adapted for proper pointing directions to maximize link gain while minimizing interference~\cite{rappaport2015pearson} for THz communication systems. Achieving optimal beamforming relies on perfect prior knowledge of the channel matrix and its singular value decomposition. Omnidirectional pilot signals are not practical in highly directional THz channels due to severe path loss and blockage, so beamforming must learn the channel to estimate the channel response and determine the best antenna beamform weightings for the current channel state.

One approach for channel estimation can involve using exhaustive beam training to explore the entire spatial search space, aiming to establish a beamforming link between transceivers. Careful codebook design can help expedite the search to identify the most suitable narrow beam pair with the highest SNR~\cite{ning2023beamforming}. As an example, IEEE 802.11ad uses a one-side exhaustive search protocol~\cite{rappaport2015pearson,ning2023beamforming} whereby a user performs an exhaustive search across all beams in the codebook, while the access point transmits with an omnidirectional beam.  Moreover, to reduce the training overhead and spatial search space associated with exhaustive beam training that searches over every \gls{aod}/\gls{aoa} pair, adaptive channel estimation algorithms utilizing codebooks with multiple spatial resolutions can accelerate the search process. Additionally, compressed sensing-based beamforming methods can take advantage of the sparsity of multipath in the channel to obtain channel state information~\cite{Alkhateeb2014jstsp}. Deep learning-based methods leveraging convolutional neural networks for channel estimation have also demonstrated promising results, making them potential candidates for accurate THz channel estimation~\cite{chary2024accurate}. Lastly, less conventional antenna designs, such as leaky-wave antennas, and new array operations, such as the joint phase-time array and the~\gls{fma}, which all exhibit unique frequency-\gls{aod} relationships, can be utilized to expedite neighbor discovery~\cite{ghasempour2020single,mannem2023reconfigurable,mannem2022mm}.

\begin{figure*}
    \centering
    \includegraphics[width=0.49\textwidth]{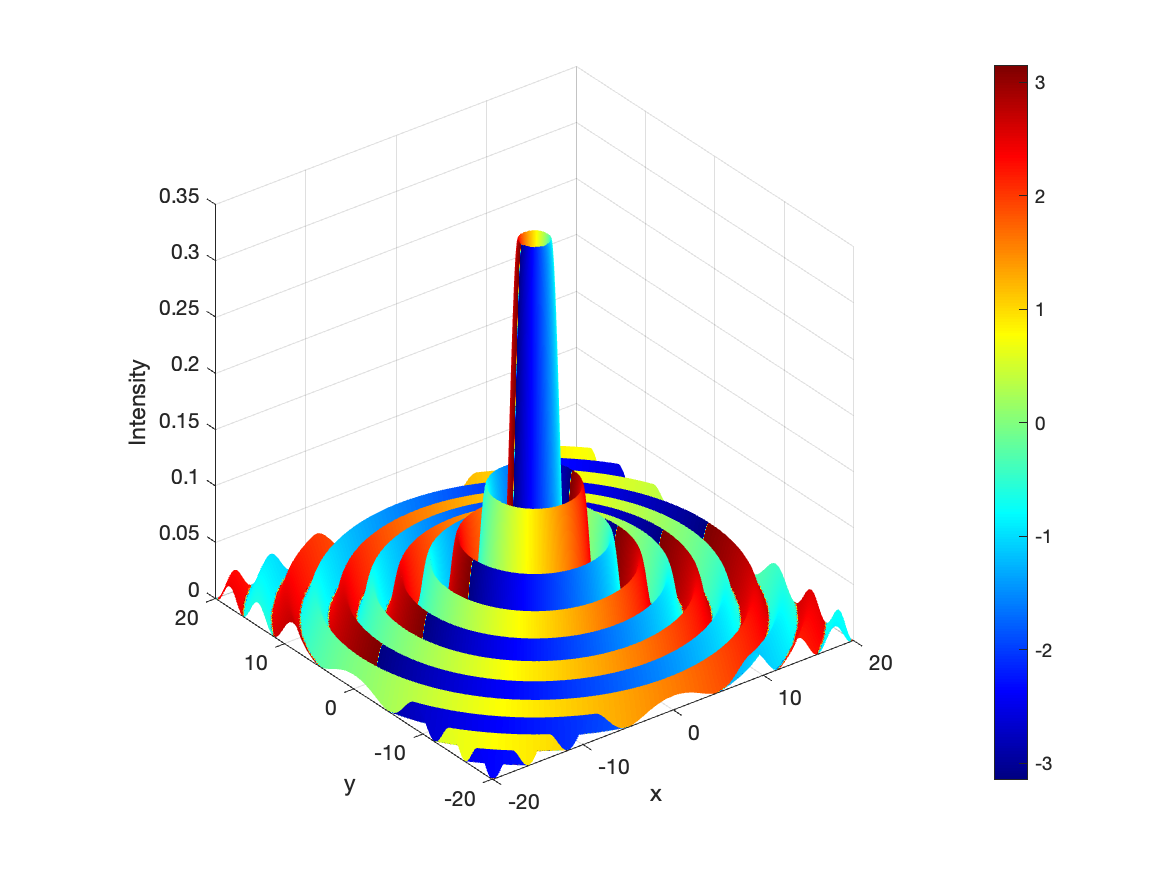}
    \includegraphics[width=0.49\textwidth]{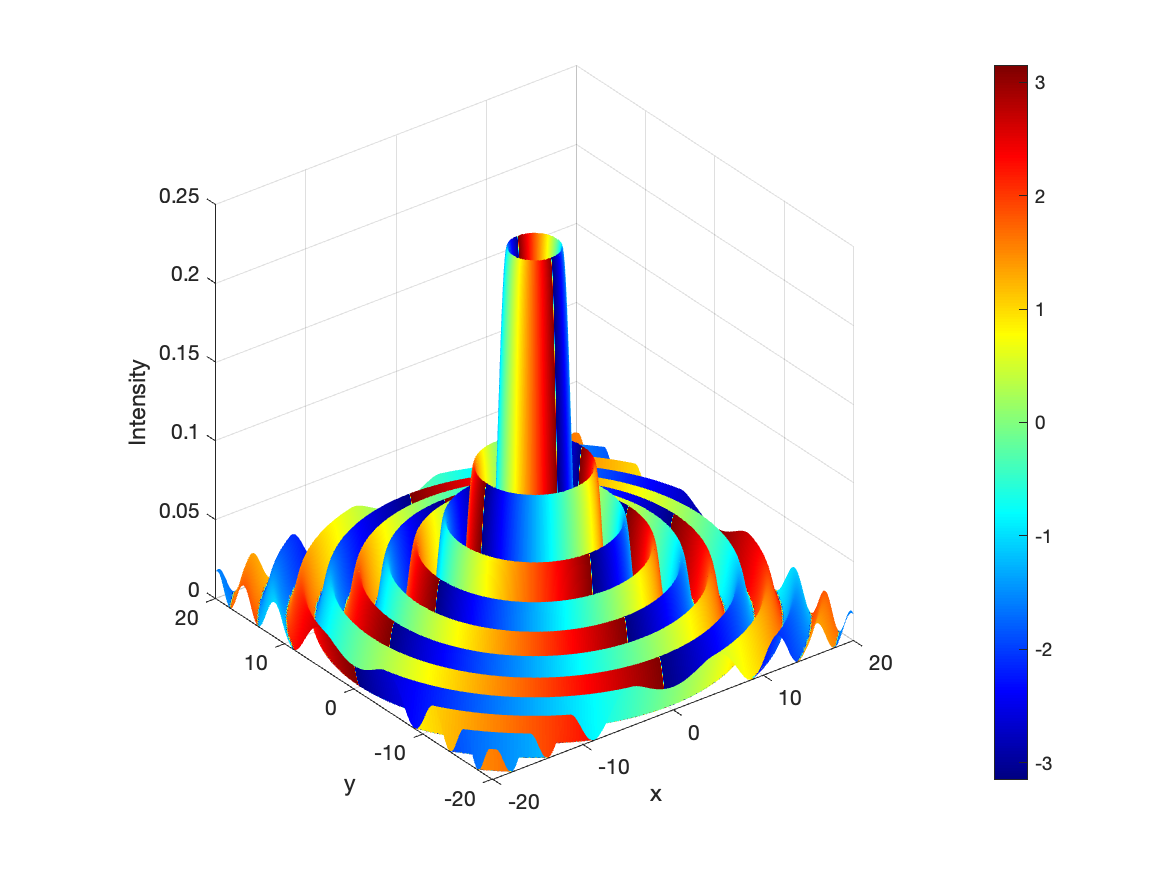}
    \caption{Intensity (height) and phase (color) profile of two different orbital angular momentum modes (helical Mode 1 on the left and helical Mode 2 on the right). The two modes are orthogonal and define an orthogonal basis that can be used for information modulation or multiplexing.}
    \label{fig:oam}
\end{figure*}

\subsubsection{Ultra-massive/XL-MIMO, Cell-free MIMO and OAM}
Up to this point, it has been mostly considered that the very large antenna arrays at terahertz frequencies will be primarily leveraged to generate highly directional narrow beams and, thus, increase the \gls{snr} of a single path and, correspondingly, the modulation order it can support. However, there are scenarios where higher gains can be achieved by exploiting spatial multiplexing. Many are the works that discuss theoretical solutions for terahertz massive, ultra-massive, and XL-MIMO (e.g.,~\cite{akyildiz2016realizing,faisal2020ultramassive,do2021terahertz,ning2023beamforming,lu2024tutorial}, only recently experimental works have been conducted. In~\cite{bodet2023characterizing}, we have recently experimentally demonstrated that there exists enough diversity in common indoor scenarios to support \gls{mimo} systems with channels over geometrically different paths. Building on this result and the aforementioned advantages of \gls{dftso}, we have designed and built a functioning MIMO system over 10~GHz of bandwidth~\cite{bodet2023directional}.

For the cases in which the channel does not naturally support orthogonal paths, diversity can be achieved using \gls{oam}. A beam that is said to have \gls{oam} manifests a spiral phase in the transverse direction, resulting in a helical wavefront and a phase singularity (a zero-intensity vortex) in the center. Overlapping beams that follow helical modes define an orthogonal basis (see Fig.~\ref{fig:oam}). This can be leveraged in different ways: different streams can be sent along different \gls{oam} modes, each one with its own amplitude and/or phase modulation, or one stream can be sent by encoding different symbols in different \gls{oam} modes.

Lastly, it is worth mentioning that very little has been studied when it comes to distributed or cell-free massive MIMO~\cite{elhoushy2021cell} at terahertz frequencies. For example, in~\cite{li2023mdd}, terahertz communications are utilized to interconnect the distributed \glspl{ap} that implement cell-free massive MIMO, but not as the access technology itself. The main reason for this would be the extremely precise synchronization that would needed to enable any form of distributed MIMO or wavefront engineering at terahertz frequencies.

\subsubsection{Network-level Integration of Advanced Physical Layer Technologies} Up to this point, we have discussed the role that different advanced physical layer solutions might play at terahertz frequencies. However, there is one last aspect that we would like to highlight: the use of extremely narrow beams, the coordination among users in distributed or cell-free massive MIMO, or the orchestration of all the network resources (including the \glspl{ris} discussed in Sec.~\ref{subsec:antennas}), can introduce significant delays, impacting the latency, the throughput and, above all, the users' \gls{qos} or \gls{qoe}. For example, if finding the optimal \gls{nlos} path between two users through a \gls{ris} requires even a few milliseconds, one should consider directly switching, even if temporarily, to a lower un-obstructed frequency band (e.g., sub-6~GHz) if multi-band radios are available~\cite{saeidi2023multi}. At this stage, while many optimization frameworks have been developed at the physical layer, few solutions consider the actual end-to-end delay, including the latency introduced by the control channel, which is ultimately what the user experiences. Now it is the time to go up in the protocol stack to ensure the success and broad adoption of the terahertz band for communications~\cite{polese2020toward}.
}

\section{Conclusion}
\label{sec:conclusion}
This article summarizes the latest progress in the field of terahertz communications and sensing, specifically focusing on the hardware aspects (such as closing the ``terahertz gap'') and the latest advancements in terahertz channel modeling. Our main conclusion is that over the last two decades, terahertz communications evolved rapidly from ``futuristic vision'' to ``forthcoming reality''. Wireless connectivity and sensing above 100~GHz are now of great interest already within the 6G timeline (2030 onward, only five years from now). While we will likely not see the full power of terahertz radios in the first 6G releases, the principal step toward adopting the terahertz spectrum for commercial radio systems and networks has clearly been made. Still, as summarized above, there are multiple research and engineering challenges to be addressed toward enabling reliable and efficient terahertz wireless systems and networks.

%\section*{Acknowledgements}

% Can use something like this to put references on a page by themselves when using endfloat and the captionsoff option.
\ifCLASSOPTIONcaptionsoff
  \newpage
\fi

\balance

\bibliographystyle{IEEEtran}
\bibliography{./Bibliography/bibliography.bib}

% \url{https://proceedingsoftheieee.ieee.org/instructions-for-authors/preparing-your-regular-paper-proposal/}

%% Here we need to add the short biographies for all the authors

\begin{IEEEbiography}[{\includegraphics[width=1in,height=1.25in,clip,keepaspectratio]{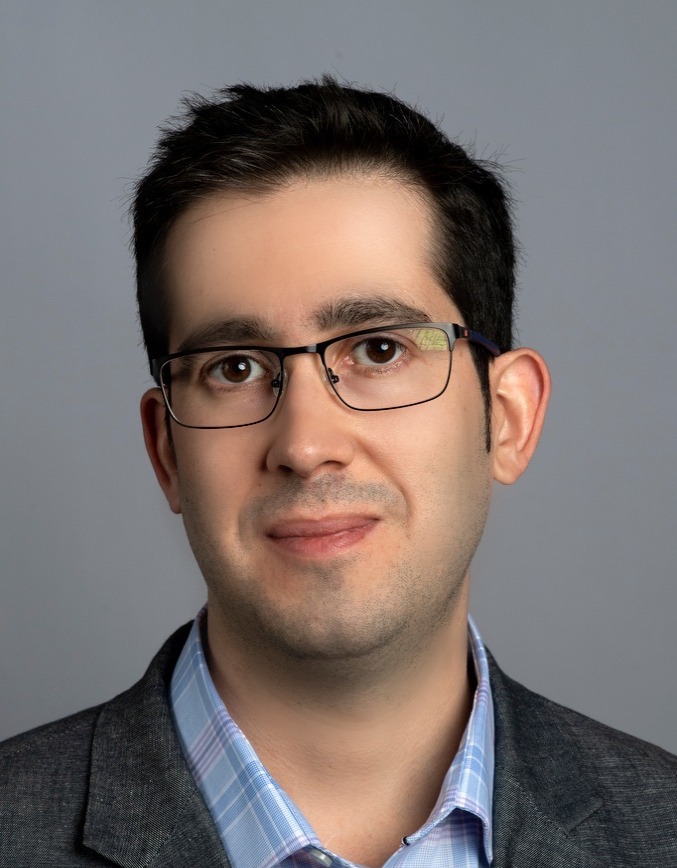}}]{Josep M. Jornet} (M'13--SM'20--F'24) is a Professor in the Department of Electrical and Computer Engineering, the director of the Ultrabroadband Nanonetworking (UN) Laboratory, and the Associate Director of the Institute for the Wireless Internet of Things at Northeastern University (NU). He received a Degree in Telecommunication Engineering and a Master of Science in Information and Communication Technologies from the Universitat Politècnica de Catalunya, Spain, in 2008. He received his Ph.D. degree in Electrical and Computer Engineering from the Georgia Institute of Technology, Atlanta, GA, in August 2013. Between August 2013 and August 2019, he was in the Department of Electrical Engineering at the University at Buffalo (UB), The State University of New York (SUNY). He is a leading expert in terahertz communications, in addition to wireless nano-bio-communication networks and the Internet of Nano-Things. In these areas, he has co-authored more than 250 peer-reviewed scientific publications, including one book, and has been granted five US patents. His work has received over 17,000 citations (h-index of 61 as of June 2024). He is serving as the lead PI on multiple grants from U.S. federal agencies including the National Science Foundation, the Air Force Office of Scientific Research, and the Air Force Research Laboratory as well as industry. He is the recipient of multiple awards, including the 2017 IEEE ComSoc Young Professional Best Innovation Award, the 2017 ACM NanoCom Outstanding Milestone Award, the NSF CAREER Award in 2019, the 2022 IEEE ComSoc RCC Early Achievement Award, and the 2022 IEEE Wireless Communications Technical Committee Outstanding Young Researcher Award, among others, as well as four best paper awards. He is a Fellow of the IEEE and an IEEE ComSoc Distinguished Lecturer (Class of 2022-2023, Extended to 2024). He is also the Editor-in-Chief of the Elsevier Nano Communication Networks journal and Editor for IEEE Transactions on Communications and Nature Scientific Reports.
\end{IEEEbiography}

\begin{IEEEbiography}[{\includegraphics[width=1in,height=1.25in,clip,keepaspectratio]{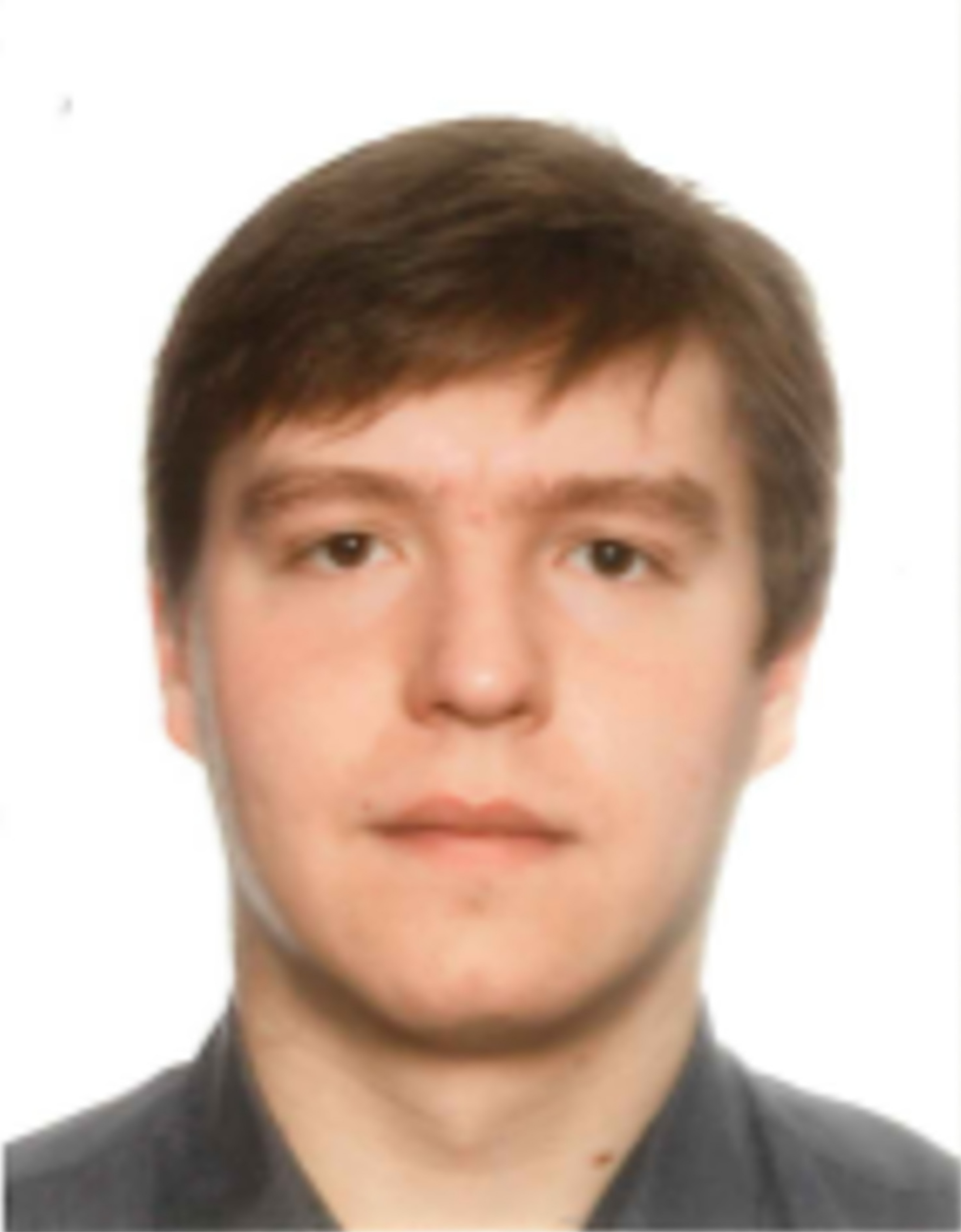}}] {Vitaly Petrov} is an Assistant Professor with the Division of Communication Systems, KTH~Royal Institute of Technology, Stockholm, Sweden. Prior to joining KTH in 2024, he was a Principal Research Scientist at Northeastern University, Boston, MA, USA (2022 -- 2024) and a Senior Standardization Specialist and a 3GPP RAN1 delegate with Nokia Bell Labs and later Nokia Standards (2020 -- 2022). Vitaly obtained his M.Sc. degree in Information Systems Security from SUAI University, St.~Petersburg, Russia, in 2011, his M.Sc. degree in IT and Communications Engineering from Tampere University of Technology, Tampere, Finland, in 2014, and his Ph.D. degree in Communications Engineering from Tampere University, Finland, in 2020. Vitaly has also been a visiting researcher with the University of Texas at Austin, Georgia Institute of Technology, and King’s College London. His current research interests include terahertz band communications and networking. He is a recipient of the Best Student Paper Award at IEEE~VTC-Fall 2015, the Best Student Poster Award at IEEE~WCNC~2017, and the Best Student Journal Paper Award from IEEE~Finland in 2019.
\end{IEEEbiography}

\begin{IEEEbiography}[{\includegraphics[width=1in,height=1.25in,clip,keepaspectratio]{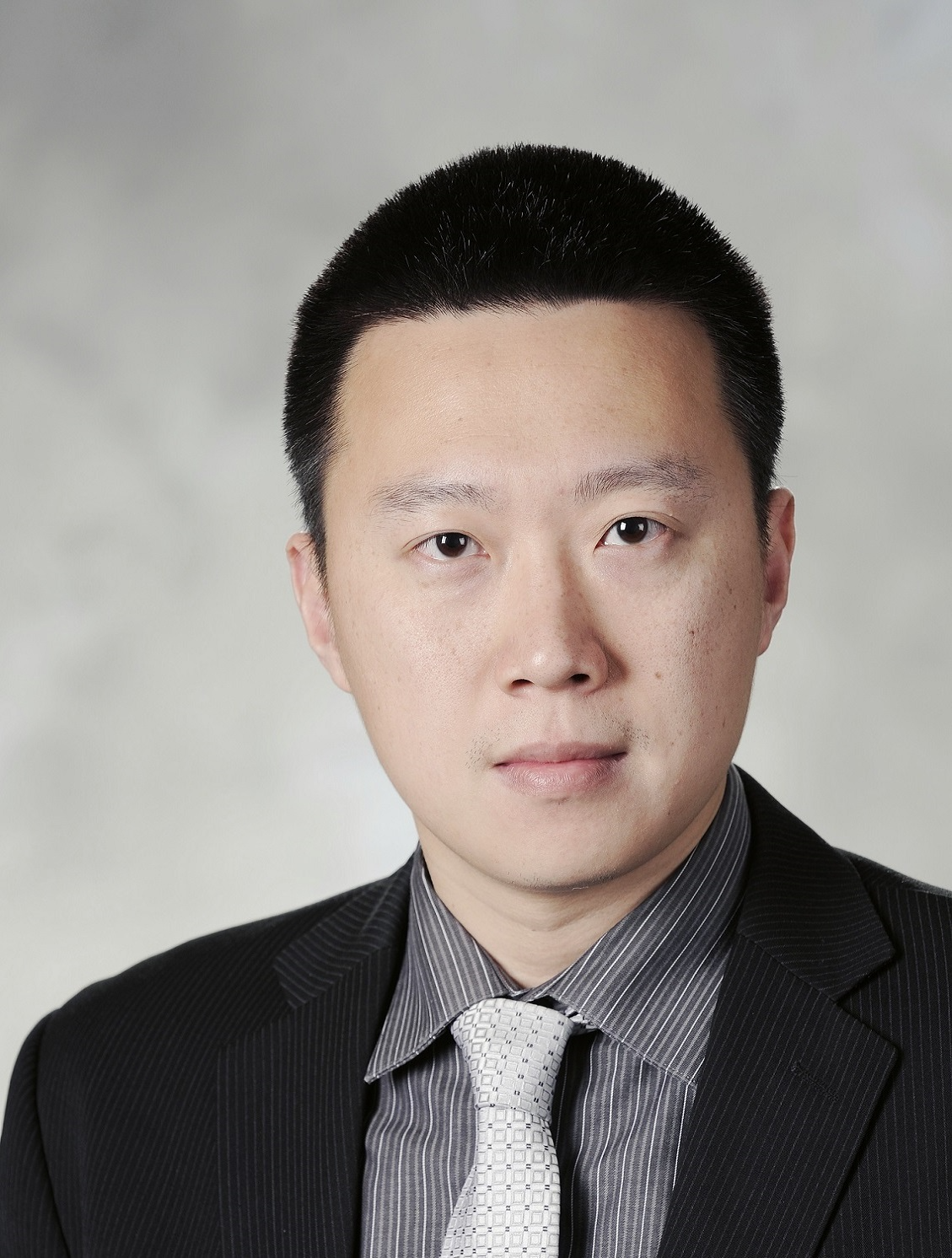}}] {Hua Wang} (Fellow, IEEE) is a Full Professor and the Chair of Electronics at Department of Information Technology and Electrical Engineering (D-ITET) of Swiss Federal Institute of Technology Zürich (ETH Zürich). He is the Institute Deputy Head of the Integrated Systems Laboratory (IIS) at ETH Zürich. He is the Director of the ETH Integrated Devices, Electronics, And Systems (IDEAS) Group. He is a faculty member of the ETH Zürich Quantum Center. Prior to that, he was a Tenured Associate Professor at the School of Electrical and Computer Engineering (ECE) at Georgia Institute of Technology, USA. He held the Demetrius T. Paris professorship at Georgia Tech. He was the director of the Georgia Tech Electronics and Micro-System (GEMS) lab. He worked at Intel Corporation and Skyworks Solutions from 2010 to 2011. He received his M.S. and Ph.D. degrees in electrical engineering from the California Institute of Technology, Pasadena, in 2007 and 2009, respectively.
Dr. Wang is interested in innovating analog, mixed-signal, RF, and mm-Wave integrated circuits and hybrid systems for wireless communication, sensing, and bioelectronics applications. He has authored or co-authored over 250 peer-reviewed journal and conference papers.
Dr. Wang is a Top Contributing Author to the IEEE International Solid-State Circuits Conference (ISSCC) of the past 70 years 1954-2023. He received the DARPA Director’s Fellowship Award in 2020 (the first awardee in Georgia Tech’s history), the DARPA Young Faculty Award in 2018, the National Science Foundation CAREER Award in 2015, the Qualcomm Faculty Award in 2020 and 2021, the IEEE MTT-S Outstanding Young Engineer Award in 2017, the Georgia Tech Sigma Xi Young Faculty Award in 2016, the Georgia Tech ECE Outstanding Junior Faculty Member Award in 2015, and the Lockheed Dean’s Excellence in Teaching Award in 2015. 
His research group has won multiple academic awards and best paper awards, including the 2019 Marconi Society Paul Baran Young Scholar, the IEEE RFIC Best Student Paper Awards (2014, 2016, 2018, and 2021), the IEEE IMS Best Student Paper Award 2021, the IEEE CICC Outstanding Student Paper Awards (2015, 2018, and 2019), the IEEE CICC Best Conference Paper Award (2017), the 2016 IEEE Microwave Magazine Best Paper Award, and the IEEE SENSORS Best Live Demo Award (2016).
Dr. Wang was a Technical Program Committee (TPC) Member for IEEE ISSCC, RFIC, CICC, and BCICTS conferences. He was a Steering Committee Member for IEEE RFIC and CICC. He was the Conference Chair for CICC 2019 and Conference General Chair for CICC 2020. He is a Distinguished Microwave Lecturer (DML) for the IEEE Microwave Theory and Techniques Society (MTT-S) for the term of 2022-2024. He was a Distinguished Lecturer (DL) for the IEEE Solid-State Circuits Society (SSCS) for the term of 2018-2019. He served as the Chair of the Atlanta’s IEEE CAS/SSCS joint chapter that won the IEEE SSCS Outstanding Chapter Award in 2014.
\end{IEEEbiography}

\begin{IEEEbiography}[{\includegraphics[width=1in,height=1.25in,clip,keepaspectratio]{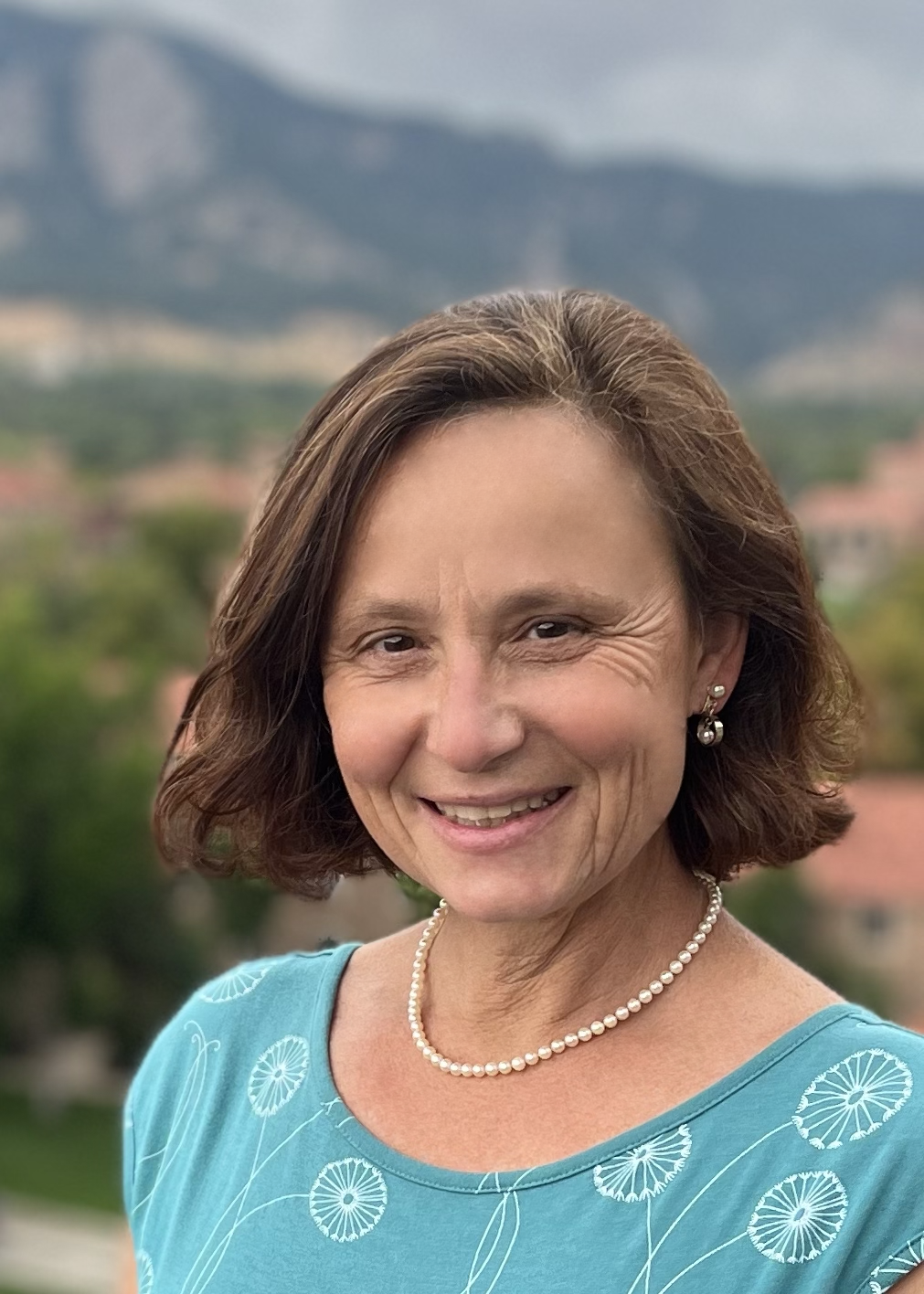}}]{Zoya Popovi\'{c}} (S'86-M'90-SM'99-F'02) is a Distinguished Professor and the Lockheed Martin Endowed Chair in Electrical Engineering at the University of Colorado, Boulder. She obtained her Dipl. Ing. degree at the University of Belgrade, Serbia, and her Ph.D. at Caltech. She was awarded Doktora Honoris Causa (honorary doctorate) in 2022 from the Carlos III University in Madrid, Spain. She was a Visiting Professor with the Technical University of Munich in 2001/03, ISAE in Toulouse, France in 2014, and was a Chair of Excellence at Carlos III University in Madrid in 2018/19.  She has graduated over 70 PhDs and currently advises 18 doctoral students. Her research interests are in high-efficiency power amplifiers and transmitters, microwave and millimeter-wave high-performance circuits for communications and radar, medical applications of microwaves, quantum sensing and metrology, and wireless powering. She is a Fellow of the IEEE and the recipient of two IEEE MTT Microwave Prizes for best journal papers, the White House NSF Presidential Faculty Fellow award, the URSI Issac Koga Gold Medal, the ASEE/HP Terman Medal and the German Alexander von Humboldt Research Award. She was elected as foreign member of the Serbian Academy of Sciences and Arts in 2006. She was named IEEE MTT Distinguished Educator in 2013 and the University of Colorado Distinguished Research Lecturer in 2015. In 2022, she was elected a Member of the National Academy of Engineering and in 2024 a Fellow of the National Academy of Inventors. 
\end{IEEEbiography}

\newpage
\begin{IEEEbiography}[{\includegraphics[width=1in,height=1.25in,clip,keepaspectratio]{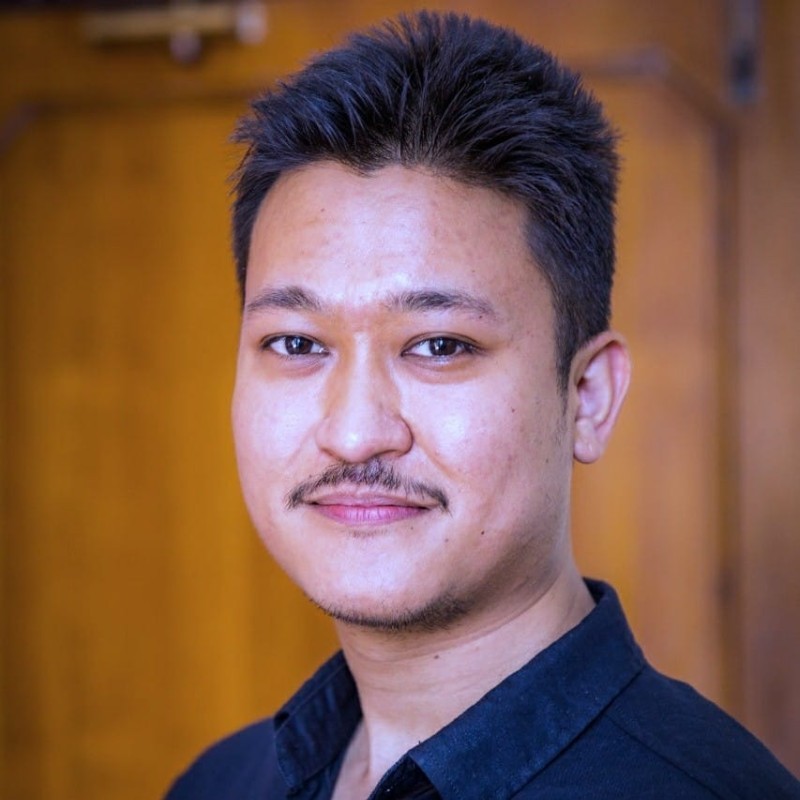}}]{Dipankar Shakya} (Graduate Student Member, IEEE) received the B.E. degree in electronics and communications from Tribhuwan University, Kirtipur, Nepal, in 2016, and the M.S. degree in electrical engineering from New York University, New York, NY, USA, in 2021. He is currently pursuing the Ph.D. degree in electrical engineering at the NYU WIRELESS Research Center, New York University Tandon School of Engineering, Brooklyn, NY, USA, under the supervision of Prof. Theodore S. Rappaport. 
He joined NYU WIRELESS Research Center in 2019 following three years of service as an Engineer for flood early warning systems in different South Asian countries. His research interests include FR1(C), FR3, millimeter-wave and terahertz radio propagation measurement systems, channel modeling, and RFIC design. 
\end{IEEEbiography}

\begin{IEEEbiography}[{\includegraphics[width=1in,height=1.25in,clip,keepaspectratio]{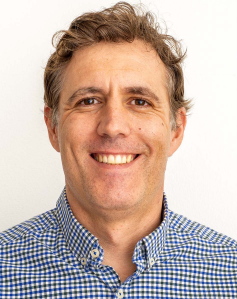}}]{Jose V. Siles} (Senior Member, IEEE) received the Ph.D. degree in electrical engineering from the Technical University of Madrid, Madrid, Spain, in 2008. From 2008 to 2010, he was a Postdoctoral Fellow with the Observatory of Paris-LERMA, Paris, France, participating in several programs funded by the CNES, the European Space Agency, and the European Commission. In 2010, he joined the Submillimeter-Wave Advanced Technology Group, NASA's Jet Propulsion Laboratory, California Institute of Technology, Pasadena, CA, USA, as a Fulbright Postdoctoral Fellow. He is currently the Project Manager and Technical Lead of NASA's ASTHROS mission, which is planned to launch from Antarctica in 2024. His Ph.D. dissertation was focused on the physics-based modeling of semiconductor devices for terahertz applications, and part of this research was performed at the University of Rome ``Tor Vergata," Rome, Italy, and the Observatory of Paris-LERMA. His research interests include the design, development, and test of solid-state power-combined multiplied local oscillator sources and receivers for high-resolution multipixel heterodyne cameras at submillimeter-wave and terahertz frequencies for astrophysics, planetary science, and radar imaging applications. Dr. Siles was the recipient of the Fulbright Postdoctoral Research Award at the NASA Jet Propulsion Laboratory in 2010–2012, the 2012 NASA/JPL Outstanding Postdoctoral Research Award, the 2019 NASA/JPL Lew Allen Award for Excellence, and the U.S. Antarctic Service Medal in 2016.
\end{IEEEbiography}

\begin{IEEEbiography}[{\includegraphics[width=1in,height=1.25in,clip,keepaspectratio]{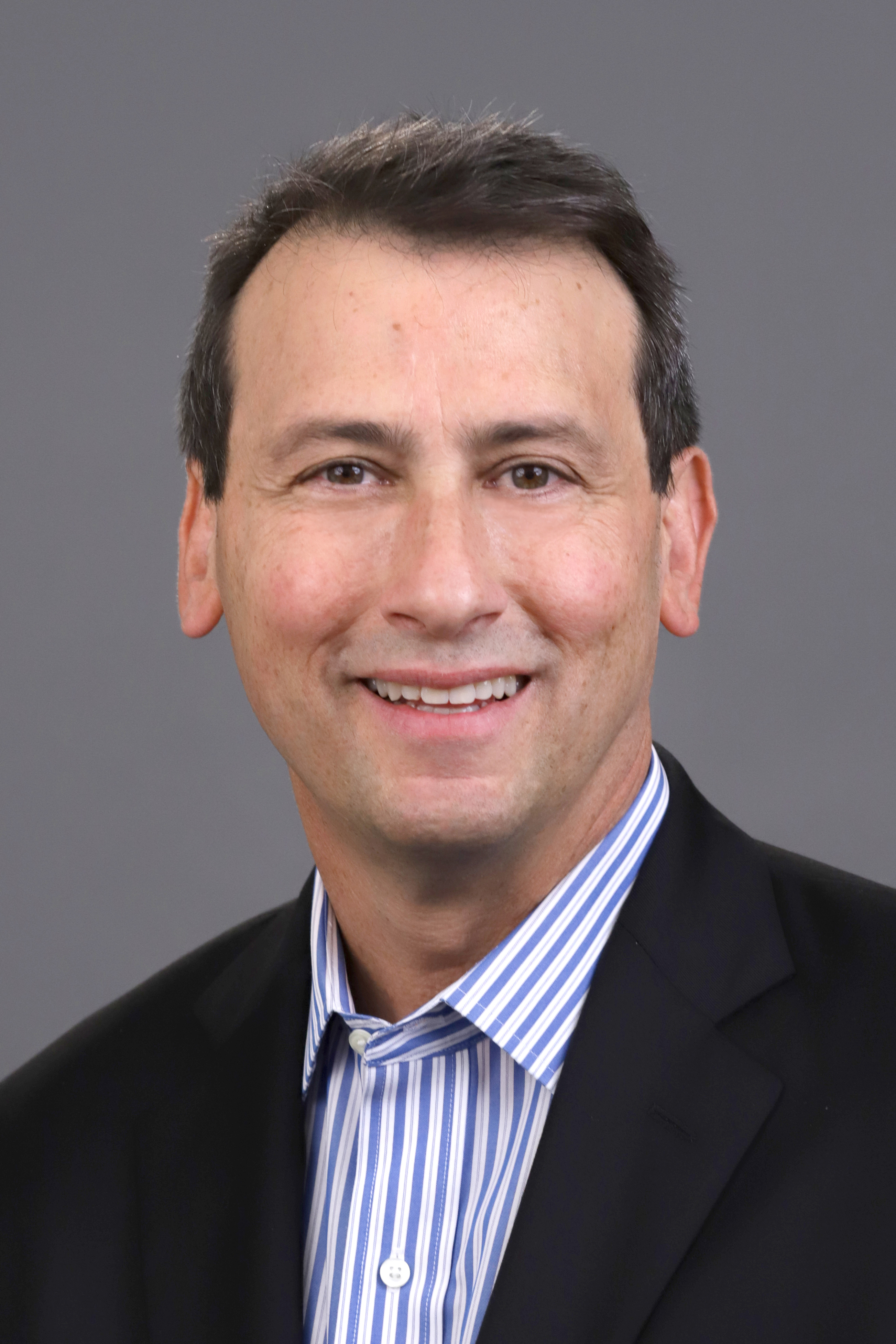}}]{Theodore S. Rappaport} 
(S’83–M’84–SM’91–F’98) is the David Lee/Ernst Weber Professor with New York University (NYU) and holds faculty appointments in the Electrical and Computer Engineering Department of the NYU Tandon School of Engineering, the Courant Computer Science Department, and the NYU Langone School of Medicine. He is the Founding Director of NYU WIRELESS, a multidisciplinary research center focused on the future of wireless communications and applications. His research has led the way for modern wireless communication systems. In 1987, his Ph.D. at Purdue University provided fundamental knowledge of indoor wireless channels used to create the first Wi-Fi standard (IEEE 802.11), and he conducted fundamental work that led to the first US Digital cellphone standards, TDMA IS-54/IS-136, and CDMA IS-95. He and his students engineered the world’s first public Wi-Fi hotspots, and his work proved the viability of millimeter waves for mobile communications. The global wireless industry adopted his vision for 5th generation (5G) millimeter wave cellphone networks. His most recent research has proven the viability of sub-terahertz wireless communications and position location for 6G, 7G and beyond. He founded three academic wireless research centers at Virginia Tech, The University of Texas, and NYU that have produced thousands of engineers and educators since 1990, and he has coauthored over 300 papers and twenty books, including the most cited books on wireless communications, adaptive antennas, wireless simulation, and millimeter wave communications. He co-founded two wireless companies, TSR Technologies and Wireless Valley Communication, which were sold to publicly traded companies, and he has advised many others. He co-founded the Virginia Tech Symposium on Wireless Communications in 1991, the Texas Wireless Summit in 2003, and the Brooklyn 5G Summit (B5GS) in 2014. He has more than 100 patents issued and pending, served on the Technological Advisory Council of the Federal Communications Commission (FCC), is a member of the National Academy of Engineering, is a member of the Wireless Hall of Fame, is a Fellow of the Radio Club of America and the National Academy of Inventors, a life member of the American Radio Relay League, a licensed professional engineer in Texas and Virginia, and an amateur radio operator (N9NB). He has received IEEE’s Eric Sumner Award, ASEE’s Terman Award, The Sir Monty Finniston Medal from the Institution of Engineering and Technology (IET), the IEEE Vehicular Technology Society’s James R. Evans Avant Garde and Stu Meyer Awards, the IEEE Education Society William E. Sayle Award for achievement in education, the IEEE Communications Society Armstrong Award, and the Armstrong Medal and Sarnoff Citation from the Radio Club of America.
\end{IEEEbiography}

% That's all folks
\end{document}